%% file: Paper.tex
\documentclass[a4paper, 10pt]{article}

\usepackage{jheppub}

\usepackage{amsthm}

\usepackage[numbers,sort&compress]{natbib}
\usepackage{dsfont}
\usepackage{slashed}
\usepackage{color}
\usepackage{empheq}

\usepackage{adjustbox}

\usepackage{bibgerm}

\usepackage{soul}

\usepackage[english]{babel}

\usepackage[utf8x]{inputenc}

\usepackage{graphicx,epsfig}
\usepackage{pstricks}

\usepackage{tikz}

\usepackage{bashful}

\usepackage{subfigure}

\usepackage{setspace}

\usepackage{booktabs}

\usepackage{float}

\usepackage{nextpage}

\usepackage{pdfpages}

\usepackage{hypcap}

\usepackage[small,bf]{caption}

\usepackage{longtable}

\usepackage{fancyhdr}

\usepackage[makeroom]{cancel}

\usepackage{microtype}

\usepackage{xparse}
\usepackage{etoolbox}

\usepackage[subfigure]{tocloft}

\usepackage{listings}

\usepackage{enumitem}

\newcommand{\p}{\partial}

\renewcommand{\Im}{\mathrm{Im}}

\renewcommand{\O}{\mathcal{O}}

\newcommand{\tr}{\mathrm{Tr}}

\renewcommand{\L}{\mathcal{L}}

\newcommand{\E}{\mathcal{E}}

\newcommand{\nn}{\nonumber\\ }

\newcommand{\q}{\mathsf{q}}

\newcommand{\msbar}{$\overline{\text{MS}}$}

\newcommand{\hc}{\mathrm{h.c.}}

\NewDocumentCommand{\g}{ O{} }{
	\ifblank{#1}{\bar g}{\bar g_{#1}}
}

\newcommand{\op}[3]{\O^{#2,#3}_{#1}}

\NewDocumentCommand{\Op}{ m m O{} o }{
	\O^{\ifblank{#3}{}{#3,}#2 }_{\IfNoValueTF{#4}{#1}{\substack{#1\\#4}}}
}
\NewDocumentCommand{\EOp}{ m m O{} o }{
	\E^{\ifblank{#3}{}{#3,}#2 }_{\IfNoValueTF{#4}{#1}{\substack{#1\\#4}}}
}
\NewDocumentCommand{\lwc}{ m m O{} o }{
	L^{\ifblank{#3}{}{#3,}#2 }_{\IfNoValueTF{#4}{#1}{\substack{#1\\#4}}}
}
\NewDocumentCommand{\kwc}{ m m O{} o }{
	K^{\ifblank{#3}{}{#3,}#2 }_{\IfNoValueTF{#4}{#1}{\substack{#1\\#4}}}
}
\NewDocumentCommand{\dlwc}{ m m O{} o }{
	{\dot L}^{\ifblank{#3}{}{#3,}#2 }_{\IfNoValueTF{#4}{#1}{\substack{#1\\#4}}}
}
\NewDocumentCommand{\cwc}{ m m O{} o }{
	C^{\ifblank{#3}{}{#3,}#2 }_{\IfNoValueTF{#4}{#1}{\substack{#1\\#4}}}
}

\newcommand{\XXint}[3]{{\setbox0=\hbox{$#1{#2#3}{\int}$}
\vcenter{\hbox{$#2#3$ }}\kern-.65\wd0}}

\definecolor{darkgreen}{rgb}{0,0.5,0}
\definecolor{darkblue}{rgb}{0,0,0.5}
\definecolor{darkred}{rgb}{0.5,0,0}
\definecolor{beige}{rgb}{0.7,0.4,0.3}

\makeatletter
  \def\my@tag@font{\normalsize}
  \def\maketag@@@#1{\hbox{\m@th\normalfont\my@tag@font#1}}
  \let\amsmath@eqref\eqref
  \renewcommand\eqref[1]{{\let\my@tag@font\relax\amsmath@eqref{#1}}}
\makeatother

\newcommand{\roverline}[1]{\mathpalette\doroverline{#1}}
\newcommand{\doroverline}[2]{\overline{#1#2}}

\makeatletter
\renewcommand\paragraph{\@startsection{paragraph}{4}{\z@}%
  {-3.25ex\@plus -1ex \@minus -.2ex}%
  {1.5ex \@plus .2ex}%
  {\normalfont\normalsize\bfseries}}
\makeatother

\cftsetindents{section}{0em}{2em}
\cftsetindents{subsection}{2em}{3em}
\cftsetindents{subsubsection}{5em}{4em}

\allowdisplaybreaks[1]

\preprint{
\mbox{}\hfill{} PSI-PR-23-24 \\
\mbox{}\hfill{} ZU-TH 66/23
}

\title{\boldmath Low-energy effective field theory below the electroweak scale: one-loop renormalization in the 't~Hooft--Veltman scheme}

\author[a,b,c]{Luca Naterop,}

\author[a,b]{Peter Stoffer}

\emailAdd{luca.naterop@physik.uzh.ch}
\emailAdd{stoffer@physik.uzh.ch}

\affiliation[a]{Physik-Institut, Universit\"at Z\"urich, Winterthurerstrasse 190, 8057 Z\"urich, Switzerland}
\affiliation[b]{Paul Scherrer Institut, 5232 Villigen PSI, Switzerland}
\affiliation[c]{Department of Physics, University of California at San Diego, 9500 Gilman Drive, \\ La Jolla, CA 92093-0319, USA}

\abstract{
The low-energy effective field theory below the electroweak scale (LEFT) describes the effects at low energies of both the weak interaction and physics beyond the Standard Model. We study the one-loop renormalization of the LEFT in the 't~Hooft--Veltman scheme, which offers an algebraically consistent definition of the Levi-Civita symbol and $\gamma_5$ in dimensional regularization. However, in connection with minimal subtraction this scheme leads to a spurious breaking of chiral symmetry in intermediate steps of the calculation. Based on the 't~Hooft--Veltman prescription, we define a renormalization scheme that restores chiral symmetry by including appropriate finite counterterms. To this end, we extend the physical LEFT operator basis by a complete set of off-shell and one-loop-evanescent operators and we perform the renormalization at one loop.  We determine the finite counterterms to the physical parameters that compensate both the insertions of evanescent operators, as well as the chiral-symmetry-breaking terms from the renormalizable part of the Lagrangian in $D$ dimensions.
Our results can be applied in next-to-leading-log calculations in the 't~Hooft--Veltman scheme: using our renormalization scheme instead of pure minimal subtraction separates the physical sector from the unphysical evanescent sector and leads to results that are manifestly free of spurious chiral-symmetry-breaking terms.
}

\numberwithin{equation}{section}

\begin{document}

	\maketitle


	\input{sections/Introduction}

	\input{sections/LEFT}

	\input{sections/Scheme}

	\input{sections/Renormalization}

	\input{sections/Results}

	\input{sections/Conclusions}

	
	\section*{Acknowledgements}
	\addcontentsline{toc}{section}{\numberline{}Acknowledgements}

	We thank T.~Engel, B.~Grinstein, S.~Kollatzsch, A.~V.~Manohar, A.~Signer, Y.~Ulrich, and M.~Zoller for valuable discussions and
	A.~V.~Manohar, D.~St\"ockinger, and J.-N.~Toelstede for useful comments on the manuscript.
	Financial support by the Swiss National Science Foundation (Project No.~PCEFP2\_194272) is gratefully acknowledged.

	
	\appendix
	
	\input{sections/Conventions}
	
	\input{sections/ConventionsSupplement}

	\input{sections/RGE}

	\input{sections/LEFT-Operators}

	\input{sections/EOMRedundancies}

	\input{sections/Evanescents}

	\clearpage

	\phantomsection
	\addcontentsline{toc}{section}{\numberline{}References}
	\bibliographystyle{utphysmod}
	\bibliography{Literature}
	
\end{document}

%% file: sections/Introduction.tex

\section{Introduction}

The absence of direct signals of physics beyond the Standard Model (SM) at the LHC has triggered increased activity in the context of indirect precision searches at low energies. The theoretical tool to systematically calculate low-energy quantum effects of heavy new physics are effective field theories (EFTs), which allow one to parametrize deviations from the SM, to combine constraints from different energy regions, and to improve perturbation theory by resumming large logarithms. At the scale of heavy new physics, the EFT can be matched to the UV model of choice that describes the underlying new physics.

Under the assumption of linear realization of the electroweak symmetry, the deviations from the SM in observables above the weak scale are described by the Standard Model Effective Field Theory (SMEFT)~\cite{Buchmuller:1985jz,Grzadkowski:2010es}, which is invariant under the full SM gauge group, see Ref.~\cite{Isidori:2023pyp} for a recent review. For observables below the electroweak scale, the heavy SM particles should first be integrated out: the top quark, the Higgs boson, as well as the electroweak gauge bosons. This results in the low-energy effective field theory below the weak scale (LEFT), which is invariant only under the QCD and QED gauge groups. A complete and non-redundant on-shell basis for the LEFT operators up to dimension six was worked out in Ref.~\cite{Jenkins:2017jig} and later has been extended up to dimension 9~\cite{Liao:2020zyx,Murphy:2020cly,Li:2020tsi}. The LEFT generalizes the well-known Fermi theory of weak interaction, which emerges as a special case if the LEFT is matched to the pure SM at the weak scale. Going beyond the SM, the complete matching of the LEFT to the SMEFT at dimension six was first worked out at tree level~\cite{Jenkins:2017jig} and later extended to one loop~\cite{Dekens:2019ept}, expressing the renormalized LEFT parameters in terms of SMEFT parameters. The dependence of the renormalized parameters on the renormalization scale is described by the renormalization-group equations (RGEs), which at one loop up to dimension six were calculated for the SMEFT in Refs.~\cite{Jenkins:2013zja,Jenkins:2013wua,Alonso:2013hga} and for the LEFT in Ref.~\cite{Jenkins:2017dyc}. Partial results for the RGEs were known previously and have been studied to higher loop orders~\cite{Altarelli:1980fi,Buras:1989xd,Dugan:1990df,Buras:1992tc,Ciuchini:1993vr,Herrlich:1994kh,Buchalla:1995vs,Ciuchini:1997bw,Buras:2000if,Misiak:2004ew,Czakon:2006ss,Cirigliano:2012ab,Dekens:2013zca,Heeck:2013rpa,Pruna:2014asa,Bhattacharya:2015rsa,Aebischer:2015fzz,Davidson:2016edt,Feruglio:2016gvd,Crivellin:2017rmk,Bordone:2017anc,Misiak:2017woa,Cirigliano:2017azj,Fuentes-Martin:2020zaz,Aebischer:2017gaw,Gonzalez-Alonso:2017iyc,Falkowski:2017pss,Panico:2018hal}.

Very strong constraints on the Wilson coefficients in the LEFT can be derived from precision observables at low energies. Of special interest are observables that are either forbidden or at least highly suppressed within the SM. One example are electric dipole moments of elementary or composite particles, which are sensitive probes of $CP$ violation beyond the SM, a necessary ingredient in explanations of the baryon asymmetry of the universe~\cite{Gavela:1993ts,Gavela:1994ds,Gavela:1994dt,Huet:1994jb}. $CP$ violation beyond the SM is described in the LEFT in terms of higher-dimension effective operators that  contain explicit factors of $\gamma_5$ or the Levi-Civita symbol. The definition of these objects cannot be continued analytically in the number of space-time dimensions, which leads to the well-known difficulties with dimensional regularization~\cite{Jegerlehner:2000dz}. The only scheme proven to be consistent to all loop orders is the original 't~Hooft--Veltman (HV) scheme~\cite{tHooft:1972tcz,Breitenlohner:1977hr}. Typically, this scheme leads to spurious symmetry-breaking terms, which can be restored by finite renormalizations. This is mandatory in the case of chiral gauge theories, where the symmetry-breaking terms violate gauge invariance. In vector-like gauge theories, global chiral symmetry is broken by the regulator: this is less severe, since symmetry-breaking terms cancel in relations between observables and do not render the theory inconsistent. However, the Ward identities following from the global symmetry are broken in the modified minimal subtraction (\msbar{}) scheme and are restored only through finite renormalizations.

In this paper, we work out the renormalization of the LEFT at one-loop order in the HV scheme up to dimension six. We extend the physical operator basis by a complete set of evanescent operators that vanish in four space-time dimensions, but are generated at one loop in the HV scheme. We compute the finite counterterms to the physical operator coefficients that compensate both the insertion of evanescent operators in one-loop diagrams, as well as the spurious symmetry-breaking terms generated by the renormalizable part of the regularized Lagrangian. In the LEFT, global chiral symmetry is broken explicitly both by the fermion mass terms as well as higher-dimension effective operators. We disentangle those physical effects from the spurious symmetry-breaking terms due to the regulator by promoting mass matrices and Wilson coefficients to spurions with appropriate chiral transformations. In addition to achieving an effective separation of the physical sector from the unphysical evanescent sector, our renormalization scheme maintains chiral spurion symmetry in one-loop calculations in the LEFT and it allows us to avoid spurious chiral-symmetry-breaking terms, e.g., in one-loop matching calculations, which otherwise only cancel in the final relations between observables. Therefore, this establishes an HV scheme suitable for calculations at next-to-leading-log (NLL) accuracy that incorporates chiral invariance and separates the physical from the evanescent sector.

The article is structured as follows. In Sect.~\ref{sec:LEFT}, we define the Lagrangian of the LEFT, discuss power counting, the appearance of redundant or nuisance operators, as well as the background-field method, which allows us to avoid gauge-variant counterterms. The explicit list of redundant operators is provided in App.~\ref{sec:RedundantOperators}. In Sect.~\ref{sec:SchemeDefinition}, we define our chirally invariant renormalization scheme based on the HV scheme. This involves the complete definition of evanescent operators, provided explicitly in App.~\ref{sec:Evanescents}, as well as the definition of finite renormalizations that compensate the evanescent insertions as well as spurious symmetry-breaking effects. In Sect.~\ref{sec:RenormalizationAndFieldRedefinitions}, we discuss the renormalization procedure as well as the non-linear field redefinitions that allow us to remove on-shell redundant operators. We also discuss the renormalization of the theta terms and check that our calculation in dimensional regularization correctly reproduces the chiral anomaly. In Sect.~\ref{sec:Results}, we discuss our results and the cross-checks that we have performed, before we conclude in Sect.~\ref{sec:Conclusions}. In addition to the operator basis, the appendices summarize our conventions. The explicit results of our calculations consist of very long expressions that are provided as supplementary material.

%% file: sections/LEFT.tex

\section{LEFT}

\label{sec:LEFT}

\subsection{Lagrangian and power counting}

The Lagrangian of the LEFT is given by\footnote{We denote the operator dimension by $d$, while $D$ stands for the number of space-time dimensions.}
\begin{align}
	\label{eq:LEFTLagrangian}
	\L_\mathrm{LEFT} = \L_\mathrm{QCD+QED} + \L_{\nu} + \sum_{d\ge5} \sum_i \lwc{i}{(d)} \O_i^{(d)} \, ,
\end{align}
where the QCD and QED part is defined by
\begin{align}
	\label{eq:qcdqed}
	\L_{\rm QCD + QED} &= - \frac14 G_{\mu \nu}^A G^{A \mu \nu} -\frac14 F_{\mu \nu} F^{\mu\nu} + \theta_{\rm QCD} \frac{g^2}{32 \pi^2} G_{\mu \nu}^A \widetilde G^{A \mu \nu} +  \theta_{\rm QED} \frac{e^2}{32 \pi^2} F_{\mu \nu} \widetilde F^{\mu \nu} \nn
		&\quad + \sum_{\psi=u,d,e}\overline \psi \left( i \slashed D - M_\psi P_L - M_\psi^\dagger P_R \right) \psi \, ,
\end{align}
with covariant derivative $D_\mu = \p_\mu + i g T^A G_\mu^A + i e Q A_\mu$, where $g$ and $e$ are the gauge couplings.\footnote{The $SU(3)$ generator $T^A$ is in the fundamental representation for $u$- and $d$-quarks and zero when acting on the leptons $e$.} The photon and gluon field-strength tensors are
\begin{align}
	F_{\mu\nu} = \p_\mu A_\nu - \p_\nu A_\mu \, , \quad G_{\mu\nu}^A = \p_\mu G_\nu^A - \p_\nu G_\mu^A - g f^{ABC} G_\mu^B G_\nu^C
\end{align}
and the dual field-strength tensors are defined by
\begin{align}
	\widetilde F^{\mu\nu} = \frac{1}{2} \epsilon^{\mu\nu\lambda\sigma} F_{\lambda\sigma} \, , \quad \widetilde G^{A\mu\nu} = \frac{1}{2} \epsilon^{\mu\nu\lambda\sigma} G^A_{\lambda\sigma}
\end{align}
with the Levi-Civita symbol normalized to $\epsilon_{0123} = +1$.
We include only left-handed neutrinos with a lepton-number-violating Majorana mass term\footnote{Since we do not explicitly specify the number of neutrino species, the LEFT also trivially covers the case of additional right-handed neutrinos~\cite{Chala:2020vqp,Li:2020lba}: these can be rewritten in terms of $\nu_{Rp}^c = C \bar\nu_{Rp}^T$, which are  left-chiral fields and can be included in the flavor vector $\nu_L$. This only affects the notion of lepton-number violation, since the charge-conjugated right-handed neutrinos carry lepton number $-1$. We thank A.~V.~Manohar for bringing this to our attention.}
\begin{align}
	\L_\nu = \bar\nu_L i \slashed \p \nu_L - \frac{1}{2} \left( \nu_L^T C M_\nu \nu_L + \bar\nu_L M_\nu^\dagger C \bar\nu_L^T \right) \, .
\end{align}
By introducing the Majorana neutrino
\begin{align}
	\nu_M := \nu_L + C \bar\nu_L^T \, , \quad  \nu_M = C \bar\nu_M^T \, ,
\end{align}
we can rewrite the neutrino Lagrangian as
\begin{align}
	\label{eq:NeutrinoLagrangian}
	\L_\nu = \frac{1}{2} \bar\nu_M \left( i \slashed \p - M_\nu P_L - M_\nu^\dagger P_R \right) \nu_M \, .
\end{align}
The LEFT Lagrangian contains an infinite tower of higher-dimension local operators in addition to the renormalizable Lagrangian of QCD and QED. The complete non-redundant set of gauge-invariant effective operators at dimension five and six was classified in Ref.~\cite{Jenkins:2017jig}. For convenience, we reproduce it in App.~\ref{sec:LEFTBasis}. By now, the operator basis is known up to mass dimension 9~\cite{Liao:2020zyx,Murphy:2020cly,Li:2020tsi}.

The organization of the LEFT Lagrangian in terms of canonical mass dimensions follows from the power counting, which is dictated by the expansion parameter $p/v$ or $m/v$, where $v$ denotes the electroweak scale, $p$ an external momentum, and $m$ a mass of the degrees of freedom retained in the theory. A graph with insertions of effective operators of dimension $d_i \ge 5$ has LEFT dimension
\begin{align}
	d = 4 + \sum_i ( d_i - 4 ) \, .
\end{align}
In the present work, we will consider effects up to dimension six in the power counting, which include single insertions of dimension-six operators, as well as double insertions of dimension-five operators~\cite{Jenkins:2017dyc}. If the LEFT is matched to the SMEFT at the electroweak scale, the SMEFT power counting is inherited, which is an expansion in the small parameter $p/\Lambda$ or $v/\Lambda$, where $\Lambda$ is the scale of new physics. At leading-log accuracy, double-insertions of dimension-five operators are of dimension 8 in the SMEFT power counting, because the tree-level matching only contributes to the dimension-five dipole-operator coefficients with terms of $\O(v/\Lambda^2)$. Therefore, double insertions of dipole operators in the LEFT are of the order
\begin{align}
	\left( \frac{v}{\Lambda^2} \right)^2 = \frac{1}{v^2} \times \frac{v^4}{\Lambda^4} \, ,
\end{align}
where the first factor reflects the LEFT dimension 6 and the second factor shows the SMEFT dimension 8. At NLL, the one-loop matching up to dimension six~\cite{Dekens:2019ept} leads to corrections to the dipole-operator coefficients of the order $\O(m/v^2)$ and $\O(m/\Lambda^2)$. Therefore, also at NLL double-dipole insertions are suppressed beyond dimension 6, either in the SMEFT or the LEFT power counting. In the following we will stay agnostic about the matching at the weak scale and treat dipole-operator coefficients as $\O(1/v)$.

\subsection{Nuisance operators}

The divergences or finite matching contributions encountered in the calculation of off-shell Green's functions do not all have the form of the canonical LEFT operator basis, but they also contain terms corresponding to operators that vanish by the classical equations of motion (EOM). These operators can be removed from the basis with appropriate field redefinitions and hence their contribution to observables are redundant. In general, the renormalization of effective operators involves three different types of counterterms~\cite{Dixon:1974ss,Kluberg-Stern:1975ebk,Joglekar:1975nu,Deans:1978wn,Collins:1984xc}:
\begin{enumerate}
	\item[I.] gauge-invariant operators without ghost fields that do not vanish by the classical EOM,
	\item[IIa.] gauge-invariant ``nuisance operators'' without ghost fields that vanish by the classical EOM,
	\item[IIb.] additional gauge-variant nuisance operators allowed by the solutions of Ward--Slavnov--Taylor identities, which can be constructed as BRST variations of operators with ghost number $-1$.
\end{enumerate}
The operators of class IIb can be avoided by making use of the background-field method~\cite{Abbott:1980hw,Abbott:1983zw}, leaving gauge-invariant EOM operators of class IIa. To linear order in the operator insertions, nuisance operators proportional to the classical EOM do not contribute to $S$-matrix elements. At higher orders in the power counting, multiple insertions of EOM operators can give non-vanishing contributions. In this case, the class-II operators are still redundant, but they should be removed by applying field redefinitions, which lead to shifts in the coefficients of higher-dimension operators.

We choose to work with a redundant set of operators, where the physical LEFT operator basis is extended by operators with additional covariant derivatives. The complete list is provided in App.~\ref{sec:RedundantOperators}. The class-IIa nuisance operators correspond to linear combinations of these derivative operators and the physical operators given in App.~\ref{sec:LEFTBasis}. With this choice of the operator basis, the identification of the divergences is more direct, but the field redefinitions that remove the redundant operators also lead to a shift in the coefficients of physical operators of the same and even lower mass dimension, as will be discussed in Sect.~\ref{sec:FieldRedefinitions}.

\subsection{Background-field method}

We will derive the one-loop counterterms by calculating diagrammatically the one-particle irreducible (1PI) off-shell Green's functions. In order to perform loop calculations in the LEFT, we need to fix the gauge, which in general only leaves BRST invariance as a residual symmetry and requires the introduction of class-IIb counterterms. In order to avoid this complication, it is convenient to employ the background-field method~\cite{Abbott:1980hw,Abbott:1983zw}: all fields are split into the sum of a classical background field $\hat F$ and a quantum field $F$ that is the integration variable in the functional integral. The gauge of the quantum gluon and photon fields are fixed by~\cite{Abbott:1980hw}
\begin{align}
	\L_\mathrm{GF}^\mathrm{QCD+QED} = - \frac{1}{2\xi_g} \left(G^A\right)^2 - \frac{1}{2\xi_\gamma} \left( \p^\mu A_\mu \right)^2 \, , \quad G^A = \p^\mu G_\mu^A - g f^{ABC} \hat G^B_\mu G^{C\mu}
\end{align}
and the corresponding ghost Lagrangian reads
\begin{align}
	\L_\mathrm{FP}^\mathrm{QCD} = - \bar\eta^A \begin{aligned}[t] &\bigg[ \Box \delta^{AB} + g \overleftarrow\p_\mu f^{ACB} ( \hat G^{C\mu} + G^{C\mu} ) \\
		& - g f^{ACB}\hat G_\mu^C \p^\mu + g^2 f^{ACE} f^{EDB} \hat G_\mu^C (\hat G^{D\mu} + G^{D\mu} ) \bigg] \eta^B \, , \end{aligned}
\end{align}
while QED ghosts decouple and can be ignored. Even after fixing the quantum-field gauge, the Lagrangian remains invariant under gauge transformations of the classical background fields. The 1PI Green's functions of background fields are manifestly background-gauge invariant and allow us to determine the counterterms for all gauge-invariant operators. Gauge-variant nuisance operators of class IIb are not required for the renormalization of background-field Green's functions. In order to arrive at the results in terms of the physical LEFT operator basis, the redundant operators need to be removed via field redefinitions. It should be noted that even when using the background-field method off-shell Green's functions are unphysical quantities: although class-IIa counterterms correspond to gauge-invariant operators, in general they can depend on the quantum-gauge parameters.

In practice, employing the background-field gauge in the LEFT merely requires the modification of the three- and four-gluon vertices as in pure QCD~\cite{Abbott:1980hw}. For the fermion and photon fields, no distinction between background and quantum fields is necessary since all vertices are unaltered, including the vertex rules for effective operators.

%% file: sections/Scheme.tex

\section{Scheme definition}
\label{sec:SchemeDefinition}

\subsection{Dimensional regularization}

We use dimensional regularization in $D = 4 - 2 \varepsilon$ space-time dimensions. The LEFT operator basis is defined in terms of chiral fermions. The chiral nature of the electroweak interaction is imprinted in the Wilson coefficients of higher-dimension operators, which lead to Feynman rules involving $\gamma_5$. In addition, the $CP$-violating three-gluon operator involves the Levi-Civita tensor $\epsilon^{\mu\nu\lambda\sigma}$. Both these symbols are intrinsically four-dimensional objects and their treatment in dimensional regularization is notoriously difficult, see Ref.~\cite{Jegerlehner:2000dz} for a review. The only scheme that is proven to be mathematically consistent to higher loop orders is the original scheme by 't~Hooft and Veltman~\cite{tHooft:1972tcz,Breitenlohner:1977hr}. In connection with minimal subtraction and related schemes (such as \msbar{}), the HV scheme leads to a spurious breaking of chiral symmetry, which however can be restored by finite renormalizations, as will be discussed in Sect.~\ref{sec:ChiralSymmetry}.

The HV scheme is defined as follows: the Levi-Civita symbol and $\gamma_5$ are treated as purely four-dimensional objects, see App.~\ref{sec:Conventions}.
The metric tensor $g^{\mu\nu}$ in $D$ dimensions is split into a four-dimensional part $\bar g^{\mu\nu}$ and a part $\hat g^{\mu\nu}$ projecting onto $-2\varepsilon$ dimensions
\begin{align}
	g^{\mu\nu} = \bar g^{\mu\nu} + \hat g^{\mu\nu} \, ,
\end{align}
which satisfy
\begin{align}
	\bar g^\mu{}_\nu \bar g^{\nu\lambda} = \bar g^{\mu\lambda} \, , \quad \hat g^\mu{}_\nu \hat g^{\nu\lambda} = \hat g^{\mu\lambda} \, , \quad \bar g^\mu{}_\nu \hat g^{\nu\lambda} = 0 \, , \quad \bar g^{\mu\nu} \bar g_{\nu\mu} = 4 \, , \quad \hat g^{\mu\nu} \hat g_{\nu\mu} = -2\varepsilon \, .
\end{align}
Projections of $D$-dimensional gamma matrices are defined by the contractions
\begin{align}
	\bar\gamma^\mu = \bar g^{\mu\nu} \gamma_\nu \, , \quad \hat\gamma^\mu = \hat g^{\mu\nu} \gamma_\nu \, ,
\end{align}
and analogous projections are used for arbitrary Lorentz vectors and tensors.

We define the LEFT Lagrangian in $D$ dimensions as follows. The renormalizable part is defined by directly promoting Eqs.~\eqref{eq:qcdqed} and~\eqref{eq:NeutrinoLagrangian} to $D$ dimensions. In particular, the kinetic terms of the gauge fields as well as the fermion gauge-kinetic terms are defined as in Eqs.~\eqref{eq:qcdqed} and~\eqref{eq:NeutrinoLagrangian} with Lorentz indices running over $D$ dimensions, leading to the standard form for propagators and gauge vertices in $D$ space-time dimensions. Importantly, the Lagrangian remains invariant under (background-field) gauge transformations in $D$ dimensions.\footnote{A different convention is possible but it would lead to gauge-symmetry-breaking terms in intermediate steps of the calculation due to a gauge-variant evanescent sector.} Due to the contractions with the Levi-Civita tensor, the Lorentz indices in the theta terms only run over four dimensions.

The higher-dimension operators are defined by keeping the physical operator basis of Ref.~\cite{Jenkins:2017jig} strictly in four space-time dimensions.\footnote{If one would keep all interactions in four space-time dimensions the scheme would have the attractive feature that factorizable graphs do not contribute to RGEs after subtraction of sub-divergences at any loop order~\cite{Jenkins:2023rtg}.} In the case of vector-type four-fermion operators, this convention automatically coincides with the definition of the operators in terms of chiral fields, since
\begin{align}
	\label{eq:VectorProjection}
	P_R \gamma^\mu P_L = \bar\gamma^\mu P_L \, .
\end{align}
In the case of tensor structures, we define the Lorentz indices in physical operators to run only over four dimensions, i.e., we replace the symbol $\sigma^{\mu\nu} = \frac{i}{2} [ \gamma^\mu, \gamma^\nu]$ by its four-dimensional counterpart $\bar\sigma^{\mu\nu} = \frac{i}{2} [ \bar\gamma^\mu, \bar\gamma^\nu]$. Defining operators with tensorial bilinears in terms of chiral fields and $\sigma^{\mu\nu}$ instead of $\bar\sigma^{\mu\nu}$ would differ from our convention by evanescent terms. The same applies to the dimension-six three-gluon operators, where we restrict the summed indices to run only over four dimensions. Our convention for the basis is explicitly given in App.~\ref{sec:LEFTBasis}. We use the same convention for the on-shell redundant operators listed in App.~\ref{sec:RedundantOperators}, i.e., all Lorentz indices only run over four space-time dimensions, as indicated by bars. Note that these scheme definitions are a convention and many different choices are possible. The final result for relations between observables is independent of these choices.

Due to the peculiarities of the HV scheme, we will {\em not} work with \msbar{}, as we will discuss in the following. The complete specification of the scheme incorporates the definition of evanescent operators, which will be given in Sect.~\ref{sec:EvanescentScheme}, as well as the definition of additional finite renormalizations as specified in Sect.~\ref{sec:ChiralSymmetry}.

\subsection{Evanescent operators}
\label{sec:EvanescentScheme}

When calculating loops in the regularized theory, one encounters divergences that correspond to evanescent operators, i.e., operators that vanish when the regulator is removed. In the HV scheme, most of the evanescent operators can be chosen to contain terms explicitly projected onto the evanescent sub-space, e.g., evanescent Dirac matrices $\hat\gamma_\mu$ or in general Lorentz indices summed over $-2\varepsilon$ dimensions. The appearance of evanescent terms has two important consequences. First, their definition is part of the renormalization scheme and affects the physical sector, starting at one loop for the finite terms and at two loops for divergent terms. Second, while tree-level matrix elements of evanescent operators vanish in four space-time dimensions, the insertions of evanescent operators into loop diagrams can lead to a physical effect: traces of terms of rank $-2\varepsilon$ can combine with a $1/\varepsilon$ divergence of a loop integral to give a finite one-loop contribution. Starting at two loops, the divergent parts are also affected. It is desirable to avoid the mixing of unphysical coefficients of evanescent operators into the coefficients of the physical operators. This is achieved by abandoning a naive pure \msbar{} scheme and by performing a finite renormalization of the coefficients of the physical operators that compensates the insertion of evanescent operators~\cite{Buras:1989xd,Dugan:1990df,Herrlich:1994kh}.

A term of the form $\varepsilon \times \O_i$ (with $\O_i$ a physical operator) is evanescent and can be used to modify the basis of evanescent operators and therefore the renormalization scheme~\cite{Herrlich:1994kh}. This does not imply that evanescent operators in general are of $\O(\varepsilon)$ (or $\O(\hbar)$ times a physical operator): since evanescent structures are of rank $-2\varepsilon$, the insertion of two evanescent operators can still lead to a finite physical one-loop effect. Evanescent operators are generated not only by renormalization but also in matching calculations (including scheme changes), where they potentially appear already at tree level~\cite{Aebischer:2022aze,Fuentes-Martin:2022vvu,Dekens:2019ept}. In order to enable a consistent perturbative treatment of evanescent terms, the coefficients of evanescent operators (but not the operators themselves) need to be suppressed by a power counting. This can be the loop expansion or the EFT power counting. In order to be as general as possible and to cover the cases of evanescent operators generated in a tree-level matching, we will assign the coefficients of evanescent operators a suppression by the LEFT power counting.

\subsubsection{Bosonic and fermion-bilinear operators}

We supplement the LEFT Lagrangian by evanescent operators $\E_i^{(d)}$ and label the corresponding coefficients as
\begin{align}
	\L_\mathrm{evan} &= \sum_{d\ge4} \sum_i K_i^{(d)} \E_i^{(d)} \, .
\end{align}
In the case of operators with at most two fermion fields, we provide an exhaustive list of evanescent operators in Tables~\ref{tab:EvanescentOperatorsD4}, \ref{tab:EvanescentOperatorsD5}, and~\ref{tab:EvanescentOperatorsD6} in App.~\ref{sec:Evanescents}. For the renormalization of the physical sector of the LEFT, it is most practical to assign a loop order to the coefficients $K_i^{(d)}$, as the evanescent operators only arise as counterterms to loops with insertions of physical operators. However, having matching calculations in mind that can potentially generate evanescent operators at tree level, we do not assign a loop order to the coefficients but only the LEFT power counting, i.e., we assume $K_i^{(d)} = \O(v^{4-d})$. An exception are the evanescent operators of mass dimension four, listed in Table~\ref{tab:EvanescentOperatorsD4}. A perturbative treatment requires a suppression of their coefficient by some power counting. We assign by hand $K_i^{(4)} = \O(v^{-1})$. This is compatible with the LEFT renormalization, where the $d=4$ evanescent counterterms are generated only if higher-dimension operators are inserted into loop diagrams. We assume that also in a matching calculation, the $d=4$ evanescent operators are generated only with the appropriate power-counting suppression.

With this power-counting assignment, we include all effects up to dimension six in the LEFT expansion, corresponding to $\O(v^{-2})$. We perform a finite renormalization of the coefficients of physical operators that compensates the finite contribution of loop diagrams with insertions of evanescent operators, in particular single insertions of dimension-six evanescent operators as well as single and double insertions of dimension-five and -four evanescent operators. Due to the large number of operators, this results in very long expressions that are provided as supplementary material; see App.~\ref{sec:ConventionsSupplement} for the conventions.

We note that when taking the physical operator basis as the starting point for the renormalization of the LEFT, one can assign a loop factor to the evanescent operator coefficients, hence single insertions of evanescent operators in one-loop diagrams correspond to a two-loop effect, while double insertions would become relevant only at the three-loop level, see App.~\ref{sec:TwoLoopRGEs}. Not all of the operators listed in Tables~\ref{tab:EvanescentOperatorsD4}, \ref{tab:EvanescentOperatorsD5}, and~\ref{tab:EvanescentOperatorsD6} are required independently as counterterms to one-loop insertions of physical operators, e.g., no divergences of the form of $\E_{\gamma'}$ or $\E_{G'}$ are generated~\cite{Belusca-Maito:2021lnk}.

\subsubsection{Evanescent four-fermion operators and Fierz relations}

In loop calculations, we encounter four-fermion structures with higher tensor products of Dirac matrices, which in four space-time dimensions could be reduced to the physical LEFT basis. Since the Dirac algebra in $D = 4-2\varepsilon$ dimensions is infinite dimensional, these tensor products give rise to an infinite set of evanescent operators. The four-fermion structures encountered in the loop calculation can be decomposed as follows. After the loop integration, all remaining contractions are between Dirac matrices in the two different Dirac chains. Dirac matrices in $D$ dimensions are split as $\gamma^\mu = \bar\gamma^\mu + \hat\gamma^\mu$, such that only contractions between two four-dimensional or two $-2\varepsilon$-dimensional Dirac matrices remain. Using the Clifford algebra, the matrices in each Dirac chain can be ordered identically, with evanescent matrices preceding the four-dimensional ones. Tensor products with more than two Dirac matrices belonging to the four-dimensional sub-space can be further simplified by making use of the Chisholm identity
\begin{align}
	\bar\gamma_\mu \bar \gamma_\nu \bar \gamma_\lambda = \bar\gamma_\mu \bar g_{\nu\lambda} - \bar\gamma_\nu \bar g_{\mu\lambda} + \bar\gamma_\lambda \bar g_{\mu\nu} - i \bar\gamma^\sigma \gamma_5 \epsilon_{\mu\nu\lambda\sigma} \, ,
\end{align}
which holds in the HV scheme~\cite{Jamin:1991dp}. Therefore, a basis of four-fermion structures is given by four-dimensional scalar, vector, and $\bar\sigma^{\mu\nu}$-tensor structures, together with the evanescent structures
\begin{align}
	\label{eq:EvanescentFourFermionStructures}
	E^{S(n),LL} &= ( P_L \hat\gamma^{\mu_1} \cdots \hat\gamma^{\mu_n} P_L ) \otimes [ P_L \hat\gamma_{\mu_1} \cdots \hat\gamma_{\mu_n} P_L ] \, , \nn
	E^{S(n),LR} &= ( P_L \hat\gamma^{\mu_1} \cdots \hat\gamma^{\mu_n} P_L ) \otimes [ P_R \hat\gamma_{\mu_1} \cdots \hat\gamma_{\mu_n} P_R ] \, , \nn
	E^{V(n),LL} &= ( P_R \hat\gamma^{\mu_1} \cdots \hat\gamma^{\mu_n} \bar\gamma^\nu P_L ) \otimes [ P_R \hat\gamma_{\mu_1} \cdots \hat\gamma_{\mu_n} \bar\gamma_\nu P_L ] \, , \nn
	E^{V(n),LR} &= ( P_R \hat\gamma^{\mu_1} \cdots \hat\gamma^{\mu_n} \bar\gamma^\nu P_L ) \otimes [ P_L \hat\gamma_{\mu_1} \cdots \hat\gamma_{\mu_n} \bar\gamma_\nu P_R ] \, , \nn
	E^{T(n),LL} &= ( P_L \hat\gamma^{\mu_1} \cdots \hat\gamma^{\mu_n} \bar\sigma^{\nu\lambda} P_L ) \otimes [ P_L \hat\gamma_{\mu_1} \cdots \hat\gamma_{\mu_n}  \bar\sigma_{\nu\lambda}P_L ] \, ,
\end{align}
as well as their parity-conjugated versions. The parentheses and brackets abbreviate Dirac indices on the fermion bilinears. The evanescent four-fermion LEFT operators can then be chosen in analogy to the four-dimensional operators given in App.~\ref{sec:LEFTBasis}, with Dirac structures replaced by the evanescent structures of Eq.~\eqref{eq:EvanescentFourFermionStructures}.

In the case of baryon-number-violating operators, we do not use Eq.~\eqref{eq:EvanescentFourFermionStructures} but instead adopt the convention of Ref.~\cite{Dugan:1990df}, which in addition involves an antisymmetrization of the evanescent Lorentz indices. Without this antisymmetrization, the projection onto the physical operators becomes more cumbersome in the $B$-violating sector, as illustrated by the following relation:
\begin{align}
	\label{eq:EvanescentBviolatingChainAlignment}
	(\psi_p^T C \hat\gamma^\mu \hat\gamma^\nu \psi_r) (\bar \psi_s \hat\gamma_\mu \hat\gamma_\nu \psi_t ) &= -(\psi_r^T C \hat\gamma^\mu \hat\gamma^\nu \psi_p) (\bar \psi_s \hat\gamma_\mu \hat\gamma_\nu \psi_t ) - 4 \varepsilon (\psi_p^T C \psi_r) (\bar \psi_s \psi_t ) \, ,
\end{align}
whereas
\begin{align}
	(\psi_p^T C \hat\gamma^{[\mu} \hat\gamma^{\nu]} \psi_r) (\bar \psi_s \hat\gamma_{[\mu} \hat\gamma_{\nu]} \psi_t ) &= -(\psi_r^T C \hat\gamma^{[\mu} \hat\gamma^{\nu]} \psi_p) (\bar \psi_s \hat\gamma_{[\mu} \hat\gamma_{\nu]} \psi_t ) \, .
\end{align}
With an antisymmetrization of the Lorentz indices, the projection onto the physical sector is simply achieved by dropping structures involving evanescent matrices $\hat\gamma_\mu$, while without antisymmetrization, this procedure is affected by the proper alignment of the $B$-violating fermion chains, taking into account Eq.~\eqref{eq:EvanescentBviolatingChainAlignment}.

In the baryon-number-conserving sector, these complications do not arise and we use directly the evanescent structures~\eqref{eq:EvanescentFourFermionStructures} for the operator basis, without antisymmetrization of the evanescent indices. This choice results in a different scheme compared to applying the convention of Ref.~\cite{Dugan:1990df} also in the $B$-conserving sector, affecting finite one-loop effects and divergences at the two-loop level~\cite{Herrlich:1994kh}.

In the physical LEFT operator basis given in App.~\ref{sec:LEFTBasis}, all redundancies are removed in four space-time dimensions, including Fierz relations between different four-fermion operators~\cite{Jenkins:2017jig}. The Fierz identities are satisfied in four space-time dimensions and in chiral notation they read
\begin{align}
	\label{eq:ChiralFierzIdentitiesD4}
	\left. \begin{aligned}
	( P_R \gamma^\mu P_L ) \otimes [ P_R \gamma_\mu P_L ] &= - ( P_R \gamma^\mu P_L ] \otimes [ P_R \gamma_\mu P_L ) \, , \\
	( P_R \gamma^\mu P_L ) \otimes [ P_L \gamma_\mu P_R ] &= 2 ( P_R ] \otimes [ P_L ) \, , \\
	( P_L \sigma^{\mu\nu} P_L ) \otimes [ P_L \sigma_{\mu\nu} P_L ] &= 8 ( P_L ] \otimes [ P_L ) -  4 ( P_L ) \otimes [ P_L ] \, , \\
	( P_L \sigma^{\mu\nu} P_L ) \otimes [ P_R \sigma_{\mu\nu} P_R ] &= 0 \, 
	\end{aligned} \qquad \right\} \text{ for $D=4$} \, ,
\end{align}
where the minus sign from anticommuting the fermion fields is not included. Analogous relations hold for opposite chirality.
Away from four space-time dimensions, the Fierz identities are not valid, even in the HV scheme. An exception is the last relation, which generalizes to
\begin{align}
	( P_L \bar\sigma^{\mu\nu} P_L ) \otimes [ P_R \bar\sigma_{\mu\nu} P_R ] &= 0 \, ,
\end{align}
and can be derived using
\begin{equation}
	\bar\sigma^{\mu\nu}\gamma_5 = - \frac{i}{2} \epsilon^{\mu\nu\alpha\beta} \bar\sigma_{\alpha\beta} \, .
\end{equation}
The other Fierz relations give rise to additional evanescent structures even in the HV scheme. We define
\begin{align}
	\label{eq:FierzEvanescents}
	( P_R \gamma^\mu P_L ) \otimes [ P_R \gamma_\mu P_L ] &= - ( P_R \gamma^\mu P_L ] \otimes [ P_R \gamma_\mu P_L ) + E_{LL}^{(F1)} \, , \nn
	( P_R \gamma^\mu P_L ) \otimes [ P_L \gamma_\mu P_R ] &= 2 ( P_R ] \otimes [ P_L ) + E_{LR}^{(F1)}  \, , \nn
	( P_L \bar\sigma^{\mu\nu} P_L ) \otimes [ P_L \bar\sigma_{\mu\nu} P_L ] &= 8 ( P_L ] \otimes [ P_L ) -  4 ( P_L ) \otimes [ P_L ] + E_{LL}^{(F2)}
\end{align}
and analogous relations with opposite chirality. Note that in contrast to the NDR definitions in Ref.~\cite{Dekens:2019ept}, all Lorentz indices in Eq.~\eqref{eq:FierzEvanescents} are restricted to the four-dimensional sub-space, even for the vector structures without explicit bars due to Eq.~\eqref{eq:VectorProjection}.

The appearance of Fierz-evanescent operators in Eq.~\eqref{eq:FierzEvanescents} might look surprising at first sight, since the repeated Lorentz indices only run over four dimensions. The reason is that the Fierz identities rely on a finite-dimensional representation of the Dirac algebra. In $D=4-2\varepsilon$ space-time dimensions, the Dirac algebra is infinite dimensional and even in the HV scheme, the Dirac matrices $\bar\gamma^\mu$ cannot be represented as $4\times4$ matrices, see Ref.~\cite{Collins:1984xc} for an explicit construction. The fact that the Fierz-evanescent operators in Eq.~\eqref{eq:FierzEvanescents} do not vanish in the HV scheme can be verified by calculating their insertions into Green's functions and the resulting finite renormalizations of physical operators. We find that these insertions do not lead to finite counterterms to fermion masses or dipole operators, but they generate a non-vanishing finite one-loop effect in four-fermion operators.

The explicit list of evanescent four-fermion operators required at one loop is given in Tables~\ref{tab:EvanescentFourFermionOperators}, \ref{tab:EvanescentFourFermionOperatorsFierz}, \ref{tab:EvanescentFourFermionOperatorsBL}, and~\ref{tab:EvanescentFourFermionOperatorsFierzBL} in App.~\ref{sec:Evanescents}. 

\subsection{Chiral symmetry}
\label{sec:ChiralSymmetry}

In four space-time dimensions, the massless QCD and QED Lagrangian~\eqref{eq:qcdqed} exhibits a symmetry under the global chiral transformation
\begin{align}
	\psi_L &\mapsto U_L^\psi \psi_L \, , \quad \psi_R \mapsto U_R^\psi \psi_R \, , \nn
	\bar\psi_L &\mapsto \bar\psi_L U_L^\psi{}^\dagger \, , \quad \bar\psi_R \mapsto \bar\psi_R  U_R^\psi{}^\dagger \, , \quad \psi = u, d, e \, ,
\end{align}
where $U_L^\psi, U_R^\psi \in U(n_\psi)$ are unitary $n_\psi \times n_\psi$ matrices. At the quantum level, the $U(1)$ axial symmetries are anomalously broken, but the flavor transformations remain a symmetry if the theta angles are transformed simultaneously to compensate the anomalous shift~\cite{Jenkins:2009dy}:
\begin{align}
	\label{eq:AnomalousThetaShifts}
	\theta_\mathrm{QCD} &\mapsto \theta_\mathrm{QCD} + \sum_{\psi=u,d} \arg\det( U_R^\psi{}^\dagger U_L^\psi ) \, , \nn
	\theta_\mathrm{QED} &\mapsto \theta_\mathrm{QED} + \sum_{\psi=u,d} 2 N_c \q_\psi^2 \arg\det( U_R^\psi{}^\dagger U_L^\psi ) + 2 \q_e^2 \arg\det( U_R^e{}^\dagger U_L^e )  \, .
\end{align}
The kinetic term of the neutrino Lagrangian is invariant under the global $U(n_\nu)$ transformation
\begin{align}
	\nu_L \mapsto U_L^\nu \nu_L \, , \quad \bar\nu_L \mapsto \bar\nu_L U_L^\nu{}^\dagger \, .
\end{align}
For non-vanishing mass matrices, chiral symmetry is explicitly broken, but it can be artificially restored by promoting the mass matrices to spurion fields with the chiral transformation
\begin{align}
	M_\psi &\mapsto U_R^\psi M_\psi U_L^\psi{}^\dagger \, , \quad M_\psi^\dagger \mapsto U_L^\psi M_\psi^\dagger U_R^\psi{}^\dagger \, , \quad \psi = u,d,e \, , \nn
	M_\nu &\mapsto U_L^\nu{}^* M_\nu U_L^\nu{}^\dagger \, , \quad M_\nu^\dagger \mapsto U_L^\nu M_\nu^\dagger U_L^\nu{}^T \, .
\end{align}
Similarly, the Lagrangian terms involving higher-dimension operators can be made chirally invariant by also promoting their Wilson coefficients to spurions with appropriate transformations, e.g.,
\begin{align}
	L_{e\gamma} &\mapsto U_L^e L_{e\gamma} U_R^e{}^\dagger \, , \nn
	\lwc{ee}{LL}[V][prst] &\mapsto \lwc{ee}{LL}[V][uvwx] U_{\substack{L \\ pu}}^e U_{\substack{L \\ vr}}^{e\dagger} U_{\substack{L \\ sw}}^e U_{\substack{L \\ xt}}^{e\dagger} \, .
\end{align}

The combined transformation of the chiral fields, theta terms, and spurions is a symmetry of the effective theory and is respected by the perturbative expansion: terms that break this symmetry need to cancel in relations between observables.

Dimensional regularization in the HV scheme leads to a violation of chiral invariance in $D$ dimensions, as is required by a scheme that reproduces the triangle anomaly. Besides the hard chiral anomaly, the HV scheme combined with \msbar{} renormalization also leads to spurious anomalies that break chiral symmetry in the spurion sense defined above. Consider the fermion gauge-kinetic term, written in terms of chiral fields:
\begin{align}
	\label{eq:FermionKineticTermsDDimensions}
	\bar\psi i \slashed D \psi = \bar\psi_L i \bar{\slashed D} \psi_L +  \bar\psi_R i \bar{\slashed D} \psi_R + \bar\psi_L i \hat{\slashed D} \psi_R + \bar\psi_R i \hat{\slashed D} \psi_L \, .
\end{align}
Each term is gauge invariant, but the evanescent contributions $\bar\psi_L i \hat{\slashed D} \psi_R$ and $\bar\psi_R i \hat{\slashed D} \psi_L$ violate chiral symmetry.\footnote{In the case of neutrinos, the evanescent kinetic terms also violate lepton number.} They contribute both to the fermion propagators and the fermion--gauge-boson vertices, resulting in spurious effects that break chiral symmetry. At one loop, contractions of evanescent symmetry-breaking terms can be multiplied by a divergence of a loop integral, resulting in a finite symmetry-breaking contribution. At the two-loop level, they also affect the $1/\varepsilon$ divergences.

As is well known~\cite{tHooft:1972tcz,Breitenlohner:1977hr,Collins:1984xc,Ferrari:1994ct,Martin:1999cc,Grassi:1999tp,Belusca-Maito:2020ala,Belusca-Maito:2021lnk,Cornella:2022hkc,Belusca-Maito:2023wah}, these spurious symmetry-breaking effects can be cured order by order in the perturbative expansion by the addition of appropriate symmetry-restoring counterterms: these finite contributions are local as they come from UV-divergent parts of loop integrals. This is similar to the procedure in chiral gauge theories, where the symmetry-breaking terms violate the Slavnov--Taylor and Ward identities that follow from gauge invariance. There, a restoration of the symmetry is mandatory in order to maintain consistency of the theory. In the present context, the problem is less severe: despite the presence of chiral fields, the LEFT is not a chiral gauge theory. The gauge interactions of the LEFT are the vector-like QCD and QED interactions, i.e., left- and right-chiral fields are in the same representation of the gauge groups. The regulator does not break gauge invariance but only affects global chiral spurion symmetry. Therefore, the restoration of the global symmetry is a pure choice of renormalization scheme and not strictly necessary for a consistent treatment of the theory, since the symmetry-breaking terms cancel in relations between observables. It is still desirable to maintain chiral symmetry in intermediate steps of the calculation in order to avoid an intricate cancellation between matching coefficients, RGEs, and matrix elements. This can be achieved by finite renormalizations of the coefficients of gauge-invariant operators, which compensate the symmetry-breaking terms and restore chiral spurion symmetry.

We define our renormalization scheme as follows: we stay as close to \msbar{} as possible by subtracting the pure $1/\varepsilon$ divergences, as well as the following two finite contributions.
\begin{enumerate}
	\item We perform a finite renormalization of the coefficients of the physical LEFT operators that cancels the explicit insertion of evanescent operators into loop diagrams, as explained in Sect.~\ref{sec:EvanescentScheme}.
	\item We apply an additional finite renormalization of the coefficients of the physical LEFT operators, which only depends on the physical operator coefficients and exactly cancels the spurious terms that break chiral symmetry.
\end{enumerate}
The symmetry-breaking terms are determined as follows. We calculate the one-loop contributions to 1PI background-field Green's functions with physical operator insertions up to finite $\O(\varepsilon^0)$ terms. We then apply a chiral spurion transformation and extract all non-invariant terms. This requires to keep the full flavor structure of the mass matrices, in particular one has to distinguish between $M_\psi$ and $M_\psi^\dagger$. Since the finite symmetry-breaking terms arise from the combination of the evanescent part of the fermion gauge-kinetic terms with a UV divergence, they are local and well-defined even for generic (non-diagonal and non-Hermitian) mass matrices, which can be treated as mass insertions. To this end, one can either directly treat the masses as interaction terms, apply a Taylor expansion to the fermion propagators, or use an exact propagator decomposition. We define the fermion propagator with generic mass matrices as
\begin{align}
	S_\psi(p) &= i \left( \slashed p - M_\psi P_L  - M_\psi^\dagger P_R  \right)^{-1} \, .
\end{align}
Denoting the loop momentum by $\ell$ and an external momentum by $p$, we use a Taylor expansion before integration:
\begin{align}
	\label{eq:TaylorExpansionFermions}
	S_\psi(l+p) & = i \Bigg[1- \frac{\slashed \ell}{\ell^2} \left(-\slashed p + M_\psi P_L + M_\psi^\dagger P_R\right) \Bigg]^{-1} \frac{\slashed \ell}{\ell^2} \nn
	& = i \Bigg[\sum_{k} \left(\frac{\slashed \ell}{\ell^2} \left(-\slashed p + M_\psi P_L + M_\psi^\dagger P_R\right)\right)^k \Bigg] \frac{\slashed \ell}{\ell^2} \, . 
\end{align}
Expanding in both masses and momenta does not affect the UV divergences of loop integrals, but introduces IR divergences and renders all integrals scaleless. However, at one loop it is straightforward to distinguish between UV and IR divergences. As an alternative, we apply the exact propagator decomposition~\cite{Misiak:1994zw,Chetyrkin:1997fm}
\begin{align}
	\label{eq:TadpoleDecompositionFermions}
	S_\psi(\ell+p) &= S_m(\ell) - i S_m(\ell) \left( M_\psi P_L  + M_\psi^\dagger P_R - m  - \slashed p \right) S_\psi(\ell+p) \, ,
\end{align}
where
\begin{align}
	S_m(p) &= i \left( \slashed p - m  \right)^{-1} = \frac{i ( \slashed p + m)}{p^2 - m^2}
\end{align}
depends on an artificial common mass parameter $m$. The decomposition~\eqref{eq:TadpoleDecompositionFermions} and the corresponding version for gauge-boson propagators can be used to reduce all one-loop integrals to massive tadpole integrals plus integrals with a smaller degree of divergence, which after recursive application of the decomposition eventually become finite. This allows one to easily extract the UV-divergent parts of loop integrals for arbitrary mass matrices. The dependence on the artificial mass parameter $m$ drops out of the final results. We checked that we obtain the same results with the Taylor expansion and the tadpole decomposition. When neglecting UV-finite parts of loop integrals, a subtlety arises with the insertion of Fierz-evanescent operators, where one needs to be careful to assign identical momentum routing for the two contributions with different Fierz ordering. 

An improved tadpole decomposition and an efficient algorithm based on Taylor expansion is described Ref.~\cite{Lang:2020nnl}. A slightly different version of the tadpole decomposition has been used in Ref.~\cite{Jenkins:2023rtg}.

%% file: sections/Renormalization.tex

\section{Renormalization and field redefinitions}
\label{sec:RenormalizationAndFieldRedefinitions}

\subsection{Renormalization procedure}
\label{sec:Renormalization}

We start with the bare LEFT Lagrangian, where we extend the basis to include EOM-redundant operators $\O_i^\text{red}$ as well as evanescent operators $\E_i$:
\begin{align}
	\L_\mathrm{LEFT} = \L_\mathrm{QCD+QED} + \L_\nu + \sum_i L_i \O_i + \sum_i L_i^\text{red} \O_i^\text{red} + \sum_i K_i \E_i \, .
\end{align}
We then compute the off-shell 1PI Green's functions of the bare background fields. In order to obtain a finite result, we renormalize the parameters of the Lagrangian
\begin{align}
	\label{eq:Renormalization}
	e &= \mu^\varepsilon Z_e e^r = \mu^\varepsilon ( e^r(\mu) + e^\mathrm{ct} ) \, , \nn
	g &= \mu^\varepsilon Z_g g^r = \mu^\varepsilon ( g^r(\mu) + g^\mathrm{ct} ) \, , \nn
	M_\psi &= M_\psi^r(\mu) + M_\psi^\mathrm{ct} \, , \nn
	X_i &= \mu^{n_i \varepsilon} ( X_i^r(\mu) + X_i^\mathrm{ct} ) \, , \quad X = L, L^\text{red}, K
\end{align}
as well as the background fields:
\begin{align}
	\label{eq:FieldRenormalization}
	A_\mu &= Z_A^{1/2} A_\mu^r \, , \quad
	G_\mu^A = Z_G^{1/2} G_\mu^{A,r} \, , \quad
	\psi_L = Z_{\psi,L}^{1/2} \psi_L^r \, \quad \psi_R = Z_{\psi,R}^{1/2} \psi_R^r \, ,
\end{align}
where $Z_{\psi,L}^{1/2}$ and $Z_{\psi,R}^{1/2}$ are Hermitian matrices in flavor space. The powers $n_i$ of the renormalization scale in Eq.~\eqref{eq:Renormalization} are fixed by requiring that the Wilson coefficients have integer mass dimension even in $D$ space-time dimensions. Schematically, for a term in the Lagrangian
\begin{align}
	\L \supset X_i ( \psi^{N_i^\psi} A_\mu^{N_i^A} \p^{N_i^\p} M^{N_i^M} ) \, ,
\end{align}
the coefficient of the operator has mass dimension
\begin{align}
	[ X_i ] = D - N_i^\psi \frac{D-1}{2} - N_i^A \frac{D-2}{2} - N_i^\p - N_i^M \, ,
\end{align}
hence $n_i = N_i^\psi + N_i^A - 2$. E.g., for the physical operator coefficients up to dimension six, we have
\begin{align}
	[ \lwc{\psi\gamma}{}[][] ] &= [ \lwc{\psi G}{}[][] ] = -1 + \varepsilon \, , \nn
	[ \lwc{G}{}[][] ] &= [ \lwc{\widetilde G}{}[][] ] = -2 + \varepsilon \, , \nn
	[ \lwc{\psi^4}{}[][] ] &= -2 + 2\varepsilon \, .
\end{align}

Since the background-field method preserves gauge invariance, one finds $Z_e Z_A^{1/2} = 1$ and $Z_g Z_G^{1/2} = 1$. There is no need to renormalize the quantum fields~\cite{Abbott:1980hw}. At this stage of the calculation, the 1PI off-shell Green's functions of renormalized background fields are finite. As described in Sects.~\ref{sec:EvanescentScheme} and~\ref{sec:ChiralSymmetry}, we do not use \msbar{}, but apply finite renormalizations that restore spurion chiral symmetry and compensate the physical effect of evanescent-operator insertions. In general, the counterterms can be written as
\begin{align}
	\label{eq:CountertermEpsilonExpansion}
	X_i^\mathrm{ct} &= \sum_{l=1}^\infty \sum_{n=0}^l \frac{1}{\varepsilon^n} \frac{1}{(16\pi^2)^l} X_i^{(l,n)}( e^r(\mu), g^r(\mu), \{M_\psi^r(\mu)\}, \{L_j^r(\mu)\}, \{ L_{j'}^\mathrm{red,r}(\mu)\}, \{ K_k^r(\mu)\} ) \, ,
\end{align}
where $l$ denotes the loop order and with $n=0$ we include finite renormalizations. In the following, we will drop the superscript ${}^r$ for simplicity.

Here, we only consider the one-loop counterterms, which we write as
\begin{align}
	\delta_\mathrm{div}( X_i ) &= \frac{1}{\varepsilon} \frac{1}{16\pi^2} X_i^{(1,1)}( e, g, \{M_\psi\}, \{L_j\}, \{ L_{j'}^\mathrm{red}\}, \{ K_k\} ) \, , \nn
	\delta_\mathrm{fin}^\chi( X_i ) &= \frac{1}{16\pi^2} X_{i,\chi}^{(1,0)}( e, g, \{M_\psi\}, \{L_j\}, \{ L_{j'}^\mathrm{red}\}, \{K_k \} ) \, , \nn
	\delta_\mathrm{fin}^\mathrm{ev}( X_i ) &= \frac{1}{16\pi^2} X_{i,\mathrm{ev}}^{(1,0)}( e, g, \{M_\psi\}, \{L_j\}, \{ L_{j'}^\mathrm{red}\}, \{ K_k\} ) \, ,
\end{align}
where the finite renormalizations $X_i^{(1,0)} = X_{i,\chi}^{(1,0)} + X_{i,\mathrm{ev}}^{(1,0)}$ are split into the terms that restore spurion chiral symmetry and terms that compensate evanescent insertions. Several comments are in order.
\begin{itemize}
	\item As will be discussed in Sect.~\ref{sec:FieldRedefinitions}, the EOM-redundant operators $\O_i^\text{red}$ can be removed by appropriate field redefinitions, which shift all the Wilson coefficients and in particular allow us to set $L_i^\text{red} = 0$. These field redefinitions can be applied order by order in the loop expansion, hence we do not need to insert EOM-redundant operators into loop diagrams and we can drop the dependence of the counterterms on $L_i^\text{red}$.
	\item The divergent physical counterterms do not depend on the coefficients of evanescent operators: one-loop insertions of evanescent operators only generate finite physical effects as well as evanescent divergences. Although these evanescent divergences exist, we do not calculate them here, as they contribute neither at fixed one-loop nor at NLL order, see App.~\ref{sec:TwoLoopRGEs}.
	\item By definition, the terms $\delta_\mathrm{fin}^\chi(X_i)$ that restore chiral symmetry do not depend on evanescent coefficients, which instead contribute to $\delta_\mathrm{fin}^\mathrm{ev}(X_i)$.
	\item Finite renormalizations of evanescent operators are possible but not required. This is a scheme choice that becomes relevant only at higher loop orders~\cite{Belusca-Maito:2020ala}.
\end{itemize}
In summary, in the present work we calculate the following LEFT counterterms:
\begin{align}
	\delta_\mathrm{div}( L_i ) &= \frac{1}{\varepsilon} \frac{1}{16\pi^2} L_i^{(1,1)}( \{L_j\} ) \, , \nn
	\delta_\mathrm{div}( L_i^\mathrm{red} ) &= \frac{1}{\varepsilon} \frac{1}{16\pi^2} L_i^{\mathrm{red},(1,1)}( \{L_j\} ) \, , \nn
	\delta_\mathrm{div}( K_i )\Big|_{\{K_k = 0\}} &= \frac{1}{\varepsilon} \frac{1}{16\pi^2} K_i^{(1,1)}( \{L_j\}, \{ K_k = 0\} ) \, , \nn
	\delta_\mathrm{fin}^\chi( L_i ) &= \frac{1}{16\pi^2} L_{i,\chi}^{(1,0)}( \{L_j\} ) \, , \nn
	\delta_\mathrm{fin}^\mathrm{ev}( L_i ) &= \frac{1}{16\pi^2} L_{i,\mathrm{ev}}^{(1,0)}( \{L_j\}, \{ K_k\} ) \, , \nn
	\delta_\mathrm{fin}^\chi( L_i^\mathrm{red} ) &= \frac{1}{16\pi^2} L_{i,\chi}^{\mathrm{red},(1,0)}( \{L_j\} ) \, , \nn
	\delta_\mathrm{fin}^\mathrm{ev}( L_i^\mathrm{red} ) &= \frac{1}{16\pi^2} L_{i,\mathrm{ev}}^{\mathrm{red},(1,0)}( \{L_j\}, \{ K_k\} ) \, ,
\end{align}
where for simplicity the physical parameters $L_i$ collectively denote couplings, mass matrices, and higher-dimension operator coefficients.

In a final step, we will remove the redundancies in the operator basis by applying field redefinitions, which will shift $L_i^\mathrm{red}$ to zero, but at the same time induce shifts in the coefficients of the physical and evanescent operators.

\subsection{Equations of motion and non-linear field redefinitions}
\label{sec:FieldRedefinitions}

The classical fermion EOM at dimension four are given by~\cite{Jenkins:2017dyc}
\begin{align}
	( i \slashed D - M_\psi P_L - M_\psi^\dagger P_R) \psi &= 0 \, , \quad \bar\psi ( i \overleftarrow{\slashed D} + M_\psi P_L + M_\psi^\dagger P_R) = 0 \, , \quad \psi = u, d, e, \nu_M \, ,
\end{align}
while the gauge-boson EOM read
\begin{align}
	(D_\mu G^{\mu\nu})^A = g j^{A\nu} \, , \quad \p_\mu F^{\mu\nu} = e j_\mathrm{em}^\nu \, ,
\end{align}
where the currents are
\begin{align}
	j^{A\mu} = \sum_{\psi=u,d} \bar\psi \gamma^\mu T^A \psi \, , \quad j_\mathrm{em}^\mu = \sum_{\psi=u,d,e} \q_\psi \, \bar\psi \gamma^\mu \psi \, ,
\end{align}
and the covariant derivative in the adjoint representation is given by $(D_\mu G^{\mu\nu})^A = \p_\mu G^{A\mu\nu} - g f^{ABC} G_\mu^B G^{C\mu\nu}$.

In contrast to a renormalizable field theory, the EFT allows non-linear field redefinitions that respect the power counting. In addition to the linear field redefinitions~\eqref{eq:FieldRenormalization} applied in the renormalization procedure, up to dimension-six effects we can perform the following redefinition of fermion fields~\cite{Dekens:2019ept}:
\begin{align}
	\label{eq:FermionFieldRedefinition}
	\psi_{L,R} &\mapsto \psi_{L,R} + A_{L,R}^\psi \psi_{L,R} + B_{L,R}^\psi i \bar{\slashed D} \psi_{R,L} + C_{L,R}^\psi (i \bar{\slashed D})^2 \psi_{L,R} \nn
		&\quad + D_{L,R}^{\psi\gamma} \bar\sigma^{\mu\nu} F_{\mu\nu} \psi_{L,R} + D_{L,R}^{\psi g} \bar\sigma^{\mu\nu} G_{\mu\nu} \psi_{L,R} \, ,
\end{align}
where $A_{L,R}$, $B_{L,R}$, $C_{L,R}$, and $D_{L,R}^{\gamma,g}$ are generic matrices in flavor space and the term involving the QCD field-strength tensor is only present for quarks, $\psi = u,d$. In the case of neutrinos, the field redefinition reads
\begin{align}
	\label{eq:NeutrinoFieldRedefinition}
	\nu_L &\mapsto \nu_L + A^\nu \nu_L + B^\nu i \bar{\slashed \p} \nu_R + C^\nu (i \bar{\slashed \p})^2 \nu_L + D^{\nu\gamma} \bar\sigma^{\mu\nu} F_{\mu\nu} \nu_L \, , \nn
	\nu_R &\mapsto \nu_R + (A^\nu)^* \nu_R + (B^\nu)^* i \bar{\slashed \p} \nu_L + (C^\nu)^* (i \bar{\slashed \p})^2 \nu_R - (D^{\nu\gamma})^* \bar\sigma^{\mu\nu} F_{\mu\nu} \nu_R \, .
\end{align}
The gauge fields can be redefined up to dimension-six effects according to
\begin{align}
	\label{eq:GaugeFieldRedefinition}
	A_\mu &\mapsto A_\mu + b^\gamma \, \p^\nu F_{\nu\mu} + \sum_{\psi=u,d,e,\nu} \bar\psi_L C_L^{\gamma\psi} \bar\gamma_\mu \psi_L + \sum_{\psi=u,d,e} \bar\psi_R C_R^{\gamma\psi} \bar\gamma_\mu \psi_R \, , \nn
	G_\mu^A &\mapsto G_\mu^A + b^g \, (D^\nu G_{\nu\mu})^A + \sum_{\psi=u,d} \left( \bar\psi_L C_L^{g\psi} \bar\gamma_\mu T^A \psi_L + \bar\psi_R C_R^{g\psi} \bar\gamma_\mu T^A \psi_R \right) \, ,
\end{align}
where $C_{L,R}^{\gamma\psi}$, $C_{L,R}^{g\psi}$ are matrices in flavor space.

These field redefinitions can be used to remove operators proportional to the classical EOM from the EFT basis. In addition to setting the coefficients of EOM operators to zero, the field redefinitions induce changes in the coefficients of operators of even higher dimension. Our basis of on-shell-redundant operators is given in App.~\ref{sec:RedundantOperators}: we keep only the derivative part of the EOM in the redundant operators, hence the field redefinitions that set their coefficients to zero also induce a shift in the physical operators of the same and lower dimensions.

We include only four-dimensional Dirac structures in the field redefinitions~\eqref{eq:FermionFieldRedefinition} and~\eqref{eq:NeutrinoFieldRedefinition}: additional evanescent field redefinitions could be used to remove EOM redundancies in the set of evanescent operators. However, these redundancies are of no consequence since the renormalization scheme prevents a mixing of the evanescent sector into the physical sector. The linear terms $A_{L,R}$ need to be included in Eq.~\eqref{eq:FermionFieldRedefinition} in order to keep the kinetic terms canonically normalized.

The coefficients of the field redefinitions start at the one-loop level and their LEFT power counting is given by
\begin{align}
	\left\{ A_{L,R}^\psi, \, B_{L,R}^\psi \right\} &= \O\left( \frac{1}{v} \right) \, , \quad
	\left\{ C_{L,R}^\psi, \, D_{L,R}^{\psi\gamma}, \, D_{L,R}^{\psi g}, \, b^{\gamma,g}, \, C_{L,R}^{\{\gamma,g\}\psi} \right\} = \O\left( \frac{1}{v^2} \right) \, .
\end{align}

The requirement that the field redefinitions shift the coefficients of the redundant operators to zero does not fix all parameters of the field redefinitions. The combinations
\begin{align}
	\label{eq:UnconstrainedChiralRotations}
	A_L^\psi - A_L^\psi{}^\dagger \, , \quad A_R^\psi - A_R^\psi{}^\dagger \, , \quad B_L^\psi - B_R^\psi{}^\dagger \, , \quad B_R^\psi - B_L^\psi{}^\dagger \, , \quad C_L^\psi - C_L^\psi{}^\dagger \, , \quad C_R^\psi - C_R^\psi{}^\dagger
\end{align}
are additional chiral transformations that can be used, e.g., to change to the basis of mass eigenstates~\cite{Jenkins:2009dy,Jenkins:2017dyc,Dekens:2019ept}. Here, we work in a basis with generic non-diagonal and non-Hermitian mass matrices, but one can always perform chiral field redefinitions and change to a basis with real diagonal mass matrices. Without this basis change, it turns out that even after removing the EOM redundancies at the one-loop level, the finite counterterms $\delta_\mathrm{fin}^\mathrm{ev}(L_i)$ to dipole operators and mass matrices that compensate evanescent-operator insertions depend on the quantum gauge parameters $\xi$ and $\xi_g$. This dependence is related to the ambiguity in the basis reflected by the unconstrained chiral transformations~\eqref{eq:UnconstrainedChiralRotations} and drops out of physical quantities. Therefore, in a final step we perform chiral field redefinitions $A_L^\psi - A_L^\psi{}^\dagger$ and $A_R^\psi - A_R^\psi{}^\dagger$ that consist of two terms, proportional to $\xi-1$ and $\xi_g-1$, respectively, which allow us to remove all gauge-parameter dependences from the one-loop counterterms. This final field redefinition vanishes in Feynman gauge, but the calculation with generic gauge parameters $\xi$ and $\xi_g$ and the cancellation of the gauge-parameter dependence in the final results provide a powerful check. It turns out that the field redefinition that removes the gauge-parameter dependence fulfills
\begin{align}
	\tr[A_L^\psi - A_L^\psi{}^\dagger] = - \tr[A_R^\psi - A_R^\psi{}^\dagger] \, ,
\end{align}
i.e., it contains an axial part that leads to a shift in the theta terms at the two-loop level.

After the field redefinitions, the counterterms that we calculate are reduced to
\begin{align}
	\label{eq:CountertermsAfterFieldRedefinitions}
	\delta_\mathrm{div}( L_i ) &= \frac{1}{\varepsilon} \frac{1}{16\pi^2} L_i^{(1,1)}( \{L_j\} ) \, , \quad
	\delta_\mathrm{div}( K_i )\Big|_{\{K_k = 0\}} = \frac{1}{\varepsilon} \frac{1}{16\pi^2} K_i^{(1,1)}( \{L_j\}, \{ K_k = 0\} ) \, , \nn
	\delta_\mathrm{fin}^\chi( L_i ) &= \frac{1}{16\pi^2} L_{i,\chi}^{(1,0)}( \{L_j\} ) \, , \nn
	\delta_\mathrm{fin}^\mathrm{ev}( L_i ) &= \frac{1}{16\pi^2} L_{i,\mathrm{ev}}^{(1,0)}( \{L_j\}, \{ K_k\} ) \, ,
\end{align}
where $L_i$ again denote couplings, mass matrices, and the coefficients of higher-dimension physical operators. The expressions for the counterterms are very long and we provide the explicit results as supplementary material. The divergent counterterms for the physical parameters are scheme independent and they determine the one-loop RGEs via
\begin{align}
	\label{eq:RGE}
	\dot L_i := 16\pi^2 \frac{d}{d\log\mu} L_i = 2 L_i^{(1,1)} = 32\pi^2 \varepsilon \, \delta_\mathrm{div}(L_i) \, ,
\end{align}
as discussed in App.~\ref{sec:TwoLoopRGEs}.

\subsection{Theta terms and anomalous axial rotations}

Since the theta terms are total derivatives, usually they are do not contribute to perturbative calculations. While the QED theta term plays a role in the presence of magnetic monopoles~\cite{Witten:1979ey}, the QCD theta term is of phenomenological interest due to its non-perturbative contribution to hadronic electric dipole moments (EDMs). The experimental bound on the neutron EDM~\cite{Abel:2020pzs} implies that the effective QCD theta parameter is tiny, which is commonly referred to as the strong $CP$ problem.

In the presence of higher-dimension operators, the QCD and QED theta parameters are renormalized. Even though the theta terms are total derivatives, this renormalization can be calculated in perturbation theory. The RGE of the theta terms was calculated in the SMEFT~\cite{Jenkins:2013zja} and LEFT~\cite{Jenkins:2017dyc}. The theta terms in the LEFT also receive a contribution from the matching to the SMEFT~\cite{Dekens:2019ept}.

We use the method of Ref.~\cite{Georgi:1980cn} to calculate the renormalization of the theta terms in ordinary perturbation theory. To this end, we multiply all sources of $CP$ violation in the LEFT Lagrangian by an artificial parameter $\zeta$ and supplement the Lagrangian by
\begin{align}
	\L_\mathrm{LEFT}(\zeta) \mapsto \L_\mathrm{LEFT}(\zeta) + \frac{\p \L_\mathrm{LEFT}(\zeta)}{\p \zeta} \delta\zeta \, ,
\end{align}
where $\delta\zeta$ is a scalar dummy field. A change in the $CP$-violating sources $\delta\zeta$ induces a change in the theta parameters $\delta\theta = \theta(\zeta) \delta\zeta$. We calculate the $\zeta$-dependent counterterms to these shifts in the theta parameters
\begin{align}
	\theta_{\rm QCD}^\mathrm{ct}(\zeta) \delta\zeta \frac{g^2}{32 \pi^2} G_{\mu \nu}^A \widetilde G^{A \mu \nu} +  \theta_{\rm QED}^\mathrm{ct}(\zeta) \delta\zeta \frac{e^2}{32 \pi^2} F_{\mu \nu} \widetilde F^{\mu \nu}
\end{align}
from the gluon-gluon-$\delta\zeta$ and photon-photon-$\delta\zeta$ three-point functions, which do not vanish in perturbation theory due to the momentum insertion into the dummy field $\delta\zeta$. Finally, the counterterms to the theta parameters are obtained from the counterterms to the shifts $\delta\theta$ via integration:
\begin{align}
	\label{eq:ThetaTermIntegral}
	\theta_\mathrm{QCD}^\mathrm{ct} = \int_0^1 d\zeta \, \theta_{\rm QCD}^\mathrm{ct}(\zeta) \, , \quad \theta_\mathrm{QED}^\mathrm{ct} = \int_0^1 d\zeta \, \theta_{\rm QED}^\mathrm{ct}(\zeta) \, ,
\end{align}
where the initial condition $\theta = 0$ is provided by the $CP$-conserving point $\zeta = 0$.

The calculation can be simplified if we notice that $CP$-even operators do not induce a theta term: therefore, the same result is obtained if all higher-dimension operators are multiplied by the parameter $\zeta$ and not only the $CP$-odd sources, as we explicitly verified. A further simplification is achieved if the $CP$-odd mass terms are not multiplied by $\zeta$: dimensional analysis implies that the $CP$-odd components of the mass matrices enter the renormalization of the theta parameters only in conjunction with coefficients of higher-dimension operators. Hence, multiplying only the effective operators by $\zeta$ leads to the same result as multiplying in addition the $CP$-odd mass term by $\zeta$, because the combinatorial factor arising from the additional couplings of the dummy field $\delta\zeta$ to the mass terms is compensated by the integral~\eqref{eq:ThetaTermIntegral}. Schematically, momentum insertion into $k$ effective operators leads to $k$ terms that add up to
\begin{align}
	k \times \int_0^1 d\zeta \, m^n \zeta^{k-1} L_{i_1} \cdots L_{i_k} = m^n L_{i_1} \cdots L_{i_k} \, ,
\end{align}
whereas if the dummy field also couples to the mass terms, there are $n$ additional topologies with momentum insertion into the masses, leading again to
\begin{align}
	(k+n) \times \int_0^1 d\zeta \, m^n \zeta^{k+n-1} L_{i_1} \cdots L_{i_k} = m^n L_{i_1} \cdots L_{i_k} \, .
\end{align}
We checked explicitly that these different methods all lead to the same result for the renormalization of the theta parameters.

As is well known, the theta parameters are not invariant under axial fermion-field redefinitions but shift due to the chiral anomaly~\cite{Adler:1969gk,Bell:1969ts}. The spurion transformation of the theta parameters~\eqref{eq:AnomalousThetaShifts} compensates this anomalous shift, turning chiral transformations into a spurion symmetry of the theory. The chiral anomaly is often derived in terms of the variation of the measure of the path integral~\cite{Fujikawa:1979ay,Fujikawa:1980eg}. In dimensional regularization, the determinant of the field redefinition is always trivial and the anomaly is instead reproduced in terms of evanescent operators~\cite{tHooft:1972tcz}, see, e.g., Refs.~\cite{Bonneau:1979jx,Bhattacharya:2015rsa}. Our calculation of the counterterms reproduces the anomaly as follows. Under a chiral transformation, the LEFT Lagrangian is invariant if the mass matrices and Wilson coefficients are assigned the spurion transformations of Sect.~\ref{sec:ChiralSymmetry}. If similar spurion transformations are assigned to the coefficients of evanescent operators, this remains true even in $D$ space-time dimensions, with the exception of the evanescent fermion kinetic terms~\eqref{eq:FermionKineticTermsDDimensions} in the Lagrangian, which induce a shift in the dimension-4 evanescent operators $\E_{\psi D}$:
\begin{align}
	\label{eq:EvanescentShift}
	K_{\psi D} \mapsto K_{\psi D} + U_L^\psi{}^\dagger U_R^\psi - 1 \, .
\end{align}
The insertions of the dimension-4 evanescent operators induce a finite contribution to the theta terms, which is compensated by the finite renormalization
\begin{align}
	\delta_\mathrm{fin}^\mathrm{ev}(\theta_\mathrm{QED}) &= \frac{1}{2i} 2 N_\psi \q_\psi^2 \left( \tr[K_{\psi D}] - \tr[K_{\psi D}^\dagger] - \frac{1}{2} \tr[K_{\psi D}K_{\psi D}] + \frac{1}{2} \tr[K_{\psi D}^\dagger K_{\psi D}^\dagger ] + \ldots \right) \, , \nn
	\delta_\mathrm{fin}^\mathrm{ev}(\theta_\mathrm{QCD}) &= \frac{1}{2i} \left( \tr[K_{\psi D}] - \tr[K_{\psi D}^\dagger] - \frac{1}{2} \tr[K_{\psi D}K_{\psi D}] + \frac{1}{2} \tr[K_{\psi D}^\dagger K_{\psi D}^\dagger ] + \ldots \right) \, ,
\end{align}
where $N_{u,d} = N_c$, $N_e = 1$, $\tr$ denotes the trace in flavor space, and in our calculation we only consider single and double insertions of $\E_{\psi D}$. The shift~\eqref{eq:EvanescentShift} in the evanescent coefficient therefore induces a shift in the renormalized theta terms
\begin{align}
	\theta_\mathrm{QED} &\mapsto \theta_\mathrm{QED} - 2 N_\psi \q_\psi^2 \Im \tr\left[ (U_R^\psi{}^\dagger U_L^\psi - 1) - \frac{1}{2} (U_R^\psi{}^\dagger U_L^\psi - 1)^2 + \ldots \right] \, , \nn
	\theta_\mathrm{QCD} &\mapsto \theta_\mathrm{QCD} - \Im \tr\left[ (U_R^\psi{}^\dagger U_L^\psi - 1) - \frac{1}{2} (U_R^\psi{}^\dagger U_L^\psi - 1)^2 + \ldots \right] \, .
\end{align}
Due to
\begin{align}
	\arg\det( U_R^\psi{}^\dagger U_L^\psi ) &= \Im \tr \log( U_R^\psi{}^\dagger U_L^\psi ) = \Im \tr\left[ (U_R^\psi{}^\dagger U_L^\psi - 1) - \frac{1}{2} (U_R^\psi{}^\dagger U_L^\psi - 1)^2 + \ldots \right] \, ,
\end{align}
this is exactly the anomalous shift that gets compensated by the spurion transformation~\eqref{eq:AnomalousThetaShifts}. If we choose a chiral transformation that renders the mass matrices real and diagonal
\begin{align}
	U_R^\psi M_\psi U_L^\psi{}^\dagger = U_L^\psi M_\psi^\dagger U_R^\psi{}^\dagger = M_\psi^\mathrm{diag} \, ,
\end{align}
then in this mass basis the theta terms become
\begin{align}
	\theta_\mathrm{QED}' &= \theta_\mathrm{QED} + 2 N_\psi \q_\psi^2 \arg\det (M_\psi) \, , \nn
	\theta_\mathrm{QCD}' &= \theta_\mathrm{QCD} + \arg\det (M_\psi) \, ,
\end{align}
which are the well-known invariant combinations of theta angles and mass phases.

%% file: sections/Results.tex

\section{Calculation and results}
\label{sec:Results}

Our main result is the complete set of one-loop LEFT counterterms as defined in Eq.~\eqref{eq:CountertermsAfterFieldRedefinitions}. Together with the operator basis in App.~\ref{sec:OperatorBasis} and the scheme definitions in Sect.~\ref{sec:SchemeDefinition}, this establishes a chirally invariant HV scheme for the LEFT at one loop and NLL that separates the physical from the evanescent sector.
In particular, $S$-matrix elements are independent of the coefficients of evanescent operators, i.e., the physical part of one-loop matrix elements computed with the LEFT Lagrangian
\begin{align}
	\label{eq:OneLoopWithEvanescents}
	\L &= \sum_i \mu^{n_i\varepsilon} L_i(\mu) \O_i + \sum_i \mu^{m_i\varepsilon} K_i(\mu) \E_i + \L_\mathrm{ct} \, , \nn
	\L_\mathrm{ct} &= \sum_i \mu^{n_i\varepsilon} \Big[ \delta_\mathrm{div}(L_i) + \delta_\mathrm{fin}^\chi(L_i) + \delta_\mathrm{fin}^\mathrm{ev}(L_i) \Big] \O_i + \sum_i \mu^{m_i\varepsilon} \delta_\mathrm{div}(K_i) \E_i
\end{align}
is independent of $K_i$, where $L_i$ and $K_i$ denote renormalized parameters. The same is true for the two-loop RGEs, see App.~\ref{sec:TwoLoopRGEs}. Up to dimension six, the counterterms $\delta_\mathrm{fin}^\mathrm{ev}(L_i)$ contain terms either linear or quadratic in $K_k$, whereas $\delta_\mathrm{div}(L_i)$ and $\delta_\mathrm{fin}^\chi(L_i)$ are independent of $K_k$. Therefore, at fixed one-loop order one would obtain the same physical results when using the Lagrangian
\begin{align}
	\label{eq:EvanescentFreeOneLoop}
	\L' = \sum_i \mu^{n_i\varepsilon} L_i(\mu) \O_i + \L_\mathrm{ct}' \, , \quad
	\L_\mathrm{ct}' = \sum_i \mu^{n_i\varepsilon} \Big[ \delta_\mathrm{div}(L_i) + \delta_\mathrm{fin}^\chi(L_i) \Big] \O_i \, .
\end{align}
Although $\delta_\mathrm{div}(K_i)|_{K_k=0}$ does not vanish, tree-level insertions of evanescent counterterms do not contribute to physical one-loop matrix elements, hence we dropped evanescent operators altogether in Eq.~\eqref{eq:EvanescentFreeOneLoop}. The renormalized parameters $L_i(\mu)$ in Eq.~\eqref{eq:EvanescentFreeOneLoop} are identical to the ones in Eq.~\eqref{eq:OneLoopWithEvanescents}. When the matching to the LEFT produces tree-level contributions to evanescent operators, it is best to use Eq.~\eqref{eq:OneLoopWithEvanescents} together with the method of regions~\cite{Beneke:1997zp}, as the naive use of Eq.~\eqref{eq:EvanescentFreeOneLoop} with the method of regions would miss the finite renormalization~\cite{Crosas:2023anw}.

In order to obtain the results for the counterterms, we made use of several tools that facilitate the loop calculation: the Feynman diagrams were generated with \texttt{qgraf}~\cite{Nogueira:1991ex} and we evaluated them using our own \texttt{Mathematica} and \texttt{FORM} routines~\cite{Vermaseren:2000nd,Ruijl:2017dtg}. In some intermediate steps of the calculation, we were making use of \texttt{FeynCalc}~\cite{Mertig:1990an,Shtabovenko:2016sxi,Shtabovenko:2016whf,Shtabovenko:2020gxv} and \texttt{\mbox{Package X}}~\cite{Patel:2015tea,Patel:2016fam}.

For the 
loop calculation, we relied on two independent implementations and we performed cross-checks of the final results for the counterterms. We were using generic quantum gauge parameters $\xi$ and $\xi_g$: after the appropriate field redefinitions, the final result is gauge-parameter independent.

We checked that the divergent counterterms for the physical operator coefficients are consistent with the LEFT RGEs of Ref.~\cite{Jenkins:2017dyc} via the relation~\eqref{eq:RGE}.\footnote{Performing these checks, we found some minor mistakes in the results of Ref.~\cite{Jenkins:2017dyc}.} The form of the RGEs obtained here differs from the result of Ref.~\cite{Jenkins:2017dyc} by a chiral field redefinition~\eqref{eq:UnconstrainedChiralRotations} that involves double-dipole insertions and only affects the RGEs of the mass matrices. As explained in Sect.~\ref{sec:FieldRedefinitions}, here we only perform a minimal field redefinition~\eqref{eq:UnconstrainedChiralRotations} that removes the gauge-parameter dependence in the finite counterterms, but our choice for the chiral rotations~\eqref{eq:UnconstrainedChiralRotations} does not contain a divergent part. As in Refs.~\cite{Jenkins:2017dyc,Dekens:2019ept}, additional field redefinitions can always be applied.

The results are provided as supplementary material: we define the notations and conventions in App.~\ref{sec:ConventionsSupplement}.
We give the results for the counterterms only for coefficients of operators listed explicitly in App.~\ref{sec:OperatorBasis}. The complete basis involves further operators that are related by Hermitian conjugation to the given operators, as indicated by ``${}+\hc$'' in the tables. The counterterms for their coefficients directly follow from the requirement that the Lagrangian be Hermitian and hence they can be easily deduced.

A large part of the expressions for the counterterms results from the contribution of double insertions of dimension-five operators, which for most phenomenological applications will be of minor relevance. If double insertions are not needed, the results can be simplified significantly. For convenience, we also provide explicitly the results for the contribution of single-operator insertions.

%% file: sections/Conclusions.tex

\section{Summary and conclusions}
\label{sec:Conclusions}

In this paper, we have introduced a renormalization scheme for the LEFT at one loop and NLL that is based on the original HV scheme, but directly implements the restoration of spurion chiral symmetry, which is broken by the regulator. We have extracted the necessary finite renormalizations by working with generic mass matrices, which together with the Wilson coefficients of higher-dimension operators are promoted to spurions with appropriate chiral transformation. In addition, we have defined a complete set of EOM-redundant operators used in the intermediate steps of the off-shell renormalization, and we have classified the full set of one-loop evanescent operators that are generated by the insertion of physical operators. With the exception of four-fermion operators, we have classified the complete list of evanescent operators in the HV scheme, including the ones that are not required as one-loop counterterms.

In our renormalization scheme, the physical effect of evanescent operators at one loop is compensated by finite counterterms, which decouples the unphysical evanescent sector from the physical one. Furthermore, our scheme has the advantage that spurion chiral symmetry is preserved even in intermediate steps of calculations and is not only restored in relations between observables. Therefore, we expect that our scheme leads to similar results as the NDR scheme in cases where the inconsistencies of the NDR scheme are not visible, differing only by chirally symmetric finite renormalizations. The symmetry-breaking contributions that arise in a pure \msbar{} HV scheme can induce spurious effects, e.g., in one-loop matching calculations. It will be interesting to compare in detail the application of our modified HV scheme to NDR results. We leave this comparison of different schemes for future work. We stress that in contrast to NDR, our scheme based on the HV definition of $\gamma_5$ and the Levi-Civita symbol is algebraically fully consistent. Its application will be of particular interest for calculations in the $CP$-odd sector of the theory, e.g., for matching calculations to schemes amenable to lattice computations in the context of the neutron EDM~\cite{Bhattacharya:2015rsa,Cirigliano:2020msr,Rizik:2020naq,Mereghetti:2021nkt,Buhler:2023gsg,Crosas:2023anw}. However, since we define a scheme for the entire LEFT up to dimension six, we expect that our results are widely applicable for fixed-order one-loop calculations in the LEFT. In addition, they represent another step towards the completion of the EFT framework at next-to-leading-log accuracy.

%% file: sections/Conventions.tex

\section{Conventions}
\label{sec:Conventions}

\subsection{Dirac algebra}

The matrix $\gamma_5$ and the chiral projectors are defined by
\begin{align}
	\gamma_5 := i \gamma^0 \gamma^1 \gamma^2 \gamma^3 = \frac{i}{4!} \epsilon_{\mu\nu\lambda\sigma} \gamma^\mu \gamma^\nu \gamma^\lambda \gamma^\sigma \, , \quad P_L = \frac{1}{2} ( 1 - \gamma_5 ) \, , \quad P_R = \frac{1}{2} ( 1 + \gamma_5 ) \, ,
\end{align}
where the Levi-Civita symbol is normalized to $\epsilon_{0123} = +1$. The matrix $\gamma_5$ fulfills the following (anti-)commutation properties:
\begin{align}
	\{ \gamma_5, \bar\gamma_\mu \} = 0 \, , \quad [ \gamma_5, \hat\gamma_\mu ] = 0 \, .
\end{align}
The charge-conjugation matrix fulfills
\begin{align}
	\label{eq:ChargeConjugationMatrix}
	C \gamma_\mu C^{-1} = - \gamma_\mu^T \, , \quad C = C^* = - C^{-1} = - C^\dagger = - C^T \, ,
\end{align}
which in 4 space-time dimensions is realized by $C = i \gamma^2 \gamma^0$. We take the relations~\eqref{eq:ChargeConjugationMatrix} to be true also in $D=4-2\varepsilon$ space-time dimensions~\cite{Belusca-Maito:2020ala,Belusca-Maito:2023wah}, in particular we use
\begin{align}
	C \bar\gamma_\mu C^{-1} = - \bar\gamma_\mu^T \, , \quad C \hat\gamma_\mu C^{-1} = - \hat\gamma_\mu^T \, .
\end{align}

\subsection{Color algebra}

For $SU(3)_c$, we use Hermitian generators
\begin{align}
	T^A = \frac{\lambda^A}{2} \, , \quad \tr[T^A T^B] = \frac{1}{2} \delta^{AB} \, ,
\end{align}
where $\lambda^A$ are the Gell-Mann matrices. The quadratic Casimir operators in the fundamental and adjoint representations are
\begin{align}
	\label{eq:SUNCasimir}
	T^A_{\alpha\beta} T^A_{\beta\gamma} = C_F \delta_{\alpha\gamma} = \frac{N_c^2 - 1}{2N_c} \delta_{\alpha\gamma} \, , \quad f^{ABC} f^{ABD} = C_A \delta^{CD} = N_c \delta^{CD} \, .
\end{align}

%% file: sections/ConventionsSupplement.tex

\section{Conventions for the supplementary material}
\label{sec:ConventionsSupplement}

The complete results for the divergent and finite counterterms after field redefinitions are provided as supplementary material, which consists of a \texttt{Mathematica} notebook and a subdirectory containing the results in the form of pure text files. The notebook allows one to easily extract selected counterterms. We also provide simplified results restricted to the contribution of only single-operator insertions.

The results are written in the form of \texttt{Mathematica} replacement rules for the counterterms. The symbols appearing in the text files are explained in detail in Table~\ref{tab:CodeVariables}. We are using a compact notation based on matrices in flavor space. The only indices are the open indices of the replacement rule, e.g.,
\begin{lstlisting}[frame=trBL]
\[Delta][KdD2LRDag[fm2_, fm1_]] -> 
	-1/8*(CF*FCHN[{LdGDag, LdGDag}, fm2, fm1])/(EpsilonUV*Pi^2) - 
	FCHN[{Ld\[Gamma]Dag, Ld\[Gamma]Dag}, fm2, fm1]/(8*EpsilonUV*Pi^2)
\end{lstlisting}
translates into
\begin{align}
	\delta_\mathrm{div}(K_{dD2}^{LR}{}^\dagger)_{pr} =  -\frac{1}{8\pi^2\varepsilon} \left[ C_F (L_{dG}^\dagger L_{dG}^\dagger)_{pr} + ( L_{d\gamma}^\dagger L_{d\gamma}^\dagger)_{pr} \right] \, .
\end{align}
For the Wilson coefficients of Hermitian conjugate operators, we follow the convention of Refs.~\cite{Jenkins:2017dyc,Dekens:2019ept}, e.g.,
\begin{align}
	\lwc{e\gamma}{\dagger}[][pr] &:= \Big(\lwc{e\gamma}{}[][rp]\Big)^* \, , \quad
	\lwc{uddu}{RR\dagger}[S1][prst] := \Big(\lwc{uddu}{RR}[S1][rpts]\Big)^* \, .
\end{align}
Furthermore, in order to use a matrix-style notation that avoids sums over repeated flavor indices, we denote rank-4 flavor tensors, e.g., by
\begin{align}
	\mathrm{FFA}(\lwc{ee}{LR}[V])_{pr} \, \mathrm{FFB}(\lwc{ee}{LR}[V])_{st} := \lwc{ee}{LR}[V][prst] \, ,
\end{align}
where the first two flavor indices are attached to the symbol $\mathrm{FFA}$ and the last two indices to the symbol $\mathrm{FFB}$, which always need to appear together in an expression. Therefore, the notation
\begin{lstlisting}[frame=trBL]
\[Delta][LVLLee[fm2_, fm1_, fm4_, fm3_]] -> 
	(e^2*FCHN[{FFB[LVLLee]}, fm4, fm3]*flTr[FFA[LVLLee]]*kd[e, fm2, fm1]*
	q[e]^2)/(96*EpsilonUV*Pi^2) + ...
\end{lstlisting}
corresponds in the index notation of Ref.~\cite{Jenkins:2017dyc} to the expression
\begin{align}
	\delta_\mathrm{div}(\lwc{ee}{LL}[V])_{prst} = \frac{e^2 \q_e^2}{96 \pi^2\varepsilon} \lwc{ee}{LL}[V][wwst] \delta_{pr} + \ldots \, .
\end{align}

\begin{table}[t]
	\centering
	\small
	\begin{tabular}{lllc}
		\toprule
		variable							& code name				& explanation						\\
		\midrule
		\midrule
		$N_c = 3$							& \verb$Nc$				& number of colors \\
		$C_F$							& \verb$CF$				& $SU(3)_c$ fundamental Casimir invariant \\
		\midrule
		$n_e$							& \verb$nf[e]$				& number of charged lepton flavors \\
		$n_u$							& \verb$nf[u]$				& number of up-type quark flavors \\
		$n_d$							& \verb$nf[d]$				& number of down-type quark flavors \\
		\midrule
		$\q_e = -1$						& \verb$q[e]$				& electron charge \\
		$\q_u = 2/3$						& \verb$q[u]$				& up-quark charge \\
		$\q_d = -1/3$						& \verb$q[d]$				& down-quark charge \\
		\midrule
		$e$, $g$							& \verb$e$, \verb$g$				& QED and QCD gauge couplings \\
		$\theta_\mathrm{QED}$, $\theta_\mathrm{QCD}$		& \verb$[\Theta]QED$, \verb$[\Theta]QCD$ 		& QED and QCD theta parameters \\
		\midrule
		$M_\nu, M_\nu^\dagger$				& \verb$M[nu]$, \verb$Mdag[nu]$	& neutrino mass matrix \\
		$M_e, M_e^\dagger$				& \verb$M[e]$, \verb$Mdag[e]$		& charged-lepton mass matrix \\
		$M_u, M_u^\dagger$				& \verb$M[u]$, \verb$Mdag[u]$		& up-quark mass matrix \\
		$M_d, M_d^\dagger$				& \verb$M[d]$, \verb$Mdag[d]$		& down-quark mass matrix \\
		\midrule
		$\varepsilon = (4-D)/2$				& \verb$EpsilonUV$ 			& dimensional regulator \\
		\midrule
		$\delta_{pr}$						& \verb$kd[f,p,r]$			& flavor Kronecker delta for fermion type $f$ \\
		$\tr[A \cdots B]$					& \verb$flTr[A,...,B]$			& trace in flavor space \\
		$(A \cdots B)_{pr}$					& \verb$FCHN[{A,...,B},p,r]$	& flavor chain: element $i,j$ of a product \\
																&&  of flavor-space matrices \\
		$(A)_{pr} (B)_{st}$					& \verb$FCHN2[{A},{B},p,r,s,t]$	& product of two flavor chains \\
		\midrule
		$\delta_\mathrm{div}(\;\cdot\;)$				& \verb$\[Delta][$ $\cdot$ \verb$]$		& divergent counterterms \\
		$\delta_\mathrm{fin}^\chi(\;\cdot\;)$			& \verb$\[Delta]\[Chi][$ $\cdot$ \verb$]$		& finite symmetry-restoring counterterms \\
		$\delta_\mathrm{fin}^\mathrm{ev}(\;\cdot\;)$	& \verb$\[Delta]ev[$ $\cdot$ \verb$]$		& finite evanescent-compensating counterterms \\
		\midrule
		$\lwc{e\gamma}{}$					& \verb$Le\[Gamma]$		& Wilson coefficients \\
		$\lwc{uu}{LR}[V8][]$					& \verb$LV8LRuu$			&  \\
		\ldots \\
		\bottomrule
	\end{tabular}
	\caption{LEFT variables appearing in the code with the one-loop counterterm results, provided as supplementary material.}
	\label{tab:CodeVariables}
\end{table}

%% file: sections/RGE.tex

\clearpage

\section{Renormalization-group equations at two loops}
\label{sec:TwoLoopRGEs}

In the present paper, we compute two types of finite counterterms: the counterterms $\delta^\chi_\mathrm{fin}(L_i)$, which restore spurion chiral symmetry, and the counterterms $\delta^\mathrm{ev}_\mathrm{fin}(L_i)$, which compensate the insertion of evanescent operators. In order to use our scheme, one has to take into account these finite renormalizations in NLL calculations, in particular in the finite parts of one-loop calculations, e.g., in matching calculations or matrix elements. It is well known that the finite renormalizations also affect NLL calculations through the two-loop RGEs~\cite{Buras:1989xd,Dugan:1990df,Herrlich:1994kh,Buras:1998raa,Fuentes-Martin:2022vvu}. Here, we review the derivation and the argument why the scheme of Refs.~\cite{Buras:1989xd,Dugan:1990df,Herrlich:1994kh} avoids a mixing of the coefficients of evanescent operators into the physical sector. We treat the generic case of multiple operator insertions and focus on the present situation in the HV scheme including symmetry-restoring counterterms.

Distinguishing non-evanescent operators $\O_i$ from evanescent operators $\E_i$, we write the Lagrangian in terms of bare parameters schematically as
\begin{align}
	\L = \sum_i L_i \O_i + \sum_i K_i \E_i
\end{align}
and as in Eq.~\eqref{eq:Renormalization} introduce renormalized parameters and counterterms according to
\begin{align}
	\L = \sum_i \mu^{n_i \varepsilon} ( L_i^r(\mu) + L_i^\mathrm{ct} ) \O_i + \sum_i \mu^{m_i \varepsilon} ( K_i^r(\mu) + K_i^\mathrm{ct} ) \E_i \, .
\end{align}
The counterterms are expanded as in Eq.~\eqref{eq:CountertermEpsilonExpansion} into a power series in $1/\varepsilon$
\begin{align}
	X_i^\mathrm{ct} &= \sum_{l=1}^\infty \sum_{n=0}^l \frac{1}{\varepsilon^n} \frac{1}{(16\pi^2)^l} X_i^{(l,n)}(\{L_j^r(\mu)\},\{K_k^r(\mu)\}) \, , \quad X = L, K \, ,
\end{align}
where gauge couplings, masses, and Wilson coefficients are treated on an equal footing. In loops with insertions of evanescent operators, the Dirac algebra produces either evanescent structures or physical structures accompanied by a factor $\varepsilon$. In the HV scheme, this factor can be traced back explicitly to the trace of an evanescent metric tensor $\hat g_{\mu}{}^\mu = - 2\varepsilon$ and it is useful to single it out in the calculation by denoting it by $\hat\varepsilon$, even though $\hat\varepsilon = \varepsilon$. To this end, we split the counterterm coefficients (with some abuse of notation) as follows:
\begin{align}
	L_i^{(l,n)}(\{L_j^r(\mu)\},\{K_k^r(\mu)\}) &= \bar L_i^{(l,n)}(\{L_j^r(\mu)\}) + \frac{\hat\varepsilon}{\varepsilon} \hat L_i^{(l,n)}(\{L_j^r(\mu)\},\{K_k^r(\mu)\}) \, , \nn
	K_i^{(l,n)}(\{L_j^r(\mu)\},\{K_k^r(\mu)\}) &= \bar K_i^{(l,n)}(\{L_j^r(\mu)\}) + \hat K_i^{(l,n)}(\{L_j^r(\mu)\},\{K_k^r(\mu)\}) \, ,
\end{align}
where
\begin{align}
	\bar X_i^{(l,n)}(\{L_j^r(\mu)\}) = X_i^{(l,n)}(\{L_j^r(\mu)\},\{K_k^r(\mu) = 0\}) \, , \quad X = L, K \, .
\end{align}
The strongest singularity of $L_i^\mathrm{ct}$ at each loop level does not depend on the coefficients of evanescent operators, $\hat L^{(l,l)}_i = 0$, or $\p L_i^{(l,l)} / \p K_j^r = 0$.

The RGEs follow from the fact that the bare parameters do not depend on $\mu$:
{\small\begin{align}
	\frac{d L_i^r(\mu) }{d\log\mu} &= - n_i \varepsilon L_i^r(\mu) - \sum_{l=1}^\infty \sum_{n=0}^l \frac{1}{\varepsilon^n} \frac{1}{(16\pi^2)^l} \left( n_i \varepsilon L_i^{(l,n)} + \sum_j  \left[ \frac{\p L_i^{(l,n)}}{\p L_j^r} \frac{d L_j^r(\mu)}{d\log\mu} + \frac{\hat\varepsilon}{\varepsilon} \frac{\p \hat L_i^{(l,n)}}{\p K_j^r} \frac{d K_j^r(\mu)}{d\log\mu} \right] \right) \, , \nn
	\frac{d K_i^r(\mu)}{d\log\mu} &= - m_i \varepsilon K_i^r(\mu) - \sum_{l=1}^\infty \sum_{n=0}^l \frac{1}{\varepsilon^n} \frac{1}{(16\pi^2)^l} \left( m_i \varepsilon K_i^{(l,n)} + \sum_j  \left[ \frac{\p K_i^{(l,n)}}{\p L_j^r} \frac{d L_j^r(\mu)}{d\log\mu} + \frac{\p \hat K_i^{(l,n)}}{\p K_j^r} \frac{d K_j^r(\mu)}{d\log\mu} \right] \right) \, .
\end{align}}%
At one loop accuracy, one obtains
\begin{align}
	\frac{d L_i^r(\mu)}{d\log\mu} = \frac{1}{16\pi^2} \left( -n_i L_i^{(1,1)} + \sum_j n_j \frac{\p L_i^{(1,1)}}{\p L_j^r} L_j^r(\mu) \right) + \O(\text{2-loop}) + \O(\varepsilon) \, .
\end{align}

For a connected graph, the topological identity
\begin{align}
	V - I + l = 1
\end{align}
holds, with $l$ the number of loops, $I$ the number of internal lines, and $V = \sum_k V_k$ the total number of vertices, where $V_k$ is the number of vertices with $k$ legs. The number of external legs is
\begin{align}
	E = \sum_k k V_k - 2 I = \sum_k (k-2) V_k - 2(l-1) \, .
\end{align}
We determine $L_i^{(l,n)}$ from the $l$-loop counterterm to 1PI diagrams with $E = n_i+2$ external legs. The counterterm is a polynomial in all the Lagrangian parameters,
\begin{align}
	L_i^{(l,n)} = \sum_g a_{g,i}^{(l,n)} \prod_{j'} (L_{j'}^r)^{V_{g,j'}} \, ,
\end{align}
where $V_{g,j'}$ is the number of insertions of vertex type $j'$ in graph $g$. Therefore
\begin{align}
	-n_i L_i^{(l,n)} &+ \sum_j  n_j L_j^r(\mu) \frac{\p L_i^{(l,n)}}{\p L_j^r} = -n_i L_i^{(l,n)} + \sum_g a_{g,i}^{(l,n)} \sum_j  n_j V_{g,j} \prod_{j'} (L_{j'}^r)^{V_{g,j'}} \nn
		 &= -n_i L_i^{(l,n)} + \sum_g a_{g,i}^{(l,n)} \sum_j  (k_j-2) V_{g,j} \prod_{j'} (L_{j'}^r)^{V_{g,j'}} \nn
		 &= -n_i L_i^{(l,n)} + (E + 2(l-1)) \sum_g a_{g,i}^{(l,n)} \prod_{j'} (L_{j'}^r)^{V_{g,j'}} = 2 l L_i^{(l,n)} \, ,
\end{align}
hence the one-loop RGE can be extracted from the $\varepsilon^{-1}$ pole of the counterterms as follows:
\begin{align}
	\frac{d L_i^r(\mu)}{d\log\mu} &= 2 \times \frac{\bar L_i^{(1,1)}}{16\pi^2}  + \O(\text{2-loop}) + \O(\varepsilon) \, ,
\end{align}
which is manifestly independent of evanescent coefficients.
Similarly, using the same topological argument, up to two-loop accuracy one obtains the RGE
\begin{align}
	\frac{d L_i^r(\mu)}{d\log\mu} &= \frac{1}{16\pi^2} 2 L_i^{(1,1)} \nn
		&\quad + \frac{1}{(16\pi^2)^2} \left[ 4 L_i^{(2,1)} - \sum_j 2 L_j^{(1,0)} \frac{\p L_i^{(1,1)}}{\p L_j^r} - \sum_j 2 L_j^{(1,1)} \frac{\p L_i^{(1,0)}}{\p L_j^r} - \sum_j 2 K_j^{(1,1)} \frac{\hat\varepsilon}{\varepsilon} \frac{\p \hat L_i^{(1,0)}}{\p K_j^r} \right] \nn
		&\quad + \frac{1}{\varepsilon} \frac{1}{(16\pi^2)^2} \left[ 4 L_i^{(2,2)} - \sum_j 2 L_j^{(1,1)} \frac{\p L_i^{(1,1)}}{\p L_j^r} \right]  + \O(\text{3-loop}) + \O(\varepsilon) \, .
\end{align}
Since the RGE needs to be finite, one obtains the consistency condition
\begin{align}
	\label{eq:TwoLoopDoublePole}
	L_i^{(2,2)} &= \frac{1}{2} \sum_j L_j^{(1,1)} \frac{\p L_i^{(1,1)}}{\p L_j^r} \, ,
\end{align}
which follows from the properties of the loop integrals~\cite{tHooft:1972tcz,tHooft:1973mfk}.
Therefore, the RGE reads
\begin{align}
	\frac{d L_i^r(\mu)}{d\log\mu} &= \frac{1}{16\pi^2} 2 L_i^{(1,1)} \nonumber\\*
		&\quad + \frac{1}{(16\pi^2)^2} \left[ 4 L_i^{(2,1)} - \sum_j 2 L_j^{(1,0)} \frac{\p L_i^{(1,1)}}{\p L_j^r} - \sum_j 2 L_j^{(1,1)} \frac{\p L_i^{(1,0)}}{\p L_j^r} - \sum_j  2 K_j^{(1,1)} \frac{\hat\varepsilon}{\varepsilon} \frac{\p \hat L_i^{(1,0)}}{\p K_j^r} \right] \nn
		&\quad + \O(\text{3-loop}) + \O(\varepsilon) \, .
\end{align}
We have not shown all appearances of $\hat\varepsilon$: the parts of the counterterm coefficients $L_i^{(2,1)}$ and $L_i^{(1,0)}$ that depend on evanescent coefficients contain factors of $\hat\varepsilon/\varepsilon$.

The requirement that the physical RGEs do not depend on the coefficients $K_i^r$ of the evanescent operators imposes a constraint on the finite renormalizations $L_i^{(1,0)}$:
\begin{align}
	\label{eq:TwoLoopRGEConstraint}
	\frac{\hat\varepsilon}{\varepsilon} \frac{\p \hat L_i^{(2,1)}}{\p K_k^r} = \frac{1}{2} \frac{\hat\varepsilon}{\varepsilon} \Biggl[ &\sum_j \frac{\p \hat L_j^{(1,0)}}{\p K_k^r} \frac{\p L_i^{(1,1)}}{\p L_j^r} + \sum_j L_j^{(1,1)} \frac{\p^2 \hat L_i^{(1,0)}}{\p K_k^r \p L_j^r} \nn
		&+ \sum_j \frac{\p K_j^{(1,1)}}{\p K_k^r} \frac{\p \hat L_i^{(1,0)}}{\p K_j^r} + \sum_j K_j^{(1,1)} \frac{\p^2 \hat L_i^{(1,0)}}{\p K_k^r \p K_j^r} \Biggr] \, .
\end{align}
Similarly to Eq.~\eqref{eq:TwoLoopDoublePole}, this follows from the property of the loop integrals~\cite{tHooft:1972tcz,tHooft:1973mfk}, which implies for evanescent insertions
\begin{align}
	\frac{\hat\varepsilon}{\varepsilon} \hat L_i^{(2,1)} = \frac{1}{2} \Biggl[ &\sum_j  \frac{\hat\varepsilon}{\varepsilon} \hat L_j^{(1,0)} \frac{\p L_i^{(1,1)}}{\p L_j^r} + \sum_j L_j^{(1,1)} \frac{\hat\varepsilon}{\varepsilon} \frac{\p \hat L_i^{(1,0)}}{\p L_j^r} + \sum_j K_j^{(1,1)} \frac{\hat\varepsilon}{\varepsilon} \frac{\p \hat L_i^{(1,0)}}{\p K_j^r} \Biggr] \, ,
\end{align}
provided that the finite renormalization $L_i^{(1,0)}$ is chosen to cancel the finite one-loop contribution from the insertion of evanescent operators~\cite{Buras:1989xd,Dugan:1990df,Herrlich:1994kh}. If Eq.~\eqref{eq:TwoLoopRGEConstraint} is not used as a cross check, the RGEs can be simplified to
\begin{align}
	\label{eq:TwoLoopRGEs}
	\frac{d L_i^r(\mu)}{d\log\mu} = \frac{1}{16\pi^2} 2 \bar L_i^{(1,1)} + \frac{1}{(16\pi^2)^2} \Bigg[ 4 \bar L_i^{(2,1)} &- \sum_j 2 \bar L_j^{(1,0)} \frac{\p \bar L_i^{(1,1)}}{\p L_j^r} - \sum_j 2 \bar L_j^{(1,1)} \frac{\p \bar L_i^{(1,0)}}{\p L_j^r} \\
			& - \sum_j  2 \bar K_j^{(1,1)} \frac{\hat\varepsilon}{\varepsilon} \frac{\p \hat L_i^{(1,0)}}{\p K_j^r} \bigg|_{K_k = 0} \Bigg]  + \O(\text{3-loop}) + \O(\varepsilon) \, . \nonumber
\end{align}
As can be seen in Eq.~\eqref{eq:TwoLoopRGEs}, the two-loop RGEs of the physical parameters are directly affected by the divergent counterterms $\delta_\mathrm{div}(K_i)$ of evanescent operator coefficients, the finite counterterms $\delta^\mathrm{ev}_\mathrm{fin}(L_i)$ of physical operator coefficients that compensate evanescent insertions, as well as the additional symmetry-restoring finite counterterms $\delta_\mathrm{fin}^\chi(L_i)$ (or any other additional finite renormalization that one might choose to perform). Double insertions of evanescent operators are not relevant for the two-loop RGEs.

%% file: sections/LEFT-Operators.tex

\clearpage

\section{LEFT operator basis}
\label{sec:OperatorBasis}

\subsection{On-shell basis}
\label{sec:LEFTBasis}

The following list of operators up to dimension six in the LEFT is reproduced from~\cite{Jenkins:2017jig}. We adapt it to the HV scheme in our convention by keeping all operators strictly in four space-time dimensions. As explained in Sect.~\ref{sec:SchemeDefinition}, explicit bars over Dirac matrices in vector bilinears are not needed, since the chiral fields automatically project to four dimensions.

\begin{table}[H]
\capstart
\begin{adjustbox}{width=0.65\textwidth,center}
%
\begin{minipage}[t]{3cm}
\renewcommand{\arraystretch}{1.51}
\small
\begin{align*}
\begin{array}[t]{c|c}
\multicolumn{2}{c}{\boldsymbol{(\nu \nu) X+\hc}} \\
\hline
\O_{\nu \gamma} & (\nu_{Lp}^T C   \bar\sigma^{\mu \nu}  \nu_{Lr})  F_{\mu \nu}  \\
\end{array}
\end{align*}
\end{minipage}
\begin{minipage}[t]{3cm}
\renewcommand{\arraystretch}{1.51}
\small
\begin{align*}
\begin{array}[t]{c|c}
\multicolumn{2}{c}{\boldsymbol{(\overline L R ) X+\hc}} \\
\hline
\O_{e \gamma} & \bar e_{Lp}   \bar\sigma^{\mu \nu} e_{Rr}\, F_{\mu \nu}  \\
\O_{u \gamma} & \bar u_{Lp}   \bar\sigma^{\mu \nu}  u_{Rr}\, F_{\mu \nu}   \\
\O_{d \gamma} & \bar d_{Lp}  \bar\sigma^{\mu \nu} d_{Rr}\, F_{\mu \nu}  \\
\O_{u G} & \bar u_{Lp}   \bar\sigma^{\mu \nu}  T^A u_{Rr}\,  G_{\mu \nu}^A  \\
\O_{d G} & \bar d_{Lp}   \bar\sigma^{\mu \nu} T^A d_{Rr}\,  G_{\mu \nu}^A \\
\end{array}
\end{align*}
\end{minipage}
\begin{minipage}[t]{3cm}
\renewcommand{\arraystretch}{1.51}
\small
\begin{align*}
\begin{array}[t]{c|c}
\multicolumn{2}{c}{\boldsymbol{X^3}} \\
\hline
\O_G     & f^{ABC} \roverline{ G_\mu^{A\nu} G_\nu^{B\rho} G_\rho^{C\mu} } \\
\O_{\widetilde G} & f^{ABC} \roverline{ \widetilde G_\mu^{A\nu} G_\nu^{B\rho} G_\rho^{C\mu} }  \\
\end{array}
\end{align*}
\end{minipage}
\end{adjustbox}
%

%

\begin{adjustbox}{width=1.05\textwidth,center}
\begin{minipage}[t]{3cm}
\renewcommand{\arraystretch}{1.51}
\small
\begin{align*}
\begin{array}[t]{c|c}
\multicolumn{2}{c}{\boldsymbol{(\overline L L)(\overline L  L)}} \\
\hline
\op{\nu\nu}{V}{LL} & (\bar \nu_{Lp}  \gamma^\mu \nu_{Lr} )(\bar \nu_{Ls} \gamma_\mu \nu_{Lt})   \\
\op{ee}{V}{LL}       & (\bar e_{Lp}  \gamma^\mu e_{Lr})(\bar e_{Ls} \gamma_\mu e_{Lt})   \\
\op{\nu e}{V}{LL}       & (\bar \nu_{Lp} \gamma^\mu \nu_{Lr})(\bar e_{Ls}  \gamma_\mu e_{Lt})  \\
\op{\nu u}{V}{LL}       & (\bar \nu_{Lp} \gamma^\mu \nu_{Lr}) (\bar u_{Ls}  \gamma_\mu u_{Lt})  \\
\op{\nu d}{V}{LL}       & (\bar \nu_{Lp} \gamma^\mu \nu_{Lr})(\bar d_{Ls} \gamma_\mu d_{Lt})     \\
\op{eu}{V}{LL}      & (\bar e_{Lp}  \gamma^\mu e_{Lr})(\bar u_{Ls} \gamma_\mu u_{Lt})   \\
\op{ed}{V}{LL}       & (\bar e_{Lp}  \gamma^\mu e_{Lr})(\bar d_{Ls} \gamma_\mu d_{Lt})  \\
\op{\nu edu}{V}{LL}      & (\bar \nu_{Lp} \gamma^\mu e_{Lr}) (\bar d_{Ls} \gamma_\mu u_{Lt})  + \hc   \\
\op{uu}{V}{LL}        & (\bar u_{Lp} \gamma^\mu u_{Lr})(\bar u_{Ls} \gamma_\mu u_{Lt})    \\
\op{dd}{V}{LL}   & (\bar d_{Lp} \gamma^\mu d_{Lr})(\bar d_{Ls} \gamma_\mu d_{Lt})    \\
\op{ud}{V1}{LL}     & (\bar u_{Lp} \gamma^\mu u_{Lr}) (\bar d_{Ls} \gamma_\mu d_{Lt})  \\
\op{ud}{V8}{LL}     & (\bar u_{Lp} \gamma^\mu T^A u_{Lr}) (\bar d_{Ls} \gamma_\mu T^A d_{Lt})   \\[-0.5cm]
\end{array}
\end{align*}
\renewcommand{\arraystretch}{1.51}
\small
\begin{align*}
\begin{array}[t]{c|c}
\multicolumn{2}{c}{\boldsymbol{(\overline R  R)(\overline R R)}} \\
\hline
\op{ee}{V}{RR}     & (\bar e_{Rp} \gamma^\mu e_{Rr})(\bar e_{Rs} \gamma_\mu e_{Rt})  \\
\op{eu}{V}{RR}       & (\bar e_{Rp}  \gamma^\mu e_{Rr})(\bar u_{Rs} \gamma_\mu u_{Rt})   \\
\op{ed}{V}{RR}     & (\bar e_{Rp} \gamma^\mu e_{Rr})  (\bar d_{Rs} \gamma_\mu d_{Rt})   \\
\op{uu}{V}{RR}      & (\bar u_{Rp} \gamma^\mu u_{Rr})(\bar u_{Rs} \gamma_\mu u_{Rt})  \\
\op{dd}{V}{RR}      & (\bar d_{Rp} \gamma^\mu d_{Rr})(\bar d_{Rs} \gamma_\mu d_{Rt})    \\
\op{ud}{V1}{RR}       & (\bar u_{Rp} \gamma^\mu u_{Rr}) (\bar d_{Rs} \gamma_\mu d_{Rt})  \\
\op{ud}{V8}{RR}    & (\bar u_{Rp} \gamma^\mu T^A u_{Rr}) (\bar d_{Rs} \gamma_\mu T^A d_{Rt})  \\
\end{array}
\end{align*}
\end{minipage}
%
%
\begin{minipage}[t]{3cm}
\renewcommand{\arraystretch}{1.51}
\small
\begin{align*}
\begin{array}[t]{c|c}
\multicolumn{2}{c}{\boldsymbol{(\overline L  L)(\overline R  R)}} \\
\hline
\op{\nu e}{V}{LR}     & (\bar \nu_{Lp} \gamma^\mu \nu_{Lr})(\bar e_{Rs}  \gamma_\mu e_{Rt})  \\
\op{ee}{V}{LR}       & (\bar e_{Lp}  \gamma^\mu e_{Lr})(\bar e_{Rs} \gamma_\mu e_{Rt}) \\
\op{\nu u}{V}{LR}         & (\bar \nu_{Lp} \gamma^\mu \nu_{Lr})(\bar u_{Rs}  \gamma_\mu u_{Rt})    \\
\op{\nu d}{V}{LR}         & (\bar \nu_{Lp} \gamma^\mu \nu_{Lr})(\bar d_{Rs} \gamma_\mu d_{Rt})   \\
\op{eu}{V}{LR}        & (\bar e_{Lp}  \gamma^\mu e_{Lr})(\bar u_{Rs} \gamma_\mu u_{Rt})   \\
\op{ed}{V}{LR}        & (\bar e_{Lp}  \gamma^\mu e_{Lr})(\bar d_{Rs} \gamma_\mu d_{Rt})   \\
\op{ue}{V}{LR}        & (\bar u_{Lp} \gamma^\mu u_{Lr})(\bar e_{Rs}  \gamma_\mu e_{Rt})   \\
\op{de}{V}{LR}         & (\bar d_{Lp} \gamma^\mu d_{Lr}) (\bar e_{Rs} \gamma_\mu e_{Rt})   \\
\op{\nu edu}{V}{LR}        & (\bar \nu_{Lp} \gamma^\mu e_{Lr})(\bar d_{Rs} \gamma_\mu u_{Rt})  +\hc \\
\op{uu}{V1}{LR}        & (\bar u_{Lp} \gamma^\mu u_{Lr})(\bar u_{Rs} \gamma_\mu u_{Rt})   \\
\op{uu}{V8}{LR}       & (\bar u_{Lp} \gamma^\mu T^A u_{Lr})(\bar u_{Rs} \gamma_\mu T^A u_{Rt})    \\ 
\op{ud}{V1}{LR}       & (\bar u_{Lp} \gamma^\mu u_{Lr}) (\bar d_{Rs} \gamma_\mu d_{Rt})  \\
\op{ud}{V8}{LR}       & (\bar u_{Lp} \gamma^\mu T^A u_{Lr})  (\bar d_{Rs} \gamma_\mu T^A d_{Rt})  \\
\op{du}{V1}{LR}       & (\bar d_{Lp} \gamma^\mu d_{Lr})(\bar u_{Rs} \gamma_\mu u_{Rt})   \\
\op{du}{V8}{LR}       & (\bar d_{Lp} \gamma^\mu T^A d_{Lr})(\bar u_{Rs} \gamma_\mu T^A u_{Rt}) \\
\op{dd}{V1}{LR}      & (\bar d_{Lp} \gamma^\mu d_{Lr})(\bar d_{Rs} \gamma_\mu d_{Rt})  \\
\op{dd}{V8}{LR}   & (\bar d_{Lp} \gamma^\mu T^A d_{Lr})(\bar d_{Rs} \gamma_\mu T^A d_{Rt}) \\
\op{uddu}{V1}{LR}   & (\bar u_{Lp} \gamma^\mu d_{Lr})(\bar d_{Rs} \gamma_\mu u_{Rt})  + \hc  \\
\op{uddu}{V8}{LR}      & (\bar u_{Lp} \gamma^\mu T^A d_{Lr})(\bar d_{Rs} \gamma_\mu T^A  u_{Rt})  + \hc \\
\end{array}
\end{align*}
\end{minipage}

\begin{minipage}[t]{3cm}
\renewcommand{\arraystretch}{1.51}
\small
\begin{align*}
\begin{array}[t]{c|c}
\multicolumn{2}{c}{\boldsymbol{(\overline L R)(\overline L R)+\hc}} \\
\hline
\op{ee}{S}{RR} 		& (\bar e_{Lp}   e_{Rr}) (\bar e_{Ls} e_{Rt})   \\
\op{eu}{S}{RR}  & (\bar e_{Lp}   e_{Rr}) (\bar u_{Ls} u_{Rt})   \\
\op{eu}{T}{RR} & (\bar e_{Lp}   \bar\sigma^{\mu \nu}   e_{Rr}) (\bar u_{Ls}  \bar\sigma_{\mu \nu}  u_{Rt})  \\
\op{ed}{S}{RR}  & (\bar e_{Lp} e_{Rr})(\bar d_{Ls} d_{Rt})  \\
\op{ed}{T}{RR} & (\bar e_{Lp} \bar\sigma^{\mu \nu} e_{Rr}) (\bar d_{Ls} \bar\sigma_{\mu \nu} d_{Rt})   \\
\op{\nu edu}{S}{RR} & (\bar   \nu_{Lp} e_{Rr})  (\bar d_{Ls} u_{Rt} ) \\
\op{\nu edu}{T}{RR} &  (\bar  \nu_{Lp}  \bar\sigma^{\mu \nu} e_{Rr} )  (\bar  d_{Ls}  \bar\sigma_{\mu \nu} u_{Rt} )   \\
\op{uu}{S1}{RR}  & (\bar u_{Lp}   u_{Rr}) (\bar u_{Ls} u_{Rt})  \\
\op{uu}{S8}{RR}   & (\bar u_{Lp}   T^A u_{Rr}) (\bar u_{Ls} T^A u_{Rt})  \\
\op{ud}{S1}{RR}   & (\bar u_{Lp} u_{Rr})  (\bar d_{Ls} d_{Rt})   \\
\op{ud}{S8}{RR}  & (\bar u_{Lp} T^A u_{Rr})  (\bar d_{Ls} T^A d_{Rt})  \\
\op{dd}{S1}{RR}   & (\bar d_{Lp} d_{Rr}) (\bar d_{Ls} d_{Rt}) \\
\op{dd}{S8}{RR}  & (\bar d_{Lp} T^A d_{Rr}) (\bar d_{Ls} T^A d_{Rt})  \\
\op{uddu}{S1}{RR} &  (\bar u_{Lp} d_{Rr}) (\bar d_{Ls}  u_{Rt})   \\
\op{uddu}{S8}{RR}  &  (\bar u_{Lp} T^A d_{Rr}) (\bar d_{Ls}  T^A u_{Rt})  \\[-0.5cm]
\end{array}
\end{align*}
\renewcommand{\arraystretch}{1.51}
\small
\begin{align*}
\begin{array}[t]{c|c}
\multicolumn{2}{c}{\boldsymbol{(\overline L R)(\overline R L) +\hc}} \\
\hline
\op{eu}{S}{RL}  & (\bar e_{Lp} e_{Rr}) (\bar u_{Rs}  u_{Lt})  \\
\op{ed}{S}{RL} & (\bar e_{Lp} e_{Rr}) (\bar d_{Rs} d_{Lt}) \\
\op{\nu edu}{S}{RL}  & (\bar \nu_{Lp} e_{Rr}) (\bar d_{Rs}  u_{Lt})  \\
\end{array}
\end{align*}
\end{minipage}
\end{adjustbox}
\setlength{\belowcaptionskip}{-3cm}
\caption{LEFT operators of dimension 
five, as well as LEFT operators of dimension six that conserve baryon and lepton number.}
\label{tab:oplist1}
\end{table}

\begin{table}[H]
\capstart
%
\centering
\begin{minipage}[t]{3cm}
\renewcommand{\arraystretch}{1.5}
\small
\begin{align*}
\begin{array}[t]{c|c}
\multicolumn{2}{c}{\boldsymbol{\Delta L = 4 + \hc}}  \\
\hline
\op{\nu\nu}{S}{LL} &  (\nu_{Lp}^T C \nu_{Lr}^{}) (\nu_{Ls}^T C \nu_{Lt}^{} )  \\
\end{array}
\end{align*}
\end{minipage}

\begin{adjustbox}{width=\textwidth,center}
\begin{minipage}[t]{3cm}
\renewcommand{\arraystretch}{1.5}
\small
\begin{align*}
\begin{array}[t]{c|c}
\multicolumn{2}{c}{\boldsymbol{\Delta L =2 + \hc}}  \\
\hline
\op{\nu e}{S}{LL}  &  (\nu_{Lp}^T C \nu_{Lr}) (\bar e_{Rs} e_{Lt})   \\
\op{\nu e}{T}{LL} &  (\nu_{Lp}^T C \bar\sigma^{\mu \nu} \nu_{Lr}) (\bar e_{Rs}\bar\sigma_{\mu \nu} e_{Lt} )  \\
\op{\nu e}{S}{LR} &  (\nu_{Lp}^T C \nu_{Lr}) (\bar e_{Ls} e_{Rt} )  \\
\op{\nu u}{S}{LL}  &  (\nu_{Lp}^T C \nu_{Lr}) (\bar u_{Rs} u_{Lt} )  \\
\op{\nu u}{T}{LL}  &  (\nu_{Lp}^T C \bar\sigma^{\mu \nu} \nu_{Lr}) (\bar u_{Rs} \bar\sigma_{\mu \nu} u_{Lt} ) \\
\op{\nu u}{S}{LR}  &  (\nu_{Lp}^T C \nu_{Lr}) (\bar u_{Ls} u_{Rt} )  \\
\op{\nu d}{S}{LL}   &  (\nu_{Lp}^T C \nu_{Lr}) (\bar d_{Rs} d_{Lt} ) \\
\op{\nu d}{T}{LL}   &  (\nu_{Lp}^T C \bar\sigma^{\mu \nu}  \nu_{Lr}) (\bar d_{Rs} \bar\sigma_{\mu \nu} d_{Lt} ) \\
\op{\nu d}{S}{LR}  &  (\nu_{Lp}^T C \nu_{Lr}) (\bar d_{Ls} d_{Rt} ) \\
\op{\nu edu}{S}{LL} &  (\nu_{Lp}^T C e_{Lr}) (\bar d_{Rs} u_{Lt} )  \\
\op{\nu edu}{T}{LL}  & (\nu_{Lp}^T C  \bar\sigma^{\mu \nu} e_{Lr}) (\bar d_{Rs}  \bar\sigma_{\mu \nu} u_{Lt} ) \\
\op{\nu edu}{S}{LR}   & (\nu_{Lp}^T C e_{Lr}) (\bar d_{Ls} u_{Rt} ) \\
\op{\nu edu}{V}{RL}   & (\nu_{Lp}^T C \gamma^\mu e_{Rr}) (\bar d_{Ls} \gamma_\mu u_{Lt} )  \\
\op{\nu edu}{V}{RR}   & (\nu_{Lp}^T C \gamma^\mu e_{Rr}) (\bar d_{Rs} \gamma_\mu u_{Rt} )  \\
\end{array}
\end{align*}
\end{minipage}
%
\begin{minipage}[t]{3cm}
\renewcommand{\arraystretch}{1.5}
\small
\begin{align*}
\begin{array}[t]{c|c}
\multicolumn{2}{c}{\boldsymbol{\Delta B = \Delta L = 1 + \hc}} \\
\hline
\op{udd}{S}{LL} &  \epsilon_{\alpha\beta\gamma}  (u_{Lp}^{\alpha T} C d_{Lr}^{\beta}) (d_{Ls}^{\gamma T} C \nu_{Lt}^{})   \\
\op{duu}{S}{LL} & \epsilon_{\alpha\beta\gamma}  (d_{Lp}^{\alpha T} C u_{Lr}^{\beta}) (u_{Ls}^{\gamma T} C e_{Lt}^{})  \\
\op{uud}{S}{LR} & \epsilon_{\alpha\beta\gamma}  (u_{Lp}^{\alpha T} C u_{Lr}^{\beta}) (d_{Rs}^{\gamma T} C e_{Rt}^{})  \\
\op{duu}{S}{LR} & \epsilon_{\alpha\beta\gamma}  (d_{Lp}^{\alpha T} C u_{Lr}^{\beta}) (u_{Rs}^{\gamma T} C e_{Rt}^{})   \\
\op{uud}{S}{RL} & \epsilon_{\alpha\beta\gamma}  (u_{Rp}^{\alpha T} C u_{Rr}^{\beta}) (d_{Ls}^{\gamma T} C e_{Lt}^{})   \\
\op{duu}{S}{RL} & \epsilon_{\alpha\beta\gamma}  (d_{Rp}^{\alpha T} C u_{Rr}^{\beta}) (u_{Ls}^{\gamma T} C e_{Lt}^{})   \\
\op{dud}{S}{RL} & \epsilon_{\alpha\beta\gamma}  (d_{Rp}^{\alpha T} C u_{Rr}^{\beta}) (d_{Ls}^{\gamma T} C \nu_{Lt}^{})   \\
\op{ddu}{S}{RL} & \epsilon_{\alpha\beta\gamma}  (d_{Rp}^{\alpha T} C d_{Rr}^{\beta}) (u_{Ls}^{\gamma T} C \nu_{Lt}^{})   \\
\op{duu}{S}{RR}  & \epsilon_{\alpha\beta\gamma}  (d_{Rp}^{\alpha T} C u_{Rr}^{\beta}) (u_{Rs}^{\gamma T} C e_{Rt}^{})  \\
\end{array}
\end{align*}
\end{minipage}
%
\begin{minipage}[t]{3cm}
\renewcommand{\arraystretch}{1.5}
\small
\begin{align*}
\begin{array}[t]{c|c}
\multicolumn{2}{c}{\boldsymbol{\Delta B = - \Delta L = 1 + \hc}}  \\
\hline
\op{ddd}{S}{LL} & \epsilon_{\alpha\beta\gamma}  (d_{Lp}^{\alpha T} C d_{Lr}^{\beta}) (\bar e_{Rs}^{} d_{Lt}^\gamma )  \\
\op{udd}{S}{LR}  & \epsilon_{\alpha\beta\gamma}  (u_{Lp}^{\alpha T} C d_{Lr}^{\beta}) (\bar \nu_{Ls}^{} d_{Rt}^\gamma )  \\
\op{ddu}{S}{LR} & \epsilon_{\alpha\beta\gamma}  (d_{Lp}^{\alpha T} C d_{Lr}^{\beta})  (\bar \nu_{Ls}^{} u_{Rt}^\gamma )  \\
\op{ddd}{S}{LR} & \epsilon_{\alpha\beta\gamma}  (d_{Lp}^{\alpha T} C d_{Lr}^{\beta}) (\bar e_{Ls}^{} d_{Rt}^\gamma ) \\
\op{ddd}{S}{RL}  & \epsilon_{\alpha\beta\gamma}  (d_{Rp}^{\alpha T} C d_{Rr}^{\beta}) (\bar e_{Rs}^{} d_{Lt}^\gamma )  \\
\op{udd}{S}{RR}  & \epsilon_{\alpha\beta\gamma}  (u_{Rp}^{\alpha T} C d_{Rr}^{\beta}) (\bar \nu_{Ls}^{} d_{Rt}^\gamma )  \\
\op{ddd}{S}{RR}  & \epsilon_{\alpha\beta\gamma}  (d_{Rp}^{\alpha T} C d_{Rr}^{\beta}) (\bar e_{Ls}^{} d_{Rt}^\gamma )  \\
\end{array}
\end{align*}
\end{minipage}
\end{adjustbox}
%
\caption{LEFT operators of dimension six that violate baryon and/or lepton number.}
\label{tab:oplist2}
\end{table}

%% file: sections/EOMRedundancies.tex

\clearpage

\subsection{On-shell-redundant operators}
\label{sec:RedundantOperators}

In Table~\ref{tab:EOMRedundantOperators}, we provide a set of operators that extend the on-shell LEFT basis in App.~\ref{sec:LEFTBasis} to an off-shell-complete basis. Together with the inclusion of evanescent operators, this extended operator basis allows us to renormalize all off-shell Green's functions.

\begin{table}[H]
\capstart
\centering
\begin{minipage}[t]{3cm}
\renewcommand{\arraystretch}{1.51}
\small
\begin{align*}
\begin{array}[t]{c|c}
\multicolumn{2}{c}{\boldsymbol{(\nu\nu)D^2 + \hc}} \\
\hline
\O_{\nu D}^{(5)} & \nu_{Lp}^T C (i\bar{\slashed \p})^2 \nu_{Lr} \\
\end{array}
\end{align*}
\end{minipage}
\begin{minipage}[t]{3cm}
\renewcommand{\arraystretch}{1.51}
\small
\begin{align*}
\begin{array}[t]{c|c}
\multicolumn{2}{c}{\boldsymbol{(\overline L R)D^2 + \hc}} \\
\hline
\O_{eD}^{(5)} & \bar e_{Lp} (i\bar{\slashed D})^2 e_{Rr} \\
\O_{uD}^{(5)} & \bar u_{Lp} (i\bar{\slashed D})^2 u_{Rr} \\
\O_{dD}^{(5)} & \bar d_{Lp} (i\bar{\slashed D})^2 d_{Rr} \\
\end{array}
\end{align*}
\end{minipage}
\begin{minipage}[t]{3cm}
\renewcommand{\arraystretch}{1.51}
\small
\begin{align*}
\begin{array}[t]{c|c}
\multicolumn{2}{c}{\boldsymbol{X^2D^2}} \\
\hline
\O_{\gamma D} & \roverline{(\p_\mu F^{\mu\nu})(\p^\lambda F_{\lambda\nu})} \\
\O_{GD} & \roverline{(D_\mu G^{\mu\nu})^A (D^\lambda G_{\lambda\nu})^A} \\
\end{array}
\end{align*}
\end{minipage}
\\
\begin{adjustbox}{width=\textwidth,center}
\begin{minipage}[t]{3cm}
\renewcommand{\arraystretch}{1.51}
\small
\begin{align*}
\begin{array}[t]{c|c}
\multicolumn{2}{c}{\boldsymbol{(\overline L L)D^3}} \\
\hline
\O_{\nu D}^{L} & \bar \nu_{Lp} (i\bar{\slashed \p})^3 \nu_{Lr} \\
\O_{eD}^{L} & \bar e_{Lp} (i\bar{\slashed D})^3 e_{Lr} \\
\O_{uD}^{L} & \bar u_{Lp} (i\bar{\slashed D})^3 u_{Lr} \\
\O_{dD}^{L} & \bar d_{Lp} (i\bar{\slashed D})^3 d_{Lr} \\
\end{array}
\end{align*}
\end{minipage}
\begin{minipage}[t]{3cm}
\renewcommand{\arraystretch}{1.51}
\small
\begin{align*}
\begin{array}[t]{c|c}
\multicolumn{2}{c}{\boldsymbol{(\overline R R)D^3}} \\
\hline
\O_{eD}^{R} & \bar e_{Rp} (i\bar{\slashed D})^3 e_{Rr} \\
\O_{uD}^{R} & \bar u_{Rp} (i\bar{\slashed D})^3 u_{Rr} \\
\O_{dD}^{R} & \bar d_{Rp} (i\bar{\slashed D})^3 d_{Rr} \\
\end{array}
\end{align*}
\end{minipage}
\begin{minipage}[t]{3cm}
\renewcommand{\arraystretch}{1.51}
\small
\begin{align*}
\begin{array}[t]{c|c}
\multicolumn{2}{c}{\boldsymbol{(\overline L L)XD}} \\
\hline
\O_{D\nu\gamma}^{L} & (\bar \nu_{Lp} i\overleftarrow{\bar{\slashed \p}} \bar\sigma_{\mu\nu} \nu_{Lr})  F^{\mu\nu}  \\
\O_{\nu D\gamma}^{L} & (\bar \nu_{Lp} \bar\sigma_{\mu\nu} i\bar{\slashed \p} \nu_{Lr})  F^{\mu\nu}  \\
\O_{\nu\gamma D}^{L} & (\bar \nu_{Lp} \bar\gamma_\nu \nu_{Lr})  (\bar\p_\mu F^{\mu\nu})  \\
\O_{De\gamma}^{L} & (\bar e_{Lp} i\overleftarrow{\bar{\slashed D}} \bar\sigma_{\mu\nu} e_{Lr})  F^{\mu\nu}  \\
\O_{eD\gamma}^{L} & (\bar e_{Lp} \bar\sigma_{\mu\nu} i\bar{\slashed D} e_{Lr})  F^{\mu\nu}  \\
\O_{e\gamma D}^{L} & (\bar e_{Lp} \bar\gamma_\nu e_{Lr})  (\bar\p_\mu F^{\mu\nu})  \\
\O_{Du\gamma}^{L} & (\bar u_{Lp} i\overleftarrow{\bar{\slashed D}} \bar\sigma_{\mu\nu} u_{Lr})  F^{\mu\nu}  \\
\O_{uD\gamma}^{L} & (\bar u_{Lp} \bar\sigma_{\mu\nu} i\bar{\slashed D} u_{Lr})  F^{\mu\nu}  \\
\O_{u\gamma D}^{L} & (\bar u_{Lp} \bar\gamma_\nu u_{Lr})  (\bar\p_\mu F^{\mu\nu})  \\
\O_{Dd\gamma}^{L} & (\bar d_{Lp} i\overleftarrow{\bar{\slashed D}} \bar\sigma_{\mu\nu} d_{Lr})  F^{\mu\nu}  \\
\O_{dD\gamma}^{L} & (\bar d_{Lp} \bar\sigma_{\mu\nu} i\bar{\slashed D} d_{Lr})  F^{\mu\nu}  \\
\O_{d\gamma D}^{L} & (\bar d_{Lp} \bar\gamma_\nu d_{Lr})  (\bar\p_\mu F^{\mu\nu})  \\
\O_{DuG}^{L} & (\bar u_{Lp} i\overleftarrow{\bar{\slashed D}} \bar\sigma_{\mu\nu} T^A u_{Lr})  G^{A\mu\nu}  \\
\O_{uDG}^{L} & (\bar u_{Lp} \bar\sigma_{\mu\nu} T^A i\bar{\slashed D} u_{Lr})  G^{A\mu\nu}  \\
\O_{uGD}^{L} & (\bar u_{Lp} \bar\gamma_\nu T^A u_{Lr})  (\bar D_\mu G^{\mu\nu})^A  \\
\O_{DdG}^{L} & (\bar d_{Lp} i\overleftarrow{\bar{\slashed D}} \bar\sigma_{\mu\nu} T^A d_{Lr})  G^{A\mu\nu}  \\
\O_{dDG}^{L} & (\bar d_{Lp} \bar\sigma_{\mu\nu} T^A i\bar{\slashed D} d_{Lr})  G^{A\mu\nu}  \\
\O_{dGD}^{L} & (\bar d_{Lp} \bar\gamma_\nu T^A d_{Lr})  (\bar D_\mu G^{\mu\nu})^A  \\
\end{array}
\end{align*}
\end{minipage}
\begin{minipage}[t]{3cm}
\renewcommand{\arraystretch}{1.51}
\small
\begin{align*}
\begin{array}[t]{c|c}
\multicolumn{2}{c}{\boldsymbol{(\overline R R)XD}} \\
\hline
\O_{De\gamma}^{R} & (\bar e_{Rp} i\overleftarrow{\bar{\slashed D}} \bar\sigma_{\mu\nu} e_{Rr})  F^{\mu\nu}  \\
\O_{eD\gamma}^{R} & (\bar e_{Rp} \bar\sigma_{\mu\nu} i\bar{\slashed D} e_{Rr})  F^{\mu\nu}  \\
\O_{e\gamma D}^{R} & (\bar e_{Rp} \bar\gamma_\nu e_{Rr})  (\bar\p_\mu F^{\mu\nu})  \\
\O_{Du\gamma}^{R} & (\bar u_{Rp} i\overleftarrow{\bar{\slashed D}} \bar\sigma_{\mu\nu} u_{Rr})  F^{\mu\nu}  \\
\O_{uD\gamma}^{R} & (\bar u_{Rp} \bar\sigma_{\mu\nu} i\bar{\slashed D} u_{Rr})  F^{\mu\nu}  \\
\O_{u\gamma D}^{R} & (\bar u_{Rp} \bar\gamma_\nu u_{Rr})  (\bar\p_\mu F^{\mu\nu})  \\
\O_{Dd\gamma}^{R} & (\bar d_{Rp} i\overleftarrow{\bar{\slashed D}} \bar\sigma_{\mu\nu} d_{Rr})  F^{\mu\nu}  \\
\O_{dD\gamma}^{R} & (\bar d_{Rp} \bar\sigma_{\mu\nu} i\bar{\slashed D} d_{Rr})  F^{\mu\nu}  \\
\O_{d\gamma D}^{R} & (\bar d_{Rp} \bar\gamma_\nu d_{Rr})  (\bar\p_\mu F^{\mu\nu})  \\
\O_{DuG}^{R} & (\bar u_{Rp} i\overleftarrow{\bar{\slashed D}} \bar\sigma_{\mu\nu} T^A u_{Rr})  G^{A\mu\nu}  \\
\O_{uDG}^{R} & (\bar u_{Rp} \bar\sigma_{\mu\nu} T^A i\bar{\slashed D} u_{Rr})  G^{A\mu\nu}  \\
\O_{uGD}^{R} & (\bar u_{Rp} \bar\gamma_\nu T^A u_{Rr})  (\bar D_\mu G^{\mu\nu})^A  \\
\O_{DdG}^{R} & (\bar d_{Rp} i\overleftarrow{\bar{\slashed D}} \bar\sigma_{\mu\nu} T^A d_{Rr})  G^{A\mu\nu}  \\
\O_{dDG}^{R} & (\bar d_{Rp} \bar\sigma_{\mu\nu} T^A i\bar{\slashed D} d_{Rr})  G^{A\mu\nu}  \\
\O_{dGD}^{R} & (\bar d_{Rp} \bar\gamma_\nu T^A d_{Rr})  (\bar D_\mu G^{\mu\nu})^A  \\
\end{array}
\end{align*}
\end{minipage}
\end{adjustbox}

\caption{Redundant LEFT operators of dimension five and six, which can be removed by field redefinitions but are required to renormalize off-shell Green's functions.}
\label{tab:EOMRedundantOperators}
\end{table}

%% file: sections/Evanescents.tex

\clearpage

\subsection{Evanescent operators}
\label{sec:Evanescents}

The LEFT Lagrangian is supplemented by evanescent operators and corresponding coefficients
\begin{align}
	\L_\mathrm{evan} &= \sum_i K_i \E_i \, .
\end{align}
In the case of operators with at most two fermion fields, we provide an exhaustive list of evanescent operators in Tables~\ref{tab:EvanescentOperatorsD4}, \ref{tab:EvanescentOperatorsD5}, and~\ref{tab:EvanescentOperatorsD6}. Not all of them are required independently as counterterms to one-loop insertions of physical operators, e.g., the operators $\E_{\gamma'}$ and $\E_{G'}$ are not required for the renormalization of the physical operators at one loop~\cite{Belusca-Maito:2021lnk}. In the case of evanescent four-fermion operators, the complete basis contains infinitely many elements and we only list the operators required as counterterms to one-loop insertions of physical operators in Tables~\ref{tab:EvanescentFourFermionOperators} and~\ref{tab:EvanescentFourFermionOperatorsBL}. Fierz-evanescent operators are listed in Tables~\ref{tab:EvanescentFourFermionOperatorsFierz} and~\ref{tab:EvanescentFourFermionOperatorsFierzBL}.

\begin{table}[H]
\capstart
\centering

\begin{minipage}[t]{3cm}
\renewcommand{\arraystretch}{1.51}
\small
\begin{align*}
\begin{array}[t]{c|c}
\multicolumn{2}{c}{\boldsymbol{(\nu \nu)D + \hc}} \\
\hline
\E_{\nu D}^{} & \nu_{Lp}^T C (i\hat{\slashed \p}) \nu_{Lr} \\
\end{array}
\end{align*}
\end{minipage}
\begin{minipage}[t]{3cm}
\renewcommand{\arraystretch}{1.51}
\small
\begin{align*}
\begin{array}[t]{c|c}
\multicolumn{2}{c}{\boldsymbol{(\overline L R)D + \hc}} \\
\hline
\E_{e D}^{} & \bar e_{Lp} (i\hat{\slashed D}) e_{Rr} \\
\E_{u D}^{} & \bar u_{Lp} (i\hat{\slashed D}) u_{Rr} \\
\E_{d D}^{} & \bar d_{Lp} (i\hat{\slashed D}) d_{Rr} \\
\end{array}
\end{align*}
\end{minipage}
\begin{minipage}[t]{3cm}
\renewcommand{\arraystretch}{1.51}
\small
\begin{align*}
\begin{array}[t]{c|c}
\multicolumn{2}{c}{\boldsymbol{X^2}} \\
\hline
\E_{\gamma} & F_{\mu\nu} F^{\mu\nu} - \roverline{F_{\mu\nu} F^{\mu\nu}} \\
\E_{G} & G_{\mu\nu}^A G^{A\mu\nu} - \roverline{G_{\mu\nu}^A G^{A\mu\nu}} \\
\E_{\gamma'} & \hat F_{\mu\nu} \hat F^{\mu\nu} \\
\E_{G'} & \hat G_{\mu\nu}^A \hat G^{A\mu\nu} \\
\end{array}
\end{align*}
\end{minipage}

\caption{Evanescent LEFT operators of dimension four in the HV scheme.}
\label{tab:EvanescentOperatorsD4}
\end{table}

\begin{table}[H]
\capstart
\centering

\begin{adjustbox}{width=0.8\textwidth,center}
\begin{minipage}[t]{3cm}
\renewcommand{\arraystretch}{1.51}
\small
\begin{align*}
\begin{array}[t]{c|c}
\multicolumn{2}{c}{\boldsymbol{(\overline L L)D^2}} \\
\hline
\E_{\nu D}^{L} & \bar \nu_{Lp} [(i\hat{\slashed \p}) , (i\bar{\slashed \p})] \nu_{Lr} \\
\E_{e D}^{L} & \bar e_{Lp} [(i\hat{\slashed D}) , (i\bar{\slashed D})] e_{Lr} \\
\E_{u D}^{L} & \bar u_{Lp} [(i\hat{\slashed D}) , (i\bar{\slashed D})] u_{Lr} \\
\E_{d D}^{L} & \bar d_{Lp} [(i\hat{\slashed D}) , (i\bar{\slashed D})] d_{Lr} \\
\end{array}
\end{align*}
\end{minipage}
\begin{minipage}[t]{3cm}
\renewcommand{\arraystretch}{1.51}
\small
\begin{align*}
\begin{array}[t]{c|c}
\multicolumn{2}{c}{\boldsymbol{(\overline R R)D^2}} \\
\hline
\E_{e D}^{R} & \bar e_{Rp} [(i\hat{\slashed D}) , (i\bar{\slashed D})] e_{Rr} \\
\E_{u D}^{R} & \bar u_{Rp} [(i\hat{\slashed D}) , (i\bar{\slashed D})] u_{Rr} \\
\E_{d D}^{R} & \bar d_{Rp} [(i\hat{\slashed D}) , (i\bar{\slashed D})] d_{Rr} \\
\end{array}
\end{align*}
\end{minipage}
\begin{minipage}[t]{3cm}
\renewcommand{\arraystretch}{1.51}
\small
\begin{align*}
\begin{array}[t]{c|c}
\multicolumn{2}{c}{\boldsymbol{(\nu \nu)D^2 + \hc}} \\
\hline
\E_{\nu D}^{RL} & \nu_{Lp}^T C (i\hat{\slashed \p})(i\hat{\slashed \p}) \nu_{Lr} \\
\end{array}
\end{align*}
\end{minipage}
\begin{minipage}[t]{3cm}
\renewcommand{\arraystretch}{1.51}
\small
\begin{align*}
\begin{array}[t]{c|c}
\multicolumn{2}{c}{\boldsymbol{(\overline L R)D^2 + \hc}} \\
\hline
\E_{e D}^{LR} & \bar e_{Lp} (i\hat{\slashed D})(i\hat{\slashed D}) e_{Rr} \\
\E_{u D}^{LR} & \bar u_{Lp} (i\hat{\slashed D})(i\hat{\slashed D}) u_{Rr} \\
\E_{d D}^{LR} & \bar d_{Lp} (i\hat{\slashed D})(i\hat{\slashed D}) d_{Rr} \\
\end{array}
\end{align*}
\end{minipage}
\end{adjustbox}

\begin{adjustbox}{width=0.97\textwidth,center}
\begin{minipage}[t]{3cm}
\renewcommand{\arraystretch}{1.51}
\small
\begin{align*}
\begin{array}[t]{c|c}
\multicolumn{2}{c}{\boldsymbol{(\overline L L)X}} \\
\hline
\E_{\nu\gamma}^{L} & (\bar\nu_{Lp} i\hat\gamma_\mu\bar\gamma_\nu \nu_{Lr})  F^{\mu\nu}  \\
\E_{e\gamma}^{L} & (\bar e_{Lp} i\hat\gamma_\mu\bar\gamma_\nu e_{Lr})  F^{\mu\nu}  \\
\E_{u\gamma}^{L} & (\bar u_{Lp} i\hat\gamma_\mu\bar\gamma_\nu u_{Lr})  F^{\mu\nu}  \\
\E_{d\gamma}^{L} & (\bar d_{Lp} i\hat\gamma_\mu\bar\gamma_\nu d_{Lr})  F^{\mu\nu}  \\
\E_{uG}^{L} & (\bar u_{Lp} i\hat\gamma_\mu\bar\gamma_\nu T^A u_{Lr})  G^{A\mu\nu}  \\
\E_{dG}^{L} & (\bar d_{Lp} i\hat\gamma_\mu\bar\gamma_\nu T^A d_{Lr})  G^{A\mu\nu}  \\
\end{array}
\end{align*}
\end{minipage}
\begin{minipage}[t]{3cm}
\renewcommand{\arraystretch}{1.51}
\small
\begin{align*}
\begin{array}[t]{c|c}
\multicolumn{2}{c}{\boldsymbol{(\overline R R)X}} \\
\hline
\E_{e\gamma}^{R} & (\bar e_{Rp} i\hat\gamma_\mu\bar\gamma_\nu e_{Rr})  F^{\mu\nu}  \\
\E_{u\gamma}^{R} & (\bar u_{Rp} i\hat\gamma_\mu\bar\gamma_\nu u_{Rr})  F^{\mu\nu}  \\
\E_{d\gamma}^{R} & (\bar d_{Rp} i\hat\gamma_\mu\bar\gamma_\nu d_{Rr})  F^{\mu\nu}  \\
\E_{uG}^{R} & (\bar u_{Rp} i\hat\gamma_\mu\bar\gamma_\nu T^A u_{Rr})  G^{A\mu\nu}  \\
\E_{dG}^{R} & (\bar d_{Rp} i\hat\gamma_\mu\bar\gamma_\nu T^A d_{Rr})  G^{A\mu\nu}  \\
\end{array}
\end{align*}
\end{minipage}
\begin{minipage}[t]{3cm}
\renewcommand{\arraystretch}{1.51}
\small
\begin{align*}
\begin{array}[t]{c|c}
\multicolumn{2}{c}{\boldsymbol{(\nu\nu)X + \hc}} \\
\hline
\E_{\nu\gamma}^{RL} & (\nu_{Lp}^T C \hat\sigma_{\mu\nu} \nu_{Lr})  F^{\mu\nu}  \\
\end{array}
\end{align*}
\end{minipage}
\begin{minipage}[t]{3cm}
\renewcommand{\arraystretch}{1.51}
\small
\begin{align*}
\begin{array}[t]{c|c}
\multicolumn{2}{c}{\boldsymbol{(\overline L R)X + \hc}} \\
\hline
\E_{e\gamma}^{LR} & (\bar e_{Lp} \hat\sigma_{\mu\nu} e_{Rr})  F^{\mu\nu}  \\
\E_{u\gamma}^{LR} & (\bar u_{Lp} \hat\sigma_{\mu\nu} u_{Rr})  F^{\mu\nu}  \\
\E_{d\gamma}^{LR} & (\bar d_{Lp} \hat\sigma_{\mu\nu} d_{Rr})  F^{\mu\nu}  \\
\E_{uG}^{LR} & (\bar u_{Lp} \hat\sigma_{\mu\nu} T^A u_{Rr})  G^{A\mu\nu}  \\
\E_{dG}^{LR} & (\bar d_{Lp} \hat\sigma_{\mu\nu} T^A d_{Rr})  G^{A\mu\nu}  \\
\end{array}
\end{align*}
\end{minipage}
\end{adjustbox}

\caption{Evanescent LEFT operators of dimension five in the HV scheme.}
\label{tab:EvanescentOperatorsD5}
\end{table}

\begin{table}[H]
\capstart
\centering

\begin{adjustbox}{width=1.05\textwidth,center}
%
\begin{minipage}[t]{4cm}
\renewcommand{\arraystretch}{1.51}
\scriptsize
\begin{align*}
\begin{array}[t]{c|c}
\multicolumn{2}{c}{\boldsymbol{(\overline L L)D^3}} \\
\hline
\E_{\nu D1}^{L} & \bar \nu_{Lp} (i\hat{\slashed \p})(i\bar{\slashed \p})(i\hat{\slashed \p}) \nu_{Lr} \\
\E_{e D1}^{L} & \bar e_{Lp} (i\hat{\slashed D})(i\bar{\slashed D})(i\hat{\slashed D}) e_{Lr} \\
\E_{u D1}^{L} & \bar u_{Lp} (i\hat{\slashed D})(i\bar{\slashed D})(i\hat{\slashed D}) u_{Lr} \\
\E_{d D1}^{L} & \bar d_{Lp} (i\hat{\slashed D})(i\bar{\slashed D})(i\hat{\slashed D}) d_{Lr} \\
\end{array}
\end{align*}
\renewcommand{\arraystretch}{1.51}
\scriptsize
\begin{align*}
\begin{array}[t]{c|c}
\multicolumn{2}{c}{\boldsymbol{(\overline R R)D^3}} \\
\hline
\E_{e D1}^{R} & \bar e_{Rp} (i\hat{\slashed D})(i\bar{\slashed D})(i\hat{\slashed D}) e_{Rr} \\
\E_{u D1}^{R} & \bar u_{Rp} (i\hat{\slashed D})(i\bar{\slashed D})(i\hat{\slashed D}) u_{Rr} \\
\E_{d D1}^{R} & \bar d_{Rp} (i\hat{\slashed D})(i\bar{\slashed D})(i\hat{\slashed D}) d_{Rr} \\
\end{array}
\end{align*}
\renewcommand{\arraystretch}{1.51}
\scriptsize
\begin{align*}
\begin{array}[t]{c|c}
\multicolumn{2}{c}{\boldsymbol{(\nu\nu)D^3 + \hc}} \\
\hline
\E_{\nu D1}^{RL} & \nu_{Lp}^T C (i\bar{\slashed \p})(i\hat{\slashed \p})(i\bar{\slashed \p}) \nu_{Lr} \\
\E_{\nu D2}^{RL} & \nu_{Lp}^T C (i\hat{\slashed \p})(i\hat{\slashed \p})(i\hat{\slashed \p}) \nu_{Lr} \\
\end{array}
\end{align*}
\renewcommand{\arraystretch}{1.51}
\scriptsize
\begin{align*}
\begin{array}[t]{c|c}
\multicolumn{2}{c}{\boldsymbol{(\overline L R)D^3 + \hc}} \\
\hline
\E_{e D1}^{LR} & \bar e_{Lp} (i\bar{\slashed D})(i\hat{\slashed D})(i\bar{\slashed D}) e_{Rr} \\
\E_{e D2}^{LR} & \bar e_{Lp} (i\hat{\slashed D})(i\hat{\slashed D})(i\hat{\slashed D}) e_{Rr} \\
\E_{u D1}^{LR} & \bar u_{Lp} (i\bar{\slashed D})(i\hat{\slashed D})(i\bar{\slashed D}) u_{Rr} \\
\E_{u D2}^{LR} & \bar u_{Lp} (i\hat{\slashed D})(i\hat{\slashed D})(i\hat{\slashed D}) u_{Rr} \\
\E_{d D1}^{LR} & \bar d_{Lp} (i\bar{\slashed D})(i\hat{\slashed D})(i\bar{\slashed D}) d_{Rr} \\
\E_{d D2}^{LR} & \bar d_{Lp} (i\hat{\slashed D})(i\hat{\slashed D})(i\hat{\slashed D}) d_{Rr} \\
\end{array}
\end{align*}
\vspace{2.15cm}
\renewcommand{\arraystretch}{1.51}
\scriptsize
\begin{align*}
\begin{array}[t]{c|c}
\multicolumn{2}{c}{\boldsymbol{X^2D^2}} \\
\hline
\E_{\gamma D1} & (\hat\p_\mu F^{\mu\nu})(\p^\lambda F_{\lambda\nu}) \\
\E_{\gamma D2} & \hat g_{\nu\sigma} (\p_\mu F^{\mu\nu})(\p_\lambda F^{\lambda\sigma}) \\
\E_{\gamma D3} & (\hat\p_\mu \hat F^{\mu\nu})(\p^\lambda F_{\lambda\nu}) \\
\E_{\gamma D4} & (\hat\p_\mu F^{\mu\nu})(\hat\p^\lambda F_{\lambda\nu}) \\
\E_{\gamma D5} & (\hat\p_\mu \hat F^{\mu\nu})(\hat\p^\lambda \hat F_{\lambda\nu}) \\
\E_{GD1} & (\hat D_\mu G^{\mu\nu})^A (D^\lambda G_{\lambda\nu})^A \\
\E_{GD2} & \hat g_{\nu\sigma} (D_\mu G^{\mu\nu})^A (D_\lambda G^{\lambda\sigma})^A \\
\E_{GD3} & (\hat D_\mu \hat G^{\mu\nu})^A (D^\lambda G_{\lambda\nu})^A \\
\E_{GD4} & (\hat D_\mu G^{\mu\nu})^A (\hat D^\lambda G_{\lambda\nu})^A \\
\E_{GD5} & (\hat D_\mu \hat G^{\mu\nu})^A (\hat D^\lambda \hat G_{\lambda\nu})^A \\
\end{array}
\end{align*}
\end{minipage}
\begin{minipage}[t]{4cm}
\renewcommand{\arraystretch}{1.51}
\scriptsize
\begin{align*}
\begin{array}[t]{c|c}
\multicolumn{2}{c}{\boldsymbol{(\overline L L)XD}} \\
\hline
\E_{D\nu\gamma 1}^{L} & (\bar \nu_{Lp} i\overleftarrow{\bar{\slashed \p}} \hat\sigma_{\mu\nu} \nu_{Lr})  F^{\mu\nu}  \\
\E_{D\nu\gamma 2}^{L} & (\bar \nu_{Lp} i\overleftarrow{\hat{\slashed \p}} i \hat \gamma_\mu \bar \gamma_\nu \nu_{Lr})  F^{\mu\nu}  \\
\E_{\nu D\gamma 1}^{L} & (\bar \nu_{Lp} \hat\sigma_{\mu\nu} i\bar{\slashed \p} \nu_{Lr})  F^{\mu\nu}  \\
\E_{\nu D\gamma 2}^{L} & (\bar \nu_{Lp} i \hat \gamma_\mu \bar \gamma_\nu i\hat{\slashed \p} \nu_{Lr})  F^{\mu\nu}  \\
\E_{\nu\gamma D}^{L} & (\bar \nu_{Lp} \bar\gamma_\nu \nu_{Lr})  (\hat\p_\mu F^{\mu\nu})  \\
\E_{De\gamma 1}^{L} & (\bar e_{Lp} i\overleftarrow{\bar{\slashed D}} \hat\sigma_{\mu\nu} e_{Lr})  F^{\mu\nu}  \\
\E_{De\gamma 2}^{L} & (\bar e_{Lp} i\overleftarrow{\hat{\slashed D}} i \hat \gamma_\mu \bar \gamma_\nu e_{Lr})  F^{\mu\nu}  \\
\end{array}
\end{align*}
\vspace{1.6cm}
\renewcommand{\arraystretch}{1.51}
\scriptsize
\begin{align*}
\begin{array}[t]{c|c}
\multicolumn{2}{c}{\boldsymbol{X^3}}  \\
\hline
\E_{G1}     & \hat g^{\rho\sigma} f^{ABC} G_\mu^{A\nu} G_{\nu\sigma}^{B} G_\rho^{C\mu}  \\
\E_{G2}     & f^{ABC} G_\mu^{A\nu} G_\nu^{B\rho} \hat G_\rho^{C\mu}  \\
\E_{G3}     & f^{ABC} \hat G_\mu^{A\nu} \hat G_\nu^{B\rho} \hat G_\rho^{C\mu}  \\
\E_{\widetilde G} & \hat g^{\rho\sigma} f^{ABC} \widetilde G_\mu^{A\nu} G_{\nu\sigma}^{B} G_\rho^{C\mu}   \\
\end{array}
\end{align*}
\end{minipage}
\begin{minipage}[t]{4cm}
\renewcommand{\arraystretch}{1.51}
\scriptsize
\begin{align*}
\begin{array}[t]{c|c}
\multicolumn{2}{c}{\boldsymbol{(\overline R R)XD}} \\
\hline
\E_{De\gamma 1}^{R} & (\bar e_{Rp} i\overleftarrow{\bar{\slashed D}} \hat\sigma_{\mu\nu} e_{Rr})  F^{\mu\nu}  \\
\E_{De\gamma 2}^{R} & (\bar e_{Rp} i\overleftarrow{\hat{\slashed D}} i \hat \gamma_\mu \bar \gamma_\nu e_{Rr})  F^{\mu\nu}  \\
\E_{eD\gamma 1}^{R} & (\bar e_{Rp} \hat\sigma_{\mu\nu} i\bar{\slashed D} e_{Rr})  F^{\mu\nu}  \\
\E_{eD\gamma 2}^{R} & (\bar e_{Rp} i \hat\gamma_\mu\bar\gamma_\nu i\hat{\slashed D} e_{Rr})  F^{\mu\nu}  \\
\E_{e\gamma D}^{R} & (\bar e_{Rp} \bar\gamma_\nu e_{Rr})  (\hat\p_\mu F^{\mu\nu})  \\
\E_{Du\gamma 1}^{R} & (\bar u_{Rp} i\overleftarrow{\bar{\slashed D}} \hat\sigma_{\mu\nu} u_{Rr})  F^{\mu\nu}  \\
\E_{Du\gamma 2}^{R} & (\bar u_{Rp} i\overleftarrow{\hat{\slashed D}} i \hat \gamma_\mu \bar \gamma_\nu u_{Rr})  F^{\mu\nu}  \\
\end{array}
\end{align*}
\renewcommand{\arraystretch}{1.51}
\scriptsize
\begin{align*}
\begin{array}[t]{c|c}
\multicolumn{2}{c}{\boldsymbol{(\nu\nu)XD + \hc}} \\
\hline
\E_{D\nu\gamma1}^{RL} & (\nu_{Lp}^T C i\overleftarrow{\hat{\slashed \p}} \bar\sigma_{\mu\nu} \nu_{Lr})  F^{\mu\nu}  \\
\E_{D\nu\gamma2}^{RL} & (\nu_{Lp}^T C i\overleftarrow{\hat{\slashed \p}} \hat\sigma_{\mu\nu} \nu_{Lr})  F^{\mu\nu}  \\
\E_{D\nu\gamma3}^{RL} & (\nu_{Lp}^T C i\overleftarrow{\bar{\slashed \p}} i \hat \gamma_\mu \bar \gamma_\nu \nu_{Lr})  F^{\mu\nu}  \\
\E_{\nu D\gamma1}^{RL} & (\nu_{Lp}^T C \bar\sigma_{\mu\nu} i\hat{\slashed \p} \nu_{Lr})  F^{\mu\nu}  \\
\E_{\nu D\gamma2}^{RL} & (\nu_{Lp}^T C \hat\sigma_{\mu\nu} i\hat{\slashed \p} \nu_{Lr})  F^{\mu\nu}  \\
\E_{\nu D\gamma3}^{RL} & (\nu_{Lp}^T C i \hat \gamma_\mu \bar \gamma_\nu i\bar{\slashed \p} \nu_{Lr})  F^{\mu\nu}  \\
\E_{\nu\gamma D1}^{RL} & (\nu_{Lp}^T C \hat\gamma_\nu \nu_{Lr})  (\p_\mu F^{\mu\nu})  \\
\E_{\nu\gamma D2}^{RL} & (\nu_{Lp}^T C \hat\gamma_\nu \nu_{Lr})  (\hat\p_\mu F^{\mu\nu})  \\
\end{array}
\end{align*}
\end{minipage}
\begin{minipage}[t]{4cm}
\renewcommand{\arraystretch}{1.51}
\scriptsize
\begin{align*}
\begin{array}[t]{c|c}
\multicolumn{2}{c}{\boldsymbol{(\overline L R)XD + \hc}} \\
\hline
\E_{De\gamma1}^{LR} & (\bar e_{Lp} i\overleftarrow{\hat{\slashed D}} \bar\sigma_{\mu\nu} e_{Rr})  F^{\mu\nu}  \\
\E_{De\gamma2}^{LR} & (\bar e_{Lp} i\overleftarrow{\hat{\slashed D}} \hat\sigma_{\mu\nu} e_{Rr})  F^{\mu\nu}  \\
\E_{De\gamma3}^{LR} & (\bar e_{Lp} i\overleftarrow{\bar{\slashed D}} i \hat \gamma_\mu \bar \gamma_\nu e_{Rr})  F^{\mu\nu}  \\
\E_{eD\gamma1}^{LR} & (\bar e_{Lp} \bar\sigma_{\mu\nu} i\hat{\slashed D} e_{Rr})  F^{\mu\nu}  \\
\E_{eD\gamma2}^{LR} & (\bar e_{Lp} \hat\sigma_{\mu\nu} i\hat{\slashed D} e_{Rr})  F^{\mu\nu}  \\
\E_{eD\gamma3}^{LR} & (\bar e_{Lp} i \hat \gamma_\mu \bar \gamma_\nu i\bar{\slashed D} e_{Rr})  F^{\mu\nu}  \\
\E_{e\gamma D1}^{LR} & (\bar e_{Lp} \hat\gamma_\nu e_{Rr})  (\p_\mu F^{\mu\nu})  \\
\end{array}
\end{align*}
\end{minipage}
\end{adjustbox}

\caption{Evanescent LEFT operators of dimension six in the HV scheme containing at most two fermion fields.}
\label{tab:EvanescentOperatorsD6}
\end{table}

\begin{table}[H]
\capstart
\centering

\begin{adjustbox}{width=1.05\textwidth,center}

\begin{minipage}[t]{3cm}
\renewcommand{\arraystretch}{1.5}
\small
\begin{align*}
\begin{array}[t]{c|c}
\multicolumn{2}{c}{\boldsymbol{(\overline L L)(\overline L L)}}  \\
\hline
\EOp{ee}{LL}[V(1)][] &  (\bar e_{Lp} \hat\gamma^\mu \bar\gamma^\nu e_{Lr}) (\bar e_{Ls} \hat\gamma_\mu \bar\gamma_\nu e_{Lt} )  \\
\EOp{ee}{LL}[V(2)][] &  (\bar e_{Lp} \hat\gamma^\mu \hat\gamma^\nu \bar\gamma^\lambda e_{Lr}) (\bar e_{Ls} \hat\gamma_\mu \hat\gamma_\nu \bar\gamma_\lambda e_{Lt} )  \\
\EOp{eu}{LL}[V(1)][] &  (\bar e_{Lp} \hat\gamma^\mu \bar\gamma^\nu e_{Lr}) (\bar u_{Ls} \hat\gamma_\mu \bar\gamma_\nu u_{Lt} )  \\
\EOp{eu}{LL}[V(2)][] &  (\bar e_{Lp} \hat\gamma^\mu \hat\gamma^\nu \bar\gamma^\lambda e_{Lr}) (\bar u_{Ls} \hat\gamma_\mu \hat\gamma_\nu \bar\gamma_\lambda u_{Lt} )  \\
\EOp{ed}{LL}[V(1)][] &  (\bar e_{Lp} \hat\gamma^\mu \bar\gamma^\nu e_{Lr}) (\bar d_{Ls} \hat\gamma_\mu \bar\gamma_\nu d_{Lt} )  \\
\EOp{ed}{LL}[V(2)][] &  (\bar e_{Lp} \hat\gamma^\mu \hat\gamma^\nu \bar\gamma^\lambda e_{Lr}) (\bar d_{Ls} \hat\gamma_\mu \hat\gamma_\nu \bar\gamma_\lambda d_{Lt} )  \\
\EOp{\nu edu}{LL}[V(1)][] &  (\bar \nu_{Lp} \hat\gamma^\mu \bar\gamma^\nu e_{Lr}) (\bar d_{Ls} \hat\gamma_\mu \bar\gamma_\nu u_{Lt} ) + \hc \\
\end{array}
\end{align*}
%
\renewcommand{\arraystretch}{1.5}
\small
\begin{align*}
\begin{array}[t]{c|c}
\multicolumn{2}{c}{\boldsymbol{(\overline R R)(\overline R R)}}  \\
\hline
\EOp{ee}{RR}[V(1)][] &  (\bar e_{Rp} \hat\gamma^\mu \bar\gamma^\nu e_{Rr}) (\bar e_{Rs} \hat\gamma_\mu \bar\gamma_\nu e_{Rt} )  \\
\EOp{ee}{RR}[V(2)][] &  (\bar e_{Rp} \hat\gamma^\mu \hat\gamma^\nu \bar\gamma^\lambda e_{Rr}) (\bar e_{Rs} \hat\gamma_\mu \hat\gamma_\nu \bar\gamma_\lambda e_{Rt} )  \\
\EOp{eu}{RR}[V(1)][] &  (\bar e_{Rp} \hat\gamma^\mu \bar\gamma^\nu e_{Rr}) (\bar u_{Rs} \hat\gamma_\mu \bar\gamma_\nu u_{Rt} )  \\
\EOp{eu}{RR}[V(2)][] &  (\bar e_{Rp} \hat\gamma^\mu \hat\gamma^\nu \bar\gamma^\lambda e_{Rr}) (\bar u_{Rs} \hat\gamma_\mu \hat\gamma_\nu \bar\gamma_\lambda u_{Rt} )  \\
\EOp{ed}{RR}[V(1)][] &  (\bar e_{Rp} \hat\gamma^\mu \bar\gamma^\nu e_{Rr}) (\bar d_{Rs} \hat\gamma_\mu \bar\gamma_\nu d_{Rt} )  \\
\EOp{ed}{RR}[V(2)][] &  (\bar e_{Rp} \hat\gamma^\mu \hat\gamma^\nu \bar\gamma^\lambda e_{Rr}) (\bar d_{Rs} \hat\gamma_\mu \hat\gamma_\nu \bar\gamma_\lambda d_{Rt} )  \\
\EOp{uu}{RR}[V1(1)][] &  (\bar u_{Rp} \hat\gamma^\mu \bar\gamma^\nu u_{Rr}) (\bar u_{Rs} \hat\gamma_\mu \bar\gamma_\nu u_{Rt} )  \\
\end{array}
\end{align*}
\end{minipage}
%
\begin{minipage}[t]{3cm}
\renewcommand{\arraystretch}{1.5}
\small
\begin{align*}
\begin{array}[t]{c|c}
\multicolumn{2}{c}{\boldsymbol{(\overline L L)(\overline R R)}}  \\
\hline
\EOp{ee}{LR}[V(1)][] &  (\bar e_{Lp} \hat\gamma^\mu \bar\gamma^\nu e_{Lr}) (\bar e_{Rs} \hat\gamma_\mu \bar\gamma_\nu e_{Rt} )  \\
\EOp{ee}{LR}[V(2)][] &  (\bar e_{Lp} \hat\gamma^\mu \hat\gamma^\nu \bar\gamma^\lambda e_{Lr}) (\bar e_{Rs} \hat\gamma_\mu \hat\gamma_\nu \bar\gamma_\lambda e_{Rt} )  \\
\EOp{eu}{LR}[V(1)][] &  (\bar e_{Lp} \hat\gamma^\mu \bar\gamma^\nu e_{Lr}) (\bar u_{Rs} \hat\gamma_\mu \bar\gamma_\nu u_{Rt} )  \\
\EOp{eu}{LR}[V(2)][] &  (\bar e_{Lp} \hat\gamma^\mu \hat\gamma^\nu \bar\gamma^\lambda e_{Lr}) (\bar u_{Rs} \hat\gamma_\mu \hat\gamma_\nu \bar\gamma_\lambda u_{Rt} )  \\
\EOp{ed}{LR}[V(1)][] &  (\bar e_{Lp} \hat\gamma^\mu \bar\gamma^\nu e_{Lr}) (\bar d_{Rs} \hat\gamma_\mu \bar\gamma_\nu d_{Rt} )  \\
\EOp{ed}{LR}[V(2)][] &  (\bar e_{Lp} \hat\gamma^\mu \hat\gamma^\nu \bar\gamma^\lambda e_{Lr}) (\bar d_{Rs} \hat\gamma_\mu \hat\gamma_\nu \bar\gamma_\lambda d_{Rt} )  \\
\EOp{ue}{LR}[V(1)][] &  (\bar u_{Lp} \hat\gamma^\mu \bar\gamma^\nu u_{Lr}) (\bar e_{Rs} \hat\gamma_\mu \bar\gamma_\nu e_{Rt} )  \\
\end{array}
\end{align*}
%
\renewcommand{\arraystretch}{1.5}
\small
\begin{align*}
\begin{array}[t]{c|c}
\multicolumn{2}{c}{\boldsymbol{(\overline L R)(\overline R L)}}  \\
\hline
\EOp{ee}{RL}[S(1)][] &  (\bar e_{Lp} \hat\gamma^\mu e_{Rr}) (\bar e_{Rs} \hat\gamma_\mu e_{Lt} )  \\
\EOp{eu}{RL}[S(1)][] &  (\bar e_{Lp} \hat\gamma^\mu e_{Rr}) (\bar u_{Rs} \hat\gamma_\mu u_{Lt} )  + \hc \\
\EOp{eu}{RL}[S(2)][] &  (\bar e_{Lp} \hat\gamma^\mu \hat\gamma^\nu e_{Rr}) (\bar u_{Rs} \hat\gamma_\mu \hat\gamma_\nu u_{Lt} )  + \hc \\
\EOp{ed}{RL}[S(1)][] &  (\bar e_{Lp} \hat\gamma^\mu e_{Rr}) (\bar d_{Rs} \hat\gamma_\mu d_{Lt} )  + \hc \\
\EOp{ed}{RL}[S(2)][] &  (\bar e_{Lp} \hat\gamma^\mu \hat\gamma^\nu e_{Rr}) (\bar d_{Rs} \hat\gamma_\mu \hat\gamma_\nu d_{Lt} )  + \hc \\
\EOp{\nu edu}{RL}[S(1)][] &  (\bar \nu_{Lp} \hat\gamma^\mu e_{Rr}) (\bar d_{Rs} \hat\gamma_\mu u_{Lt} )  + \hc \\
\EOp{\nu edu}{RL}[S(2)][] &  (\bar \nu_{Lp} \hat\gamma^\mu \hat\gamma^\nu e_{Rr}) (\bar d_{Rs} \hat\gamma_\mu \hat\gamma_\nu u_{Lt} )  + \hc \\
\EOp{uu}{RL}[S1(1)][] &  (\bar u_{Lp} \hat\gamma^\mu u_{Rr}) (\bar u_{Rs} \hat\gamma_\mu u_{Lt} )  \\
\EOp{uu}{RL}[S8(1)][] &  (\bar u_{Lp} \hat\gamma^\mu T^A u_{Rr}) (\bar u_{Rs} \hat\gamma_\mu T^A u_{Lt} )  \\
\EOp{ud}{RL}[S1(1)][] &  (\bar u_{Lp} \hat\gamma^\mu u_{Rr}) (\bar d_{Rs} \hat\gamma_\mu d_{Lt} ) + \hc \\
\EOp{ud}{RL}[S8(1)][] &  (\bar u_{Lp} \hat\gamma^\mu T^A u_{Rr}) (\bar d_{Rs} \hat\gamma_\mu T^A d_{Lt} ) + \hc  \\
\EOp{dd}{RL}[S1(1)][] &  (\bar d_{Lp} \hat\gamma^\mu d_{Rr}) (\bar d_{Rs} \hat\gamma_\mu d_{Lt} )  \\
\EOp{dd}{RL}[S8(1)][] &  (\bar d_{Lp} \hat\gamma^\mu T^A d_{Rr}) (\bar d_{Rs} \hat\gamma_\mu T^A d_{Lt} )  \\
\EOp{uddu}{RL}[S1(1)][] &  (\bar u_{Lp} \hat\gamma^\mu d_{Rr}) (\bar d_{Rs} \hat\gamma_\mu u_{Lt} ) + \hc \\
\EOp{uddu}{RL}[S8(1)][] &  (\bar u_{Lp} \hat\gamma^\mu T^A d_{Rr}) (\bar d_{Rs} \hat\gamma_\mu T^A u_{Lt} ) + \hc  \\
\end{array}
\end{align*}
\end{minipage}
%
\begin{minipage}[t]{3cm}
\renewcommand{\arraystretch}{1.5}
\small
\begin{align*}
\begin{array}[t]{c|c}
\multicolumn{2}{c}{\boldsymbol{(\overline L R)(\overline L R) + \hc}}  \\
\hline
\EOp{ee}{RR}[S(1)][] &  (\bar e_{Lp} \hat\gamma^\mu e_{Rr}) (\bar e_{Ls} \hat\gamma_\mu e_{Rt} )  \\
\EOp{ee}{RR}[S(2)][] &  (\bar e_{Lp} \hat\gamma^\mu \hat\gamma^\nu e_{Rr}) (\bar e_{Ls} \hat\gamma_\mu \hat\gamma_\nu e_{Rt} )  \\
\EOp{ee}{RR}[T(1)][] &  (\bar e_{Lp} \hat\gamma^\mu \bar\sigma^{\nu\lambda} e_{Rr}) (\bar e_{Ls} \hat\gamma_\mu \bar\sigma_{\nu\lambda} e_{Rt} )  \\
\EOp{ee}{RR}[T(2)][] &  (\bar e_{Lp} \hat\gamma^\mu \hat\gamma^\nu \bar\sigma^{\lambda\rho} e_{Rr}) (\bar e_{Ls} \hat\gamma_\mu \hat\gamma_\nu \bar\sigma_{\lambda\rho} e_{Rt} )  \\
\EOp{eu}{RR}[S(1)][] &  (\bar e_{Lp} \hat\gamma^\mu e_{Rr}) (\bar u_{Ls} \hat\gamma_\mu u_{Rt} )  \\
\EOp{eu}{RR}[S(2)][] &  (\bar e_{Lp} \hat\gamma^\mu \hat\gamma^\nu e_{Rr}) (\bar u_{Ls} \hat\gamma_\mu \hat\gamma_\nu u_{Rt} )  \\
\EOp{eu}{RR}[T(1)][] &  (\bar e_{Lp} \hat\gamma^\mu \bar\sigma^{\nu\lambda} e_{Rr}) (\bar u_{Ls} \hat\gamma_\mu \bar\sigma_{\nu\lambda} u_{Rt} )  \\
\end{array}
\end{align*}
\end{minipage}
\end{adjustbox}

\caption{Evanescent four-fermion LEFT operators other than Fierz-evanescent ones that appear at one loop in the HV scheme and conserve baryon and lepton number.}
\label{tab:EvanescentFourFermionOperators}
\end{table}

\begin{table}[H]
\capstart
\centering

\begin{adjustbox}{width=0.65\textwidth,center}

\begin{minipage}[t]{3cm}
\renewcommand{\arraystretch}{1.5}
\small
\begin{align*}
\begin{array}[t]{c|c}
\multicolumn{2}{c}{\boldsymbol{(\overline L L)[\overline L L] - (\overline L L][\overline L L)}}  \\
\hline
\EOp{ee}{LL}[(F1)][] &  (\bar e_{Lp} \gamma^\mu e_{Lr}) (\bar e_{Ls} \gamma_\mu e_{Lt}) - (\bar e_{Lp} \gamma^\mu e_{Lt}) (\bar e_{Ls} \gamma_\mu e_{Lr})  \\
\EOp{uu}{LL}[(F1)][] & \displaystyle (\bar u_{Lp} \gamma^\mu u_{Lr}) (\bar u_{Ls} \gamma_\mu u_{Lt}) - 2 (\bar u_{Lp} \gamma^\mu T^A u_{Lt}) (\bar u_{Ls} \gamma_\mu T^A u_{Lr}) - \frac{1}{N_c} (\bar u_{Lp} \gamma^\mu u_{Lt}) (\bar u_{Ls} \gamma_\mu u_{Lr})  \\
\EOp{dd}{LL}[(F1)][] & \displaystyle (\bar d_{Lp} \gamma^\mu d_{Lr}) (\bar d_{Ls} \gamma_\mu d_{Lt}) - 2 (\bar d_{Lp} \gamma^\mu T^A d_{Lt}) (\bar d_{Ls} \gamma_\mu T^A d_{Lr}) - \frac{1}{N_c} (\bar d_{Lp} \gamma^\mu d_{Lt}) (\bar d_{Ls} \gamma_\mu d_{Lr})  \\
\color{lightgray} \EOp{ud}{LL}[(F1)][] & \color{lightgray} \displaystyle (\bar u_{Lp} \gamma^\mu u_{Lr}) (\bar d_{Ls} \gamma_\mu d_{Lt}) - 2 (\bar u_{Lp} \gamma^\mu T^A d_{Lt}) (\bar d_{Ls} \gamma_\mu T^A u_{Lr}) - \frac{1}{N_c} (\bar u_{Lp} \gamma^\mu d_{Lt}) (\bar d_{Ls} \gamma_\mu u_{Lr})  \\
\color{lightgray} \EOp{uddu}{LL}[(F1)][] & \color{lightgray} \displaystyle (\bar u_{Lp} \gamma^\mu d_{Lr}) (\bar d_{Ls} \gamma_\mu u_{Lt}) - 2 (\bar u_{Lp} \gamma^\mu T^A u_{Lt}) (\bar d_{Ls} \gamma_\mu T^A d_{Lr}) - \frac{1}{N_c} (\bar u_{Lp} \gamma^\mu u_{Lt}) (\bar d_{Ls} \gamma_\mu d_{Lr})  \\
\end{array}
\end{align*}

\renewcommand{\arraystretch}{1.5}
\small
\begin{align*}
\begin{array}[t]{c|c}
\multicolumn{2}{c}{\boldsymbol{(\overline R R)[\overline R R] - (\overline R R][\overline R R)}}  \\
\hline
\EOp{ee}{RR}[(F1)][] &  (\bar e_{Rp} \gamma^\mu e_{Rr}) (\bar e_{Rs} \gamma_\mu e_{Rt}) - (\bar e_{Rp} \gamma^\mu e_{Rt}) (\bar e_{Rs} \gamma_\mu e_{Rr})  \\
\EOp{uu}{RR}[(F1)][] & \displaystyle (\bar u_{Rp} \gamma^\mu u_{Rr}) (\bar u_{Rs} \gamma_\mu u_{Rt}) - 2 (\bar u_{Rp} \gamma^\mu T^A u_{Rt}) (\bar u_{Rs} \gamma_\mu T^A u_{Rr}) - \frac{1}{N_c} (\bar u_{Rp} \gamma^\mu u_{Rt}) (\bar u_{Rs} \gamma_\mu u_{Rr})  \\
\EOp{dd}{RR}[(F1)][] & \displaystyle (\bar d_{Rp} \gamma^\mu d_{Rr}) (\bar d_{Rs} \gamma_\mu d_{Rt}) - 2 (\bar d_{Rp} \gamma^\mu T^A d_{Rt}) (\bar d_{Rs} \gamma_\mu T^A d_{Rr}) - \frac{1}{N_c} (\bar d_{Rp} \gamma^\mu d_{Rt}) (\bar d_{Rs} \gamma_\mu d_{Rr})  \\
\color{lightgray} \EOp{ud}{RR}[(F1)][] & \color{lightgray} \displaystyle (\bar u_{Rp} \gamma^\mu u_{Rr}) (\bar d_{Rs} \gamma_\mu d_{Rt}) - 2 (\bar u_{Rp} \gamma^\mu T^A d_{Rt}) (\bar d_{Rs} \gamma_\mu T^A u_{Rr}) - \frac{1}{N_c} (\bar u_{Rp} \gamma^\mu d_{Rt}) (\bar d_{Rs} \gamma_\mu u_{Rr})  \\
\color{lightgray} \EOp{uddu}{RR}[(F1)][] & \color{lightgray} \displaystyle (\bar u_{Rp} \gamma^\mu d_{Rr}) (\bar d_{Rs} \gamma_\mu u_{Rt}) - 2 (\bar u_{Rp} \gamma^\mu T^A u_{Rt}) (\bar d_{Rs} \gamma_\mu T^A d_{Rr}) - \frac{1}{N_c} (\bar u_{Rp} \gamma^\mu u_{Rt}) (\bar d_{Rs} \gamma_\mu d_{Rr})  \\
\end{array}
\end{align*}

\renewcommand{\arraystretch}{1.5}
\small
\begin{align*}
\begin{array}[t]{c|c}
\multicolumn{2}{c}{\boldsymbol{(\overline L L)[\overline R R] - (\overline L R][\overline R L)}}  \\
\hline
\EOp{ee}{LR}[(F1)][] &  (\bar e_{Lp} \gamma^\mu e_{Lr}) (\bar e_{Rs} \gamma_\mu e_{Rt} ) + 2 (\bar e_{Lp} e_{Rt}) (\bar e_{Rs} e_{Lr} )  \\
\EOp{uu}{LR}[(F1)1][] & \displaystyle (\bar u_{Lp} \gamma^\mu u_{Lr}) (\bar u_{Rs} \gamma_\mu u_{Rt}) + 4 (\bar u_{Lp} T^A u_{Rt}) (\bar u_{Rs} T^A u_{Lr}) + \frac{2}{N_c} (\bar u_{Lp} u_{Rt}) (\bar u_{Rs} u_{Lr})  \\
\EOp{uu}{LR}[(F1)8][] & \displaystyle (\bar u_{Lp} \gamma^\mu T^A u_{Lr}) (\bar u_{Rs} \gamma_\mu T^A u_{Rt}) - \frac{2}{N_c} (\bar u_{Lp} T^A u_{Rt}) (\bar u_{Rs} T^A u_{Lr}) + \frac{2C_F}{N_c} (\bar u_{Lp} u_{Rt}) (\bar u_{Rs} u_{Lr})  \\
\EOp{dd}{LR}[(F1)1][] & \displaystyle (\bar d_{Lp} \gamma^\mu d_{Lr}) (\bar d_{Rs} \gamma_\mu d_{Rt}) + 4 (\bar d_{Lp} T^A d_{Rt}) (\bar d_{Rs} T^A d_{Lr}) + \frac{2}{N_c} (\bar d_{Lp} d_{Rt}) (\bar d_{Rs} d_{Lr})  \\
\EOp{dd}{LR}[(F1)8][] & \displaystyle (\bar d_{Lp} \gamma^\mu T^A d_{Lr}) (\bar d_{Rs} \gamma_\mu T^A d_{Rt}) - \frac{2}{N_c} (\bar d_{Lp} T^A d_{Rt}) (\bar d_{Rs} T^A d_{Lr}) + \frac{2C_F}{N_c} (\bar d_{Lp} d_{Rt}) (\bar d_{Rs} d_{Lr})  \\
\color{lightgray} \EOp{ud}{LR}[(F1)1][] & \color{lightgray} \displaystyle (\bar u_{Lp} \gamma^\mu u_{Lr}) (\bar d_{Rs} \gamma_\mu d_{Rt}) + 4 (\bar u_{Lp} T^A d_{Rt}) (\bar d_{Rs} T^A u_{Lr}) + \frac{2}{N_c} (\bar u_{Lp} d_{Rt}) (\bar d_{Rs} u_{Lr})  \\
\color{lightgray} \EOp{ud}{LR}[(F1)8][] & \color{lightgray} \displaystyle (\bar u_{Lp} \gamma^\mu T^A u_{Lr}) (\bar d_{Rs} \gamma_\mu T^A d_{Rt}) - \frac{2}{N_c} (\bar u_{Lp} T^A d_{Rt}) (\bar d_{Rs} T^A u_{Lr}) + \frac{2C_F}{N_c} (\bar u_{Lp} d_{Rt}) (\bar d_{Rs} u_{Lr})  \\
\end{array}
\end{align*}
\end{minipage}
\end{adjustbox}

\begin{adjustbox}{width=0.95\textwidth,center}

\begin{minipage}[t]{3cm}
\renewcommand{\arraystretch}{1.5}
\small
\begin{align*}
\begin{array}[t]{c|c}
\multicolumn{2}{c}{\boldsymbol{(\overline L R)[\overline L R] - (\overline L R][\overline L R) + \hc}}  \\
\hline
\EOp{ee}{RR}[(F2)][] &  (\bar e_{Lp} \bar\sigma^{\mu\nu} e_{Rr}) (\bar e_{Ls} \bar\sigma_{\mu\nu} e_{Rt}) + 4 (\bar e_{Lp} e_{Rr}) (\bar e_{Ls} e_{Rt}) + 8 (\bar e_{Lp} e_{Rt}) (\bar e_{Ls} e_{Rr})  \\
\EOp{uu}{RR}[(F2)1][] & \displaystyle (\bar u_{Lp} \bar\sigma^{\mu\nu} u_{Rr}) (\bar u_{Ls} \bar\sigma_{\mu\nu} u_{Rt}) + 4 (\bar u_{Lp} u_{Rr}) (\bar u_{Ls} u_{Rt}) + 16 (\bar u_{Lp} T^A u_{Rt}) (\bar u_{Ls} T^A u_{Rr}) + \frac{8}{N_c} (\bar u_{Lp} u_{Rt}) (\bar u_{Ls} u_{Rr})  \\
\EOp{uu}{RR}[(F2)8][] & \displaystyle (\bar u_{Lp} \bar\sigma^{\mu\nu} T^A u_{Rr}) (\bar u_{Ls} \bar\sigma_{\mu\nu} T^A u_{Rt}) + 4 (\bar u_{Lp} T^A u_{Rr}) (\bar u_{Ls} T^A u_{Rt}) + \frac{8C_F}{N_c} (\bar u_{Lp} u_{Rt}) (\bar u_{Ls} u_{Rr}) - \frac{8}{N_c} (\bar u_{Lp} T^A u_{Rt}) (\bar u_{Ls} T^A u_{Rr})  \\
\EOp{dd}{RR}[(F2)1][] & \displaystyle (\bar d_{Lp} \bar\sigma^{\mu\nu} d_{Rr}) (\bar d_{Ls} \bar\sigma_{\mu\nu} d_{Rt}) + 4 (\bar d_{Lp} d_{Rr}) (\bar d_{Ls} d_{Rt}) + 16 (\bar d_{Lp} T^A d_{Rt}) (\bar d_{Ls} T^A d_{Rr}) + \frac{8}{N_c} (\bar d_{Lp} d_{Rt}) (\bar d_{Ls} d_{Rr})  \\
\EOp{dd}{RR}[(F2)8][] & \displaystyle (\bar d_{Lp} \bar\sigma^{\mu\nu} T^A d_{Rr}) (\bar d_{Ls} \bar\sigma_{\mu\nu} T^A d_{Rt}) + 4 (\bar d_{Lp} T^A d_{Rr}) (\bar d_{Ls} T^A d_{Rt}) + \frac{8C_F}{N_c} (\bar d_{Lp} d_{Rt}) (\bar d_{Ls} d_{Rr}) - \frac{8}{N_c} (\bar d_{Lp} T^A d_{Rt}) (\bar d_{Ls} T^A d_{Rr})  \\
\EOp{ud}{RR}[(F2)1][] & \displaystyle (\bar u_{Lp} \bar\sigma^{\mu\nu} u_{Rr}) (\bar d_{Ls} \bar\sigma_{\mu\nu} d_{Rt}) + 4 (\bar u_{Lp} u_{Rr}) (\bar d_{Ls} d_{Rt}) + 16 (\bar u_{Lp} T^A d_{Rt}) (\bar d_{Ls} T^A u_{Rr}) + \frac{8}{N_c} (\bar u_{Lp} d_{Rt}) (\bar d_{Ls} u_{Rr})  \\
\end{array}
\end{align*}
\end{minipage}
\end{adjustbox}

\caption{Fierz-evanescent LEFT operators that appear at one loop in the HV scheme and conserve baryon and lepton number. Light gray operators are not needed as divergent counterterms and we do not insert them into loops, but it is convenient to keep them for the extraction of finite counterterms.}
\label{tab:EvanescentFourFermionOperatorsFierz}
\end{table}

\begin{table}[H]
\capstart
\centering

\begin{adjustbox}{width=1.05\textwidth,center}
\begin{minipage}[t]{3cm}
\renewcommand{\arraystretch}{1.5}
\small
\begin{align*}
\begin{array}[t]{c|c}
\multicolumn{2}{c}{\boldsymbol{\Delta L =2 + \hc}}  \\
\hline
\EOp{\nu e}{LL}[T(1)][] &  (\nu_{Lp}^T C \hat\gamma^\mu \bar\sigma^{\nu\lambda} \nu_{Lr}) (\bar e_{Rs} \hat\gamma_\mu \bar\sigma_{\nu\lambda} e_{Lt} )  \\
\EOp{\nu u}{LL}[T(1)][] &  (\nu_{Lp}^T C \hat\gamma^\mu \bar\sigma^{\nu\lambda} \nu_{Lr}) (\bar u_{Rs} \hat\gamma_\mu \bar\sigma_{\nu\lambda} u_{Lt} )  \\
\EOp{\nu d}{LL}[T(1)][] &  (\nu_{Lp}^T C \hat\gamma^\mu \bar\sigma^{\nu\lambda} \nu_{Lr}) (\bar d_{Rs} \hat\gamma_\mu \bar\sigma_{\nu\lambda} d_{Lt} )  \\
\EOp{\nu edu}{LL}[S(1)][] &  (\nu_{Lp}^T C \hat\gamma^\mu e_{Lr}) (\bar d_{Rs} \hat\gamma_\mu u_{Lt} )  \\
\EOp{\nu edu}{LL}[S(2)][] &  (\nu_{Lp}^T C \hat\gamma^\mu \hat\gamma^\nu e_{Lr}) (\bar d_{Rs} \hat\gamma_\mu \hat\gamma_\nu u_{Lt} )  \\
\EOp{\nu edu}{LL}[T(1)][] &  (\nu_{Lp}^T C \hat\gamma^\mu \bar\sigma^{\nu\lambda} e_{Lr}) (\bar d_{Rs} \hat\gamma_\mu \bar\sigma_{\nu\lambda} u_{Lt} )  \\
\EOp{\nu edu}{LL}[T(2)][] &  (\nu_{Lp}^T C \hat\gamma^\mu \hat\gamma^\nu \bar\sigma^{\lambda\rho} e_{Lr}) (\bar d_{Rs} \hat\gamma_\mu \hat\gamma_\nu \bar\sigma_{\lambda\rho} u_{Lt} )  \\
\EOp{\nu edu}{LR}[S(1)][] &  (\nu_{Lp}^T C \hat\gamma^\mu e_{Lr}) (\bar d_{Ls} \hat\gamma_\mu u_{Rt} )  \\
\EOp{\nu edu}{LR}[S(2)][] &  (\nu_{Lp}^T C \hat\gamma^\mu \hat\gamma^\nu e_{Lr}) (\bar d_{Ls} \hat\gamma_\mu \hat\gamma_\nu u_{Rt} )  \\
\EOp{\nu edu}{RL}[V(1)][] &  (\nu_{Lp}^T C \hat\gamma^\mu \bar\gamma^\nu e_{Rr}) (\bar d_{Ls} \hat\gamma_\mu \bar\gamma_\nu u_{Lt} )  \\
\EOp{\nu edu}{RL}[V(2)][] &  (\nu_{Lp}^T C \hat\gamma^\mu \hat\gamma^\nu \bar\gamma^\lambda e_{Rr}) (\bar d_{Ls} \hat\gamma_\mu \hat\gamma_\nu \bar\gamma_\lambda u_{Lt} )  \\
\EOp{\nu edu}{RR}[V(1)][] &  (\nu_{Lp}^T C \hat\gamma^\mu \bar\gamma^\nu e_{Rr}) (\bar d_{Rs} \hat\gamma_\mu \bar\gamma_\nu u_{Rt} )  \\
\EOp{\nu edu}{RR}[V(2)][] &  (\nu_{Lp}^T C \hat\gamma^\mu \hat\gamma^\nu \bar\gamma^\lambda e_{Rr}) (\bar d_{Rs} \hat\gamma_\mu \hat\gamma_\nu \bar\gamma_\lambda u_{Rt} )  \\
\end{array}
\end{align*}
\end{minipage}
%
\begin{minipage}[t]{3cm}
\renewcommand{\arraystretch}{1.5}
\small
\begin{align*}
\begin{array}[t]{c|c}
\multicolumn{2}{c}{\boldsymbol{\Delta B = \Delta L = 1 + \hc}} \\
\hline
\EOp{udd}{LL}[S(T)][] &  \epsilon_{\alpha\beta\gamma}  (u_{Lp}^{\alpha T} C \hat\sigma^{\mu\nu} d_{Lr}^{\beta}) (d_{Ls}^{\gamma T} C \hat\sigma_{\mu\nu} \nu_{Lt}^{})   \\
\EOp{udd}{LL}[V(1)][] &  \epsilon_{\alpha\beta\gamma}  (u_{Rp}^{\alpha T} C \hat\gamma^\mu \bar\gamma^\nu d_{Lr}^{\beta}) (d_{Rs}^{\gamma T} C \hat\gamma_\mu \bar\gamma_\nu \nu_{Lt}^{})   \\
\EOp{dud}{LL}[V(1)][] & \epsilon_{\alpha\beta\gamma}  (d_{Rp}^{\alpha T} C \hat\gamma^\mu \bar\gamma^\nu u_{Lr}^{\beta}) (d_{Rs}^{\gamma T} C  \hat\gamma_\mu \bar\gamma_\nu \nu_{Lt}^{})   \\
\EOp{duu}{LL}[S(T)][] & \epsilon_{\alpha\beta\gamma}  (d_{Lp}^{\alpha T} C \hat\sigma^{\mu\nu} u_{Lr}^{\beta}) (u_{Ls}^{\gamma T} C \hat\sigma_{\mu\nu} e_{Lt}^{})  \\
\EOp{duu}{LL}[V(1)][] & \epsilon_{\alpha\beta\gamma}  (d_{Rp}^{\alpha T} C \hat\gamma^\mu \bar\gamma^\nu u_{Lr}^{\beta}) (u_{Rs}^{\gamma T} C \hat\gamma_\mu \bar\gamma_\nu e_{Lt}^{})  \\
\EOp{uud}{LR}[V(1)][] & \epsilon_{\alpha\beta\gamma}  (u_{Rp}^{\alpha T} C \hat\gamma^\mu \bar\gamma^\nu u_{Lr}^{\beta}) (d_{Ls}^{\gamma T} C \hat\gamma_\mu \bar\gamma_\nu e_{Rt}^{})  \\
\EOp{duu}{LR}[S(T)][] & \epsilon_{\alpha\beta\gamma}  (d_{Lp}^{\alpha T} C \hat\sigma^{\mu\nu} u_{Lr}^{\beta}) (u_{Rs}^{\gamma T} C \hat\sigma_{\mu\nu} e_{Rt}^{})   \\
\EOp{duu}{LR}[V(1)][] & \epsilon_{\alpha\beta\gamma}  (d_{Rp}^{\alpha T} C \hat\gamma^\mu \bar\gamma^\nu u_{Lr}^{\beta}) (u_{Ls}^{\gamma T} C \hat\gamma_\mu \bar\gamma_\nu e_{Rt}^{})   \\
\EOp{uud}{RL}[V(1)][] & \epsilon_{\alpha\beta\gamma}  (u_{Lp}^{\alpha T} C \hat\gamma^\mu \bar\gamma^\nu u_{Rr}^{\beta}) (d_{Rs}^{\gamma T} C \hat\gamma_\mu \bar\gamma_\nu e_{Lt}^{})   \\
\EOp{duu}{RL}[S(T)][] & \epsilon_{\alpha\beta\gamma}  (d_{Rp}^{\alpha T} C \hat\sigma^{\mu\nu} u_{Rr}^{\beta}) (u_{Ls}^{\gamma T} C \hat\sigma_{\mu\nu} e_{Lt}^{})   \\
\EOp{duu}{RL}[V(1)][] & \epsilon_{\alpha\beta\gamma}  (d_{Lp}^{\alpha T} C \hat\gamma^\mu \bar\gamma^\nu u_{Rr}^{\beta}) (u_{Rs}^{\gamma T} C \hat\gamma_\mu \bar\gamma_\nu e_{Lt}^{})   \\
\EOp{dud}{RL}[S(T)][] & \epsilon_{\alpha\beta\gamma}  (d_{Rp}^{\alpha T} C \hat\sigma^{\mu\nu} u_{Rr}^{\beta}) (d_{Ls}^{\gamma T} C  \hat\sigma_{\mu\nu} \nu_{Lt}^{})   \\
\EOp{ddu}{RL}[V(1)][] & \epsilon_{\alpha\beta\gamma}  (d_{Lp}^{\alpha T} C \hat\gamma^\mu \bar\gamma^\nu d_{Rr}^{\beta}) (u_{Rs}^{\gamma T} C \hat\gamma_\mu \bar\gamma_\nu \nu_{Lt}^{})   \\
\EOp{duu}{RR}[S(T)][]  & \epsilon_{\alpha\beta\gamma}  (d_{Rp}^{\alpha T} C \hat\sigma^{\mu\nu} u_{Rr}^{\beta}) (u_{Rs}^{\gamma T} C \hat\sigma_{\mu\nu} e_{Rt}^{})  \\
\EOp{duu}{RR}[V(1)][]  & \epsilon_{\alpha\beta\gamma}  (d_{Lp}^{\alpha T} C \hat\gamma^\mu \bar\gamma^\nu u_{Rr}^{\beta}) (u_{Ls}^{\gamma T} C \hat\gamma_\mu \bar\gamma_\nu e_{Rt}^{})  \\
\end{array}
\end{align*}
\end{minipage}
%
\begin{minipage}[t]{3cm}
\renewcommand{\arraystretch}{1.5}
\small
\begin{align*}
\begin{array}[t]{c|c}
\multicolumn{2}{c}{\boldsymbol{\Delta B = - \Delta L = 1 + \hc}}  \\
\hline
\EOp{ddd}{LL}[V(1)][] & \epsilon_{\alpha\beta\gamma}  (d_{Rp}^{\alpha T} C \hat\gamma^\mu \bar\gamma^\nu d_{Lr}^{\beta}) (\bar e_{Ls}^{} \hat\gamma_\mu \bar\gamma_\nu d_{Lt}^\gamma )  \\
\EOp{udd}{LR}[S(T)][]  & \epsilon_{\alpha\beta\gamma}  (u_{Lp}^{\alpha T} C \hat\sigma^{\mu\nu} d_{Lr}^{\beta}) (\bar \nu_{Ls}^{} \hat\sigma_{\mu\nu} d_{Rt}^\gamma )  \\
\EOp{ud\nu d}{LL}[V(1)][]  & \epsilon_{\alpha\beta\gamma}  (u_{Rp}^{\alpha T} C \hat\gamma^\mu \bar\gamma^\nu d_{Lr}^{\beta}) (\bar \nu_{Ls}^{} \hat\gamma_\mu \bar\gamma_\nu d_{Lt}^\gamma )  \\
\EOp{ddu}{LL}[V(1)][] & \epsilon_{\alpha\beta\gamma}  (d_{Rp}^{\alpha T} C \hat\gamma^\mu \bar\gamma^\nu d_{Lr}^{\beta})  (\bar \nu_{Ls}^{} \hat\gamma_\mu \bar\gamma_\nu u_{Lt}^\gamma )  \\
\EOp{ddd}{LR}[V(1)][] & \epsilon_{\alpha\beta\gamma}  (d_{Rp}^{\alpha T} C \hat\gamma^\mu \bar\gamma^\nu d_{Lr}^{\beta}) (\bar e_{Rs}^{} \hat\gamma_\mu \bar\gamma_\nu d_{Rt}^\gamma )  \\
\EOp{udd}{RR}[S(T)][]  & \epsilon_{\alpha\beta\gamma}  (u_{Rp}^{\alpha T} C \hat\sigma^{\mu\nu} d_{Rr}^{\beta}) (\bar \nu_{Ls}^{} \hat\sigma_{\mu\nu} d_{Rt}^\gamma )  \\
\EOp{ud\nu d}{RL}[V(1)][]  & \epsilon_{\alpha\beta\gamma}  (u_{Lp}^{\alpha T} C \hat\gamma^\mu \bar\gamma^\nu d_{Rr}^{\beta}) (\bar \nu_{Ls}^{} \hat\gamma_\mu \bar\gamma_\nu d_{Lt}^\gamma )  \\
\end{array}
\end{align*}
\end{minipage}
\end{adjustbox}

\caption{Evanescent four-fermion LEFT operators other than Fierz-evanescent ones that appear at one loop in the HV scheme and violate baryon and/or lepton number.}
\label{tab:EvanescentFourFermionOperatorsBL}
\end{table}

\begin{table}[H]
\capstart
\centering

\begin{adjustbox}{width=0.75\textwidth,center}
\begin{minipage}[t]{3cm}
\renewcommand{\arraystretch}{1.5}
\small
\begin{align*}
\begin{array}[t]{c|c}
\multicolumn{2}{c}{\boldsymbol{\Delta B = \Delta L = 1 + \hc}}  \\
\hline
\color{lightgray} \EOp{ddu}{LL}[(F)][] & \color{lightgray} \epsilon_{\alpha\beta\gamma} \Big[ ( d_{Lp}^{\alpha T} C d^\beta_{Lr}) (u_{Ls}^{\gamma T} C \nu_{Lt}) - (u_{Ls}^{\alpha T} C d_{Lr}^\beta) (d_{Lp}^{\gamma T} C \nu_{Lt}) + (u_{Ls}^{\alpha T} C d_{Lp}^\beta) (d_{Lr}^{\gamma T} C \nu_{Lt}) \Big]  \\
\color{lightgray} \EOp{uud}{LL}[(F)][] & \color{lightgray} \epsilon_{\alpha\beta\gamma} \Big[ ( u_{Lp}^{\alpha T} C u^\beta_{Lr}) (d_{Ls}^{\gamma T} C e_{Lt}) - (d_{Ls}^{\alpha T} C u_{Lr}^\beta) (u_{Lp}^{\gamma T} C e_{Lt}) + (d_{Ls}^{\alpha T} C u_{Lp}^\beta) (u_{Lr}^{\gamma T} C e_{Lt}) \Big]  \\
\color{lightgray} \EOp{uud}{RR}[(F)][] & \color{lightgray} \epsilon_{\alpha\beta\gamma} \Big[ ( u_{Rp}^{\alpha T} C u^\beta_{Rr}) (d_{Rs}^{\gamma T} C e_{Rt}) - (d_{Rs}^{\alpha T} C u_{Rr}^\beta) (u_{Rp}^{\gamma T} C e_{Rt}) + (d_{Rs}^{\alpha T} C u_{Rp}^\beta) (u_{Rr}^{\gamma T} C e_{Rt}) \Big]  \\
\color{lightgray} \EOp{ddu}{LL}[(F2)][] & \color{lightgray} \epsilon_{\alpha\beta\gamma} \Big[ ( d_{Lp}^{\alpha T} C \bar\sigma^{\mu\nu} d^\beta_{Lr}) (u_{Ls}^{\gamma T} C \bar\sigma_{\mu\nu} \nu_{Lt}) - 4 (u_{Ls}^{\alpha T} C d_{Lr}^\beta) (d_{Lp}^{\gamma T} C \nu_{Lt}) - 4 (u_{Ls}^{\alpha T} C d_{Lp}^\beta) (d_{Lr}^{\gamma T} C \nu_{Lt}) \Big]  \\
\color{lightgray} \EOp{uud}{LL}[(F2)][] & \color{lightgray} \epsilon_{\alpha\beta\gamma} \Big[ ( u_{Lp}^{\alpha T} C \bar\sigma^{\mu\nu} u^\beta_{Lr}) (d_{Ls}^{\gamma T} C \bar\sigma_{\mu\nu} e_{Lt}) - 4 (d_{Ls}^{\alpha T} C u_{Lr}^\beta) (u_{Lp}^{\gamma T} C e_{Lt}) - 4 (d_{Ls}^{\alpha T} C u_{Lp}^\beta) (u_{Lr}^{\gamma T} C e_{Lt}) \Big]  \\
\color{lightgray} \EOp{uud}{RR}[(F2)][] & \color{lightgray} \epsilon_{\alpha\beta\gamma} \Big[ ( u_{Rp}^{\alpha T} C \bar\sigma^{\mu\nu} u^\beta_{Rr}) (d_{Rs}^{\gamma T} C \bar\sigma_{\mu\nu} e_{Rt}) - 4 (d_{Rs}^{\alpha T} C u_{Rr}^\beta) (u_{Rp}^{\gamma T} C e_{Rt}) - 4 (d_{Rs}^{\alpha T} C u_{Rp}^\beta) (u_{Rr}^{\gamma T} C e_{Rt}) \Big]  \\
\EOp{duu}{LL}[(F2)][] & \epsilon_{\alpha\beta\gamma} \Big[ ( d_{Lp}^{\alpha T} C \bar\sigma^{\mu\nu} u^\beta_{Lr}) (u_{Ls}^{\gamma T} C \bar\sigma_{\mu\nu} e_{Lt}) - 4 (d_{Lp}^{\alpha T} C u_{Lr}^\beta) (u_{Ls}^{\gamma T} C e_{Lt}) + 8 (d_{Lp}^{\alpha T} C u_{Ls}^\beta ) (u_{Lr}^{\gamma T} C e_{Lt}) \Big]  \\
\EOp{duu}{RR}[(F2)][] & \epsilon_{\alpha\beta\gamma} \Big[ ( d_{Rp}^{\alpha T} C \bar\sigma^{\mu\nu} u^\beta_{Rr}) (u_{Rs}^{\gamma T} C \bar\sigma_{\mu\nu} e_{Rt}) - 4 (d_{Rp}^{\alpha T} C u_{Rr}^\beta) (u_{Rs}^{\gamma T} C e_{Rt}) + 8 (d_{Rp}^{\alpha T} C u_{Rs}^\beta ) (u_{Rr}^{\gamma T} C e_{Rt}) \Big]  \\
\EOp{udd}{LL}[(F2)][] & \epsilon_{\alpha\beta\gamma} \Big[ ( u_{Lp}^{\alpha T} C \bar\sigma^{\mu\nu} d^\beta_{Lr}) (d_{Ls}^{\gamma T} C \bar\sigma_{\mu\nu} \nu_{Lt}) - 4 (u_{Lp}^{\alpha T} C d_{Lr}^\beta) (d_{Ls}^{\gamma T} C \nu_{Lt}) + 8 (u_{Lp}^{\alpha T} C d_{Ls}^\beta) (d_{Lr}^{\gamma T} C \nu_{Lt}) \Big]  \\
\end{array}
\end{align*}
\end{minipage}
\end{adjustbox}

\begin{adjustbox}{width=0.75\textwidth,center}
\begin{minipage}[t]{3cm}
\renewcommand{\arraystretch}{1.5}
\small
\begin{align*}
\begin{array}[t]{c|c}
\multicolumn{2}{c}{\boldsymbol{\Delta B = - \Delta L = 1 + \hc}}  \\
\hline
\color{lightgray} \EOp{ddu}{RR}[(F)][] & \color{lightgray} \epsilon_{\alpha\beta\gamma} \Big[ ( d_{Rp}^{\alpha T} C d^\beta_{Rr}) (\bar\nu_{Ls} u_{Rt}^\gamma ) - (u_{Rt}^{\alpha T} C d_{Rr}^\beta) (\bar\nu_{Ls} d_{Rp}^\gamma ) + (u_{Rt}^{\alpha T} C d_{Rp}^\beta) (\bar\nu_{Ls} d_{Rr}^\gamma) \Big]  \\
\color{lightgray} \EOp{ddd}{LL}[(F)][] & \color{lightgray} \epsilon_{\alpha\beta\gamma} \Big[ ( d_{Lp}^{\alpha T} C d^\beta_{Lr}) (\bar e_{Rs} d_{Lt}^\gamma ) - (d_{Lt}^{\alpha T} C d_{Lr}^\beta) (\bar e_{Rs} d_{Lp}^\gamma ) + (d_{Lt}^{\alpha T} C d_{Lp}^\beta) (\bar e_{Rs} d_{Lr}^\gamma) \Big]  \\
\color{lightgray} \EOp{ddd}{RR}[(F)][] & \color{lightgray} \epsilon_{\alpha\beta\gamma} \Big[ ( d_{Rp}^{\alpha T} C d^\beta_{Rr}) (\bar e_{Ls} d_{Rt}^\gamma ) - (d_{Rt}^{\alpha T} C d_{Rr}^\beta) (\bar e_{Ls} d_{Rp}^\gamma ) + (d_{Rt}^{\alpha T} C d_{Rp}^\beta) (\bar e_{Ls} d_{Rr}^\gamma) \Big]  \\
\color{lightgray} \EOp{ddu}{RR}[(F2)][] & \color{lightgray} \epsilon_{\alpha\beta\gamma} \Big[ ( d_{Rp}^{\alpha T} C \bar\sigma^{\mu\nu} d^\beta_{Rr}) (\bar \nu_{Ls} \bar\sigma_{\mu\nu} u_{Rt}^\gamma) + 4 (u_{Rt}^{\alpha T} C d_{Rr}^\beta) (\bar \nu_{Ls} d_{Rp}^\gamma) + 4 (u_{Rt}^{\alpha T} C d_{Rp}^\beta ) (\bar \nu_{Ls} d_{Rr}^\gamma) \Big]  \\
\color{lightgray} \EOp{ddd}{LL}[(F2)][] & \color{lightgray} \epsilon_{\alpha\beta\gamma} \Big[ ( d_{Lp}^{\alpha T} C \bar\sigma^{\mu\nu} d^\beta_{Lr}) (\bar e_{Rs} \bar\sigma_{\mu\nu} d_{Lt}^\gamma) - 4 (d_{Lr}^{\alpha T} C d_{Lt}^\beta) (\bar e_{Rs} d_{Lp}^\gamma) - 4 (d_{Lp}^{\alpha T} C d_{Lt}^\beta ) (\bar e_{Rs} d_{Lr}^\gamma) \Big]  \\
\color{lightgray} \EOp{ddd}{RR}[(F2)][] & \color{lightgray} \epsilon_{\alpha\beta\gamma} \Big[ ( d_{Rp}^{\alpha T} C \bar\sigma^{\mu\nu} d^\beta_{Rr}) (\bar e_{Ls} \bar\sigma_{\mu\nu} d_{Rt}^\gamma) - 4 (d_{Rr}^{\alpha T} C d_{Rt}^\beta) (\bar e_{Ls} d_{Rp}^\gamma) - 4 (d_{Rp}^{\alpha T} C d_{Rt}^\beta ) (\bar e_{Ls} d_{Rr}^\gamma) \Big]  \\
\EOp{udd}{RR}[(F2)][] & \epsilon_{\alpha\beta\gamma} \Big[ ( u_{Rp}^{\alpha T} C \bar\sigma^{\mu\nu} d^\beta_{Rr}) (\bar \nu_{Ls} \bar\sigma_{\mu\nu} d_{Rt}^\gamma) + 4 (u_{Rp}^{\alpha T} C d_{Rr}^\beta) (\bar\nu_{Ls} d_{Rt}^\gamma) - 8 (u_{Rp}^{\alpha T} C d_{Rt}^\beta ) (\bar\nu_{Ls} d_{Rr}^\gamma) \Big]  \\
\end{array}
\end{align*}
\end{minipage}
\end{adjustbox}

\caption{Fierz-evanescent LEFT operators that appear at one loop in the HV scheme and violate baryon and/or lepton number. Light gray operators are not needed as divergent counterterms and we do not insert them into loops, but it is convenient to keep them for the extraction of finite counterterms.}
\label{tab:EvanescentFourFermionOperatorsFierzBL}
\end{table}